%% file: main.tex
\documentclass[a4paper, 11pt, oneside]{book}

\usepackage[backend=biber, style=phys, biblabel=brackets, maxbibnames  = 8, minbibnames  = 8, eprint=true]{biblatex}
\usepackage{dsfont}
\usepackage{hyperref}
\usepackage{xcolor}
\usepackage{physics}
\usepackage{graphicx}
\usepackage{framed} 
\usepackage{circuitikz}
\usetikzlibrary{decorations.pathreplacing}
\usepackage{amssymb}
\usepackage{relsize}
\usepackage{subcaption}
\usepackage{centernot}
\usepackage{import}
\usepackage{pgf}
\usepackage{emptypage}

\usepackage[includeheadfoot]{geometry}
\usepackage{fancyhdr} 
\fancypagestyle{plain}{%
\fancyhf{}%
\fancyfoot[LE,RO]{\thepage}%
\renewcommand{\headrulewidth}{0pt}%
\renewcommand{\footrulewidth}{0pt}}
\usepackage[bf]{titlesec}
\usepackage{tocloft}

\titleformat{\chapter}[hang]{\normalfont\huge\bfseries}{\newline\newline\thechapter.}{1em}{\huge}
\titlespacing{\chapter}{0pt}{100pt}{40pt}

\DeclareMathOperator{\diag}{diag}
\DeclareMathOperator{\myRe}{Re}
\DeclareMathOperator{\myIm}{Im}

\begin{document}


\frontmatter
\input{frontmatter/titlepage}
\newgeometry{
    twoside,
    top=1in,
    bottom=1in,
    outer=1in,
    inner=1in}
\setcounter{page}{1}
\input{frontmatter/abstract.tex}


\pagestyle{fancy}
\fancyhf{}
\fancyhead[RO]{\nouppercase{\rightmark}}
\fancyhead[le]{\nouppercase{\thechapter. \leftmark}}
\renewcommand{\chaptermark}[1]{\markboth{#1}{}}
\fancyfoot[ro, le]{\thepage}
\renewcommand{\headrulewidth}{0.4pt}
\renewcommand{\footrulewidth}{0.4pt}

\tableofcontents


\mainmatter
\input{mainmatter/ch1/chapter1.tex}
\input{mainmatter/ch2/chapter2.tex}
\input{mainmatter/ch3/chapter3.tex}

\input{mainmatter/ch4/chapter4.tex}
\input{mainmatter/ch5/chapter5.tex}


\input{backmatter/acknowledgements.tex}
\input{backmatter/appendix.tex}


\newpage
\printbibliography

\end{document}

%% file: frontmatter/titlepage.tex
\newgeometry{
    left=1.5cm,
    top=1.15cm,
    right=1.5cm,
    bottom=1.5cm,
    headheight=0cm,
    headsep=0cm
}

\begin{titlepage}
    \begin{center}

    \end{center}
    \begin{center}
    \vspace*{\fill}
    {\Huge \textbf{Efficiently Building and Characterizing Electromagnetic Models of Multi-Qubit Superconducting Circuits} \par }
    \vspace{2cm}
    {\large by \par}
    \vspace{1cm}
    {\LARGE Fadi Wassaf \par}
    \vspace{2.5cm}
    {\huge \textsc{Master's Thesis in Physics} \par}
    \vspace{2.5cm}
    {\large presented to \par }
    \vspace{0.5cm}
    {\large The Faculty of Mathematics, Computer Science \par and Natural Sciences \par}
    \vspace{0.5cm}
    {\large at \par}
    \vspace{0.5cm}
    {\Large \textsc{RWTH Aachen University} \par}
    \vspace{2cm}
    {\large Institute for Quantum Information \par}
    {\Large April 2024 \par}
    \vspace{1cm}
    {\large supervised by \par}
    \vspace{0.5cm}
    {\large Prof. Dr. David P. \textsc{DiVincenzo} \par Prof. Dr. Rami \textsc{Barends}}
    \vspace*{\fill}    
    \end{center}
\end{titlepage}

\restoregeometry
\newpage
\thispagestyle{empty}
\mbox{}
\newpage

%% file: frontmatter/abstract.tex
\section*{Abstract}
\thispagestyle{empty}

In an attempt to better leverage superconducting quantum computers, scaling efforts have become the central concern. These efforts have been further exacerbated by the increased complexity of these circuits. The added complexity can introduce parasitic couplings and resonances, which may hinder the overall performance and scalability of these devices. We explore a method of modeling and characterization based on multiport impedance functions that correspond to multi-qubit circuits. By combining vector fitting techniques with a novel method for interconnecting rational impedance functions, we are able to efficiently construct Hamiltonians for multi-qubit circuits using electromagnetic simulations. Our methods can also be applied to circuits that contain both lumped and distributed element components. The constructed Hamiltonians account for all the interactions within a circuit that are described by the impedance function. We then present characterization methods that allow us to estimate effective qubit coupling rates, state-dependent dispersive shifts of resonant modes, and qubit relaxation times. 

\newpage
\thispagestyle{empty}
\mbox{}
\newpage

%% file: mainmatter/ch1/chapter1.tex
\chapter{Introduction}

Over the past two decades, superconducting circuits have shown much promise toward the realization of viable quantum computers. Current efforts are dedicated to scaling these devices so that they may one day outperform classical computers in solving complex problems such as simulating quantum systems \cite{Feynman1982} or prime factorization \cite{prime}. 

Enhancing the capabilities of these devices necessitates an increase in the number of qubits used, while simultaneously ensuring that qubit lifetimes and gate fidelities remain high. Additional efforts of scaling superconducting quantum computers have resulted in the construction of qubit lattices, which aim to test quantum simulation methods \cite{evidence_utility,karamlou2023probing,Barends2016} and quantum error correcting codes with the intent to protect against physical errors \cite{google_surface, surface17}. Scaling these devices, however, will ultimately increase their complexity, leading to unexpected parasitic qubit interactions and decay channels that have not previously caused problems. For this reason, it is important to use modeling methods that expose and quantify the potential effects caused by parasitic coupling present in a circuit.

In this thesis, we explore methods for modeling and characterizing superconducting devices that utilize multiport impedance functions. The multiport impedance function contains all of the information about the coupling within a network. Naturally, the characterization of multi-qubit models derived from an impedance function provides us with the coupling rates between all the qubits and resonant modes within a circuit. This will include the effects of any parasitic resonances, stray coupling between qubits, and any contributions to control crosstalk. Using the impedance has the additional benefit that our characterization methods can be applied to electromagnetic simulations, and to circuit models containing both lumped elements (e.g. capacitors and inductors) and distributed elements (e.g. ideal transmission lines).

In Chapter \ref{chapter:lossless_impedance}, we will discuss the lossless reciprocal impedance function and its synthesis. The synthesis then leads to a method that can be used to generally interconnect rational impedance functions. This then allows us to build circuit models for qubit networks given an arbitrary impedance function. Chapter \ref{chapter:analysis_impedance} will explain how to characterize the resulting multi-qubit models by computing effective qubit coupling rates, dispersive shifts in resonant modes and qubit relaxation times. This characterization process is rather simple after obtaining the impedance function. Finally, in Chapter \ref{chapter:examples}, we examine how these methods are used for lumped and distributed element models, in addition to the full electromagnetic simulations. We also showcase a ``brick building” approach for electromagnetic modeling that can help decrease simulation time, thus making the simulation and characterization process more efficient. The models, data, and code implementation used for the examples within Chapter \ref{chapter:examples} are all available at \url{https://github.com/fadi-wassaf/msc-thesis}.

%% file: mainmatter/ch2/chapter2.tex
\chapter{Lossless Impedance Functions}\label{chapter:lossless_impedance}
We will begin by discussing immittance and scattering parameters that describe multiport networks. We focus mostly on the rational lossless impedance parameter, its synthesis, and network interconnections. We will also discuss a routine that can be used for reconstructing these rational parameters from electromagnetic simulation results.

\input{mainmatter/ch2/immittance_scattering.tex}
\input{mainmatter/ch2/lossless_impedance.tex}
\input{mainmatter/ch2/interconnects.tex}

\input{mainmatter/ch2/vector_fitting.tex}

%% file: mainmatter/ch2/immittance_scattering.tex
\section{Immittance and Scattering Response of Multiport Networks}

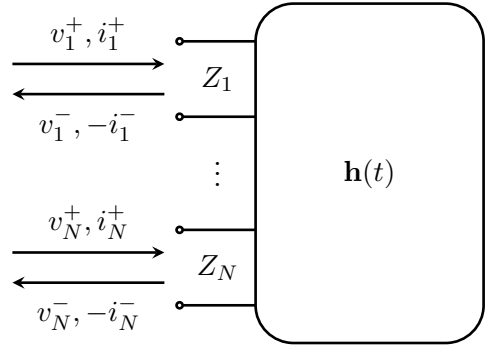
\begin{figure}[h!]
    \centering
    \begin{circuitikz}[line width=1pt]
        \ctikzset{american}
        \ctikzset{bipoles/thickness=1, bipoles/length=1cm}
        \ctikzset { label/align = straight }
        
        \draw[rounded corners=.5cm] (0,2) -- (0,4.5) -- (3,4.5) -- (3,0) -- (0,0) -- (0,2);
        \node at (1.5,2.25) {$\vb{h}(t)$};

        \draw (0,0.5) to[short, -o] (-1,0.5);
        \draw (0,1.5) to[short, -o] (-1,1.5);
        \node at (-.5,1) {$Z_N$};
        
        \draw [-stealth](-3.2, 1.2) -- (-1.2, 1.2);
        \node at (-2.2,1.6) {$v_N^+, i_N^+$};
        \draw [stealth-](-3.2, .8) -- (-1.2, .8);
        \node at (-2.2,.4) {$v_N^-, -i_N^-$};

        \draw (0,3) to[short, -o] (-1,3);
        \draw (0,4) to[short, -o] (-1,4);
        \node at (-.5,3.5) {$Z_1$};

        \draw [-stealth](-3.2, 3.7) -- (-1.2, 3.7);
        \node at (-2.2,4.1) {$v_1^+, i_1^+$};
        \draw [stealth-](-3.2, 3.3) -- (-1.2, 3.3);
        \node at (-2.2,2.9) {$v_1^-, -i_1^-$};

        \node at (-0.5,2.35) {$\vdots$};

    \end{circuitikz}
    \caption{A general $N$-port network with incident and reflected voltage and current waves. Each port consists of a two-terminal pair. The network response can be described by the $N \times N$ matrix $\vb{h}(t)$ which can be an impedance, admittance or scattering impulse response. The impedance and admittance relate the port voltages $v_k=v_k^+ + v_k^-$ and port currents $i_k=i_k^+ - i_k^-$. Each port also has a characteristic impedance $Z_k$ defined through the relationship between the port voltages and currents: $ Z_k = v_k^+/i_k^+ = v_k^-/i_k^-$.}
    \label{fig:general_multiport}
\end{figure}

In the following, we will be concerned with causal linear time-invariant (LTI) networks and their descriptions as used in \cite{passive_macromodeling, newcomb}. These networks are defined by the following properties:
\begin{itemize}
    \item {\bf\textit{Linearity}}: Linear combinations of inputs yield the same linear combinations for the output. Consider a set of signals $\vb{a}_i(t)$ with corresponding outputs $\vb{b}_i(t)$. For an input $\vb{a}(t) = \sum_i \alpha_i \vb{a}_i(t)$, the corresponding output will be $\vb{b}(t) = \sum_i \alpha_i \vb{b}_i(t)$.
    \item {\bf\textit{Time-Invariance}}: A time shift in the input results in the same time shift at the output. If you have an input $\vb{a}(t)$ with a known output $\vb{b}(t)$, you immediately know that for an input $\vb{a}(t-t_0)$, the output will be $\vb{b}(t-t_0)$.
    \item {\bf\textit{Causality}}: The output of the system only depends on the current and past inputs. In other words, the output at a given time $\vb{b}(t_0)$ will only be a function of $\vb{a}(t), \forall t \leq t_0$.
\end{itemize}
A multiple-input multiple-output (MIMO) system, can be described by the following input-output relation:
\begin{equation}\label{eq:lti_system}
    \vb{b}(t) = (\vb{h} \ast \vb{a})(t) = \int_{-\infty}^{+\infty} \vb{h}(t-\tau) \vb{a}(\tau)\; d\tau
\end{equation}
for some input vector $\vb{a}(t) \in \mathbb{R}^{n_a}$, output vector $\vb{b}(t) \in \mathbb{R}^{n_b}$ and an impulse response matrix $\vb{h}(t) \in \mathbb{R}^{n_b \times n_a}$. Note that for our electrical multiports, the number of inputs and outputs will be the equal so the impulse response matrix will be square. It is clear that (\ref{eq:lti_system}) describes an LTI system, but it does not enforce that the network is causal. This can be enforced by placing the following restriction on the impulse response vector:
\begin{equation}
    \vb{h}(t) = \vb{0},\quad \forall t < 0
\end{equation}
The above definitions allow us to describe multiport electrical networks using various impulse response functions. The most common representations of electrical multiport networks are the immittance (impedance and admittance) and scattering responses. Consider the general $N$-port electrical multiport structure pictured in Fig.\ \ref{fig:general_multiport}. The currents and voltages at the two-terminal ports can be related by the $N \times N$ impedance $\vb{z}(t)$ or admittance $\vb{y}(t)$ impulse response matrices such that:
\begin{align}
    \label{eq:z_time}\vb{v}(t) &= (\vb{z} \ast \vb{i})(t) \\
    \label{eq:y_time}\vb{i}(t) &= (\vb{y} \ast \vb{v})(t)
\end{align}
Alternatively, we can relate the incident ($\vb{v^+}$) and reflected ($\vb{v^-}$) voltage waves at the ports using the scattering response (the S-parameter can also be defined in terms of power waves; for more on this see \cite[Chapter 2.6.3]{passive_macromodeling} \cite[Chapter 4.3]{Pozar_2011}):
\begin{align}
    \label{eq:s_time}\vb{v^-}(t) &= (\vb{s} \ast \vb{v^+})(t)
\end{align}

In practice, when describing electrical multiports, the time-dependent impulse response matrices are not commonly used. Instead we look at the response matrices in the Laplace domain. This is defined by the bilateral Laplace transform
\begin{equation}
    X(s) = \int_{-\infty}^{\infty} x(t) e^{-st}\; dt
\end{equation}
with the complex frequency $s=\sigma + i\omega$. Under this transform, Eqs. (\ref{eq:z_time})-(\ref{eq:s_time}) become:
\begin{align}
    \label{eq:Z_s}\vb{V}(s) &= \vb{Z}(s)\vb{I}(s) \\
    \label{eq:Y_s}\vb{I}(s) &= \vb{Y}(s)\vb{V}(s) \\
    \label{eq:S_s}\vb{V}^-(s) &= \vb{S}(s)\vb{V}^+(s) 
\end{align}
When $\sigma=0$, we have $s=i\omega$ and in that case, the transformed impulse response functions $\vb{H}(i\omega)$ describe the AC response of the network (Fourier domain). The concept of complex frequency will be useful in understanding the form of these response functions. For example, the complex poles of an impedance function tell us about the resonant frequencies and decay rates of the resonant modes present in a circuit. 

The matrix elements of the Z-parameter can be understood as the ratio between the voltage measured at a port $i$ while driving a current at port $j$ while all other ports are open-circuited:
\begin{equation}
    Z_{ij}(s) = \frac{V_i(s)}{I_j(s)}, \quad I_k(s)=0, \quad\forall k\neq j
\end{equation}
This also suggests that if one has an impedance parameter or function, ports can be ``left open'' by neglecting the corresponding row and column of the Z-parameter. The matrix elements of the Y-parameter have a similar definition but instead the ratio is defined between the current measured at one port while driving another port with a voltage while all other ports are short circuited. The matrix elements of the S-parameter are defined by the ratio between a measured reflected voltage wave and an incident voltage wave such that
\begin{equation}
    S_{jk} = \frac{V^-_i(s)}{V^+_j(s)}, \quad V^+_k(s)=0, \quad\forall k\neq j
\end{equation}
We can also convert between the Z, Y and S-parameters when needed. From equations (\ref{eq:Z_s}) and (\ref{eq:Y_s}) it is clear that $\vb{Z}(s)=\vb{Y}^{-1}(s)$. When the characteristic impedances of the ports are all $Z_0$, we can obtain the S-parameter from the Z-parameter using: $\vb{S}(s) = (\vb{Z}(s) + Z_0\mathds{1})^{-1} (\vb{Z}(s) - Z_0\mathds{1}) $. For more on these conversions and cases when the characteristic impedances of the ports are not all equal, see \cite[Chapter 4]{Pozar_2011}.

For our circuit models, it will also be important to consider the following properties:
\begin{enumerate}
    \item {\bf \textit{Passivity}}: We can define an instantaneous power flow into a multiport using port voltages and currents:
    \begin{equation}
        p(t) = \vb{i}(t)^T \vb{v}(t)
    \end{equation}
    If $E(t)$ is the amount of energy stored in the network at a given time, a network will be passive if for all time intervals \cite[Chapter 9.1]{passive_macromodeling},
    \begin{equation}\label{eq:passivity_condition}
        E(t_1) \leq E(t_0) + \int_{t_0}^{t_1} p(t)\; dt
    \end{equation}
    \item {\bf \textit{Losslessness}}: A network will be lossless if (\ref{eq:passivity_condition}) holds with equality. The immittance matrices for lossless networks are always purely imaginary. Additionally, the scattering matrix of a lossless network will be unitary.
    \item {\bf \textit{Reciprocity}}: Immittance or scattering responses will be symmetric for reciprocal systems: $\vb{H}(s)=\vb{H}^T(s)$. From the Reciprocity theorem \cite[Chapter 16.4]{desoer_kuh}, which is a consequence of Tellegen's thereom, an electrical circuit with only capacitors, inductors, ideal transformers and resistors will be reciprocal.
\end{enumerate}

In this thesis, we may use combinations of lumped and distributed elements to make models of superconducting circuits. Alternatively, the models may be generated by electromagnetic simulations. Either way, we will assume that the models are lossless. Our electromagnetic models will consist of superconducting (or perfectly conducting) thin films on a dielectric substrate to implement capacitive or inductive circuit elements. This is done so that we can characterize the ideal interactions between qubits in a given model. We will later on only be considering losses to the environment through the external ports of the system. Any other losses are assumed to be small (i.e. quasiparticle, two-level systems in dielectrics (TLS), and radiation loss \cite{disentangling_losses}). Furthermore, we will not be including any permanent magnetic elements in our electromagnetic models that could potentially be used in isolator or circulator components. This will result in reciprocal models that we can synthesize using capacitors, inductors, ideal transformers (and resistors when including loss to the environment) \cite{tellegen_gyrator}.

%% file: mainmatter/ch2/lossless_impedance.tex
\newpage
\section{Lossless Impedance Functions}
\subsection{Hamiltonians Derived from Lossless Reciprocal Impedance Functions}\label{section:impedance_hamiltonian}

We will primarily be studying the partial fraction expansion of a general lossless impedance function and its synthesis that is presented in \cite[Chapter 7]{newcomb}. In the Laplace domain with $s=\sigma+i\omega$, the partial fraction expansion of a multiport lossless impedance function is given by
\begin{equation}\label{eq:general_impedance}
    \vb{Z}(s) = \frac{\vb{A}_0}{s} + s\vb{A}_\infty + \vb{B}_\infty + \sum_{k=1}^{M} \frac{s\vb{A}_k + \vb{B}_k}{s^2 + \omega_{R_k}^2}
\end{equation}
where the $\vb{A}$ matrices are real, positive semidefinite, and symmetric, and the $\vb{B}$ matrices are real and antisymmetric. The resonant modes of the network have frequency $\omega_{R_k}$. In our modeling of superconducting circuits, we will make the following assumptions that will put some restrictions on (\ref{eq:general_impedance}) which reduce it to a simpler form:
\begin{enumerate}
    \item Our models will have a reciprocal response such that all matrices $\vb{B}_i = 0$. It is clear that in this case, $\vb{Z}(s)^T=\vb{Z}(s)$.
    \item Each port of our network will have a non-zero shunt capacitance. This is physically motivated and the same assumption is made in \cite{solgun_sirf,sherbrooke_sirf}. The result is that the DC residue $\vb{A}_0$ is of full-rank and positive definite.
    \item The poles of the network at infinite frequency can be reasonably well approximated by finite frequency poles that are outside of the relevant frequency range of our devices. Thus, the residue $\vb{A}_\infty$ can be treated as a finite frequency pole. Inclusion of the infinite frequency residue will lead to a circuit Lagrangian with a singular capacitance matrix for which we cannot write a Hamiltonian (see Appendix \ref{appendix:infty_freq_residue}).
    \item A residue $\vb{A}_k$ corresponding to resonant mode $k$ is restricted to being rank-1. If degenerate resonant modes are present, multiple rank-1 residues can be associated with a single resonant frequency. Generally, parasitic or nonsymmetric coupling in physical models will break these degeneracies (see Appendix \ref{appendix:degen_res_mode}).
\end{enumerate}
The assumptions listed are based on the physical models of the devices that we will consider as explained in the previous section. With the above assumptions in mind, we now have the following lossless reciprocal impedance function:
\begin{equation}\label{eq:impedance}
    \vb{Z}(s) = \frac{\vb{R}_0}{s} + \sum_{k=1}^M \frac{s \vb{R}_k}{s^2 + \omega_{R_k}^2}
\end{equation}
An $N$-port impedance function (\ref{eq:impedance}) can be synthesized using a multiport canonical Cauer circuit as shown in Fig.\ \ref{fig:cauer_circuit} in the same way as \cite{solgun_sirf}. The following relationships link the residues and resonant frequencies of the impedance function (\ref{eq:impedance}) to the lumped elements of the synthesized circuit in Fig.\ \ref{fig:cauer_circuit}:
\begin{align}
    \vb{C}_0 &= \diag(C_1, \dots, C_N) \\
    \vb{R}_0 &=  \vb{U} \vb{C}_0^{-1} \vb{U}^T \label{eq:DC_residue}\\
    \vb{r}_k &= (r_{k1}, \dots, r_{kN}) \label{eq:row_R}\\
    \vb{R}_k &= \vb{r}_k^T \vb{r}_k \\
    L_{R_k} &= 1/\omega_{R_k}^2 \\
    C_{R_k} &= 1 \label{eq:C_R}
\end{align}

\definecolor{nodecolor}{HTML}{990000}
\definecolor{junctioncolor}{HTML}{224466}
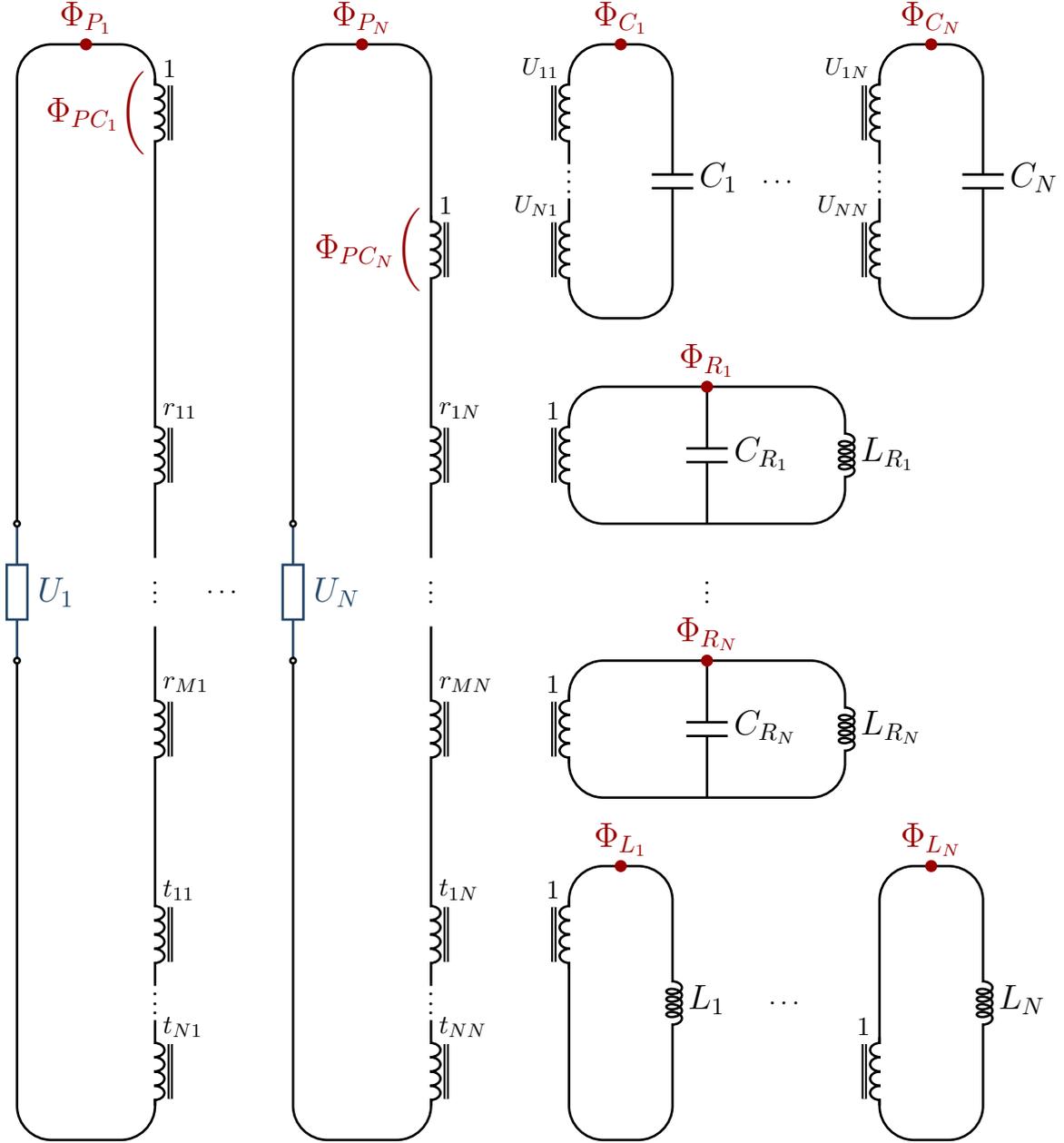
\begin{figure}[!p]
    \centering
    \begin{circuitikz}[line width=1pt]
    \ctikzset{american}
    \ctikzset{bipoles/thickness=1, bipoles/length=1cm}
    \ctikzset { label/align = straight }

    {\color{junctioncolor} 
    \draw (0,-1) to[generic, color=junctioncolor] (0,1);
    \node[anchor=west] at (0.15,0) {\Large $U_1$}; 
    }
    \draw (2,7.5) to[L, name=TJC1] (2,6.5) -- (2,2.5) to[L, name=r11] (2,1.5) -- (2,0.5);
    \draw (2,-0.5) -- (2,-1.5) to[L, name=rm1] (2,-2.5) -- (2,-4.5) to[L, name=t11] (2,-5.5) -- (2,-5.75);
    \draw (2,-6.25) -- (2,-6.5) to[L, name=tn1] (2,-7.5);
    \draw[rounded corners=.5cm] (2,7.4) -- (2,8) -- (0,8) to[short, -o] (0,1);
    \draw[rounded corners=.5cm] (2,-7.4) -- (2,-8) -- (0,-8) to[short, -o] (0,-1);
    
    \node[circle, fill=nodecolor, inner sep=0pt,minimum size=5pt, label={[label distance=-0.1cm]above:{\Large \color{nodecolor} $\Phi_{P_1}$}}] at (1,8) {};
    
    \node[anchor=east] at (2,7) {\Large\color{nodecolor} $\Phi_{PC_1}\bigg($};
    
    \draw[thick, double] (TJC1.core east) -- (TJC1.core west);
    \draw[thick, double] (r11.core east) -- (r11.core west);
    \draw[thick, double] (rm1.core east) -- (rm1.core west);
    \draw[thick, double] (t11.core east) -- (t11.core west);
    \draw[thick, double] (tn1.core east) -- (tn1.core west);
    
    \node at (2,.1) {$\vdots$};
    \node at (2,-5.9) {$\vdots$};
    \node at (3,0) {\dots};
    \node[anchor=west] at (1.97,7.65) {1};
    \node[anchor=west] at (1.97,2.65) {$r_{11}$};
    \node[anchor=west] at (1.97,-1.35) {$r_{M1}$};
    \node[anchor=west] at (1.97,-4.35) {$t_{11}$};
    \node[anchor=west] at (1.97,-6.35) {$t_{N1}$};

    {\color{junctioncolor}
    \draw (4,-1) to[generic, color=junctioncolor] (4,1);
    \node[anchor=west] at (4.15,0) {\Large $U_N$};}
    \draw (6,5.5) to[L, name=TJCN] (6,4.5) -- (6,2.5) to[L, name=r1n] (6,1.5) -- (6,0.5);
    \draw (6,-0.5) -- (6,-1.5) to[L, name=rmn] (6,-2.5) -- (6,-4.5) to[L, name=t1n] (6,-5.5) -- (6, -5.75);
    \draw (6, -6.25) -- (6,-6.5) to[L, name=tnn] (6,-7.5);
    \draw[rounded corners=.5cm] (6,5.4) -- (6,8) -- (4,8) to[short, -o] (4,1);
    \draw[rounded corners=.5cm] (6,-7.4) -- (6,-8) -- (4,-8) to[short, -o] (4,-1);
    
    \node[circle, fill=nodecolor, inner sep=0pt,minimum size=5pt, label={[label distance=-0.1cm]above:{\Large \color{nodecolor} $\Phi_{P_N}$}}] at (5,8) {};
    
    \node[anchor=east] at (6,5) {\Large\color{nodecolor} $\Phi_{PC_N}\bigg($};

    \draw[thick, double] (TJCN.core east) -- (TJCN.core west);
    \draw[thick, double] (r1n.core east) -- (r1n.core west);
    \draw[thick, double] (rmn.core east) -- (rmn.core west);
    \draw[thick, double] (t1n.core east) -- (t1n.core west);
    \draw[thick, double] (tnn.core east) -- (tnn.core west);

    \node at (6,.1) {$\vdots$};
    \node at (6,-5.9) {$\vdots$};
    \node[anchor=west] at (5.97,5.65) {1};
    \node[anchor=west] at (5.97,2.65) {$r_{1N}$};
    \node[anchor=west] at (5.97,-1.35) {$r_{MN}$};
    \node[anchor=west] at (5.97,-4.35) {$t_{1N}$};
    \node[anchor=west] at (5.97,-6.35) {$t_{NN}$};

    \draw (8,6.25) -- (8,6.5) to[L, name=u11] (8, 7.5);
    \draw (8,5.75) -- (8,5.5) to[L, name=un1, mirror] (8,4.5);
    \draw[rounded corners=.5cm] (8, 7.4) -- (8,8) -- (9.5,8) to[C={\Large $C_1$}, name=c1] (9.5,4) -- (8,4) -- (8,4.6);
    \node[circle, fill=nodecolor, inner sep=0pt,minimum size=5pt, label={[label distance=-0.1cm]above:{\Large \color{nodecolor} $\Phi_{C_1}$}}] at (8.75,8) {};
    \node[anchor=east] at (8,7.65) {\small $U_{11}$};
    \node[anchor=east] at (8,5.65) {\small $U_{N1}$};
    \draw[thick, double] (u11.core east) -- (u11.core west);
    \draw[thick, double] (un1.core east) -- (un1.core west);
    \node at (8,6.1) {\vdots};

    \node at (11.05,6) {\dots};

    \draw (12.5,6.25) -- (12.5,6.5) to[L, name=u11] (12.5, 7.5);
    \draw (12.5,5.75) -- (12.5,5.5) to[L, name=un1, mirror] (12.5,4.5);
    \draw[rounded corners=.5cm] (12.5, 7.4) -- (12.5,8) -- (14,8) to[C={\Large $C_N$}, name=c1] (14,4) -- (12.5,4) -- (12.5,4.6);
    \node[circle, fill=nodecolor, inner sep=0pt,minimum size=5pt, label={[label distance=-0.1cm]above:{\Large \color{nodecolor} $\Phi_{C_N}$}}] at (13.25,8) {};
    \node[anchor=east] at (12.5,7.65) {\small $U_{1N}$};
    \node[anchor=east] at (12.5,5.65) {\small $U_{NN}$};
    \draw[thick, double] (u11.core east) -- (u11.core west);
    \draw[thick, double] (un1.core east) -- (un1.core west);
    \node at (12.5,6.1) {\vdots};

    \draw (8,1.5) to[L, name=res1] (8,2.5);
    {
    \ctikzset{cute}
    \draw[rounded corners=.5cm] (8,2.4) -- (8,3) -- (11,3) -- (12,3) to[L={\Large $L_{R_1}$}] (12,1) -- (8,1) -- (8,1.6);
    }
    \draw (10,3) to[C={\Large $C_{R_1}$}] (10,1);
    \node[circle, fill=nodecolor, inner sep=0pt,minimum size=5pt, label={[label distance=-0.1cm]above:{\Large \color{nodecolor} $\Phi_{R_1}$}}] at (10,3) {};

    \draw[thick, double] (res1.core west) -- (res1.core east);
    \node[anchor=east] at (8.0125,2.65) {1};
    
    \draw (8,-2.5) to[L, name=resN] (8,-1.5);
    {
    \ctikzset{cute}
    \draw[rounded corners=.5cm] (8,-1.6) -- (8,-1) -- (12,-1) to[L={\Large $L_{R_N}$}] (12,-3) -- (11,-3) -- (8,-3) -- (8,-2.4);
    }
    \draw (10,-1) to[C={\Large $C_{R_N}$}] (10,-3);
    \node[circle, fill=nodecolor, inner sep=0pt,minimum size=5pt, label={[label distance=-0.1cm]above:{\Large \color{nodecolor} $\Phi_{R_N}$}}] at (10,-1) {};

    \draw[thick, double] (resN.core west) -- (resN.core east);
    \node[anchor=east] at (8.0125,-1.35) {1};
    \node at (10,.1) {$\vdots$}; 

    \draw (8, -5.5) to[L, name=L1] (8, -4.5);
    {
    \ctikzset{cute}
    \draw[rounded corners=.5cm] (8,-5.4) -- (8,-8) -- (9.5,-8) to[L,l_={\Large $L_1$}, mirror] (9.5,-4) -- (8,-4) -- (8,-4.6);
    }
    \draw[thick, double] (L1.core east) -- (L1.core west);
    \node[circle, fill=nodecolor, inner sep=0pt,minimum size=5pt, label={[label distance=-0.1cm]above:{\Large \color{nodecolor} $\Phi_{L_1}$}}] at (8.75,-4) {};
    \node[anchor=east] at (8.0125,-4.35) {1};
    \node at (11.15,-6) {\dots};

    \draw (12.5, -7.5) to[L, name=LN] (12.5, -6.5);
    {
    \ctikzset{cute}
    \draw[rounded corners=.5cm] (12.5,-7.4) -- (12.5,-8) -- (14,-8) to[L,l_={\Large $L_N$}, mirror] (14,-4) -- (12.5,-4) -- (12.5,-6.6);
    }
    \draw[thick, double] (LN.core east) -- (LN.core west);
    \node[circle, fill=nodecolor, inner sep=0pt,minimum size=5pt, label={[label distance=-0.1cm]above:{\Large \color{nodecolor} $\Phi_{L_N}$}}] at (13.25,-4) {};
    \node[anchor=east] at (12.5125,-6.35) {1};
\end{circuitikz}
\caption{The canonical Cauer circuit that represents the synthesis of the impedance function (\ref{eq:impedance}) with the ports shunted by potentials that are functions of the port fluxes: $U_i(\Phi_{J_i})$. The multiport transformer with turns ratios in matrix $\vb{U}$ couple the ports of the network to the purely capacitive stage that is related to the DC residue $\vb{R}_0$. The turns ratio matrix $\vb{R}$ represents the multiport transformer that couples the ports to the resonant modes. The residue at infinite frequency corresponds to the purely inductive stage which is included here to help motivate why it can be neglected. In practice, parasitic capacitances will bring the infinite frequency pole to finite frequency and thus this purely inductive stage can be included in the resonant stage. In Appendix \ref{appendix:infty_freq_residue}, we can also see that if one attempts to construct a Hamiltonian when the purely inductive stage is present, the resulting capacitance matrix is singular.}
\label{fig:cauer_circuit}
\end{figure}

$\vb{U}$ and $\vb{R}$ correspond to the turns ratios of the multiport Belevitch transformers that correspond to the purely capacitive and resonant stages, respectively, as shown in Fig.\ \ref{fig:cauer_circuit}. Furthermore, $\vb{r}_k$ in (\ref{eq:row_R}) is defined as row $k$ of the turns ratio matrix $\vb{R}$. This definition of the residues $\vb{R}_k$ guarantees that they are rank-1.

Our goal is to make a quantum mechanical model of the lumped element circuit that corresponds to the impedance function. Before doing this, it is first worth discussing how this is generally done for lumped element circuits in the context of circuit-QED \cite{vool_devoret,cqed_lecture_notes}. 

\begin{figure}[!h]
    \centering
    \begin{subfigure}{.4\textwidth}
        \centering
        \begin{circuitikz}[line width=1.5pt]
            \ctikzset{american}
            \ctikzset{bipoles/thickness=1, bipoles/length=2cm}
            \ctikzset { label/align = straight }

            \draw[color=white] (0,4) to (0,-0.25);

            \draw (0,3) to[generic] (0,0);

            \node[circle, fill=nodecolor, inner sep=0pt,minimum size=5pt] at (0,3) {};
            \node[circle, fill=nodecolor, inner sep=0pt,minimum size=5pt] at (0,0) {};

            \node[color=nodecolor] at (-1,3) {\LARGE $\boldsymbol{+}$};
            \node[color=nodecolor] at (-1,0) {\LARGE $\boldsymbol{-}$};
            \node[color=nodecolor] at (-1,1.5) {\Large $v_{b}$};
            
            \draw [color=nodecolor] [-stealth] (0.6,2.25) to (0.6,0.75);
            \node[color=nodecolor] at (1,1.5) {\Large $i_{b}$};

            \node[color=nodecolor] at (0,1.5) {$b$};

        \end{circuitikz}
        \caption{}
        \label{fig:branch_element}
    \end{subfigure}%
    \begin{subfigure}{.4\textwidth}
        \centering
        \begin{circuitikz}[line width=1.5pt]
            \ctikzset{american}
            \ctikzset{bipoles/thickness=1, bipoles/length=2cm}
            \ctikzset { label/align = straight }
            \draw[color=white] (0,-1.75) to (0,2.5);
            \draw (0,0) node[transformer core] (T) {};
            \node[circle, fill=nodecolor, inner sep=0pt,minimum size=5pt, label={[label distance=0cm]above:{\Large \color{nodecolor} $i_{b_1}$}}] at (-1.5,1.5) {};
            \draw[color=nodecolor] [-stealth](-1.15,2) to (-.75,2);
            \node[circle, fill=nodecolor, inner sep=0pt,minimum size=5pt] at (-1.5,-1.5) {};
            \node[circle, fill=nodecolor, inner sep=0pt,minimum size=5pt, label={[label distance=0cm]above:{\Large \color{nodecolor} $i_{b_2}$}}] at (1.5,1.5) {};
            \node[circle, fill=nodecolor, inner sep=0pt,minimum size=5pt] at (1.5,-1.5) {};
            \draw[color=nodecolor] [-stealth](1.15,2) to (.75,2);
            
            \node[color=nodecolor] at (-1.5,1) {\LARGE $\boldsymbol{+}$};
            \node[color=nodecolor] at (-1.5,-1) {\LARGE $\boldsymbol{-}$};
            \node[color=nodecolor] at (-1.5,0) {\Large $v_{b_1}$};    

            \node[color=nodecolor] at (1.5,1) {\LARGE $\boldsymbol{+}$};
            \node[color=nodecolor] at (1.5,-1) {\LARGE $\boldsymbol{-}$};
            \node[color=nodecolor] at (1.5,0) {\Large $v_{b_2}$};

            \node at (0,1.1) {\scriptsize $N_1:N_2$};

        \end{circuitikz}
        \caption{}
        \label{fig:ideal_transformer}
    \end{subfigure}
    \caption{(a) General lumped-element branch $b$. (b) Two-branch ideal transformer element with turns ratio $t=N_2/N_1$.}
    \label{fig:branch_and_transformer}
\end{figure}
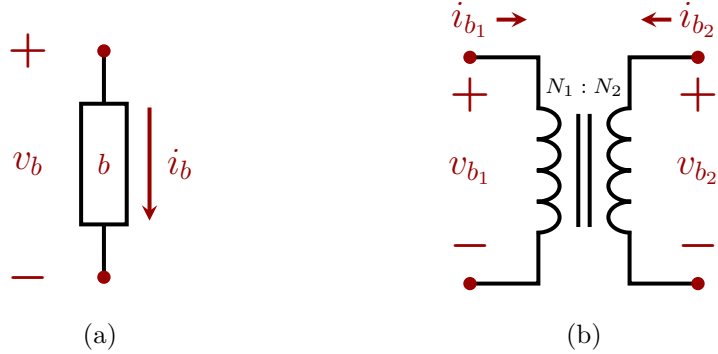

Consider a general lumped element as pictured in Fig.\ \ref{fig:branch_element} with the voltage across the branch $v_b(t)$. This is defined as the differences between the node voltages at the top and bottom of the pictured branch. To make a quantum mechanical model of a circuit with branch elements as shown, we would like to obtain a Hamiltonian that we can quantize. To do this, we will first generally find a Lagrangian for our circuit to which we apply a Legendre transformation. The degrees of freedom for this Lagrangian are flux variables that are defined in terms of branch or node voltages:
\begin{equation}
    \Phi_b(t) = \int_{-\infty}^t v_b(t')\; dt'
\end{equation}
We assume that the electromagnetic fields are not present for $t \rightarrow -\infty$. Using the flux variables corresponding to branches of a lumped element circuit, we can write a Lagrangian $\mathcal{L} = T - U$ that contains kinetic and potential energy terms for various elements. Throughout this thesis we will primarily be concerned with three two-terminal elements: capacitors, inductors, and Josephson junctions. The capacitor and inductor are linear elements, while the Josephson junction is nonlinear. The kinetic and potential energies of branches with a capacitor $C$, inductor $L$ and Josephson junction are defined in terms of branch fluxes as:
\begin{align}
    T_C(t) &= \frac{1}{2}C\dot{\Phi}_b^2(t) \\
    U_L(t) &= \frac{1}{2L}\Phi_b^2(t) \\ 
    U_J(t) &= -E_J \cos(\frac{2\pi}{\Phi_0} \Phi_b)
\end{align}
where $\Phi_0=h/2e$ is the superconducting flux quantum, and $E_J=\Phi_0 I_c/2\pi$ is the Josephson junction energy defined in terms of its critical current. In our lumped element circuits, we will also consider ideal transformers as shown in Fig.\ \ref{fig:ideal_transformer}. This element constrains the two branch voltages across its terminals such that $v_{b_2} = tv_{b_1}$ where $t=N_2/N_1 \in \mathbb{R}$ is the turns ratio of the ideal transformer. This places a similar constraint on the branch fluxes: $\Phi_{b_2} = t\Phi_{b_1}$.

Using the above potential and kinetic energies for capacitors and inductors, as well as the relationship between two branches coupled with an ideal transformer, we can find a circuit Hamiltonian that will later on allow us to characterize the behavior of qubit networks. We will arrive at the same Hamiltonian presented in \cite{solgun_sirf} where a different method was used to treat the multiport transformers \cite{burkard_circuit_2005,multiport_impedance_quantization}. To start, we write the Lagrangian for the synthesized circuit of in Fig.\ \ref{fig:cauer_circuit} using the node fluxes of the purely capacitive stage $\Phi_{C_i}$, the resonant stage $\Phi_{R_i}$, and the elements shunting the ports $\Phi_{P_i}$:
\begin{equation}\label{eq:impedance_lagrangian}
    \mathcal{L} = \frac{1}{2}\dot{\vb{\Phi}}_C^T \vb{C}_0 \dot{\vb{\Phi}}_C^{\phantom{^T}} + \frac{1}{2}\dot{\vb{\Phi}}_R^T \vb{C}_R \dot{\vb{\Phi}}_R^{\phantom{^T}} - \frac{1}{2}\vb{\Phi}_R^T \vb{M}_R \vb{\Phi}_R^{\phantom{^T}} - \sum_{i=1}^N U_i(\Phi_{P_i})
\end{equation}
where $\vb{C}_R = \diag(C_{R_1}, \dots, C_{R_M})$, $\vb{M}_R = \diag(L_{R_1}^{-1}, \dots, L_{R_M}^{-1})$, and the elements of the $\vb{\Phi}$ vectors contain the node fluxes of capacitive and resonant stages in the Cauer circuit. The potential functions $U_i(\Phi_{P_i})$ correspond to lumped elements that shunt the ports of the impedance with some arbitrary potential. This arbitrary potential can correspond to both linear or nonlinear elements. The next step is to make a substitution that returns a Lagrangian dependent only on the port fluxes $\Phi_{P_i}$ and the resonant stage fluxes $\Phi_{R_i}$. The substitution that we make invokes the constraint that an ideal transformer puts on the two branches that it couples. For example, the constraint between the flux variable of the branch $\Phi_{C_1}$ and the ideal transformer branches at the ports gives:
\begin{equation}
    \Phi_{C_1} = U_{11}\Phi_{PC_1} + \dots + U_{N1}\Phi_{PC_N}
\end{equation}
Using this reasoning for the rest of the branches coupled using ideal transformers, we find the following constraints for the branch and node flux vectors:
\begin{align}
    \vb{\Phi}_C &= \vb{U}^T \vb{\Phi}_{PC} \\
    \vb{\Phi}_{PC} &= \vb{\Phi}_P - \vb{R}^T \vb{\Phi}_R
\end{align}
This allows for the substitution $\vb{\Phi}_C \rightarrow \vb{U}^T (\vb{\Phi}_P - \vb{R}^T\vb{\Phi}_R)$, so we can eliminate $\vb{\Phi}_C$ from the first term in (\ref{eq:impedance_lagrangian}):
\begin{equation}
    \dot{\vb{\Phi}}_C^T \vb{C}_0 \dot{\vb{\Phi}}_C^{\phantom{^T}} = \dot{\vb{\Phi}}_P^T \vb{R}_0^{-1} \dot{\vb{\Phi}}_P^{\phantom{^T}} - \dot{\vb{\Phi}}_P^{T} \vb{R}_0^{-1} \vb{R}^T \dot{\vb{\Phi}}_R^{\phantom{^T}} - \dot{\vb{\Phi}}_R^{T} \vb{R}^T \vb{R}_0^{-1} \dot{\vb{\Phi}}_P^{\phantom{T}} + \dot{\vb{\Phi}}_R^{T} \vb{R} \vb{R}_0^{-1} \vb{R}^T \dot{\vb{\Phi}}_R^{\phantom{^T}}
\end{equation}
Above we have also eliminated $\vb{C}_0$ using (\ref{eq:DC_residue}). Applying a Legendre transformation to the resulting Lagrangian will yield the following circuit Hamiltonian in agreement with \cite{solgun_sirf}:
\begin{equation}\label{eq:impedance_hamiltonian}
    \mathcal{H} = \frac{1}{2} \vb{Q}^T \vb{C}^{-1} \vb{Q} + \frac{1}{2} \vb{\Phi}^T \vb{M} \vb{\Phi} + \sum_{i=1}^N U_i(\Phi_{P_i})
\end{equation}
where we have now defined the conjugate variables $Q_{P_i} = \partial \mathcal{L}/\partial \Phi_{P_i}$ and the following vectors and matrices:
\begin{align}
    \vb{\Phi} &= (\Phi_{P_1}, \dots, \Phi_{P_N}, \Phi_{R_1}, \dots, \Phi_{R_M})^T \\
    \vb{Q} &= (Q_{P_1}, \dots, Q_{P_N}, Q_{R_1}, \dots, Q_{R_M})^T \\
    \vb{C} &= \mqty( \vb{R}_0^{-1} & -\vb{R}_0^{-1}\vb{R}^T \\ -\vb{R} \vb{R}_0^{-1} & \vb{C}_R + \vb{R}\vb{R}_0^{-1}\vb{R}^T) \label{eq:impedance_cap}\\
    \vb{M} &= \mqty(\vb{0}_{N\times N} & \vb{0}_{N\times M} \\ \vb{0}_{M\times N} & \vb{M}_R)
\end{align}
We expect that a capacitance matrix should always be positive definite \cite{pos_cap}. For our choice of $\vb{C_R} = \mathds{1}_{M\times M}$ in (\ref{eq:C_R}), $\vb{C}$ will always be positive definite ($\vb{C} > 0$). This is the case since $\vb{R}_0 > 0$ and the Schur complement $\vb{C}/\vb{R}_0^{-1}=\mathds{1}_{M\times M} > 0$ \cite[Theorem 1.12]{schur_comp}. Also, since $\vb{C} > 0$, it is invertible, which allows us to write the final Hamiltonian (\ref{eq:impedance_hamiltonian}). Computing the inverse of $\vb{C}$, we find \cite[Proposition 2.8.7]{matrix_mathematics}:
\begin{equation}\label{eq:impedance_cap_inverse}
    \vb{C}^{-1} = \mqty( 
        \vb{R}_0 + \vb{R}^T \vb{C}_R^{-1} \vb{R} & \vb{R}^T \vb{C}_R^{-1} \\ 
        \vb{C}_R^{-1} \vb{R} & \vb{C}_R^{-1}
    ) = 
    \mqty( 
        \vb{R}_0 + \vb{R}^T \vb{R} & \vb{R}^T  \\ 
         \vb{R} & \mathds{1}_{M\times M}
    )
\end{equation}

\newpage
\subsection{Cascade Synthesis of the Lossless Reciprocal Impedance Function}\label{section:cascade_synthesis}
\begin{figure}[h!]
    \centering
    \begin{circuitikz}[line width=1pt]
        \ctikzset{bipoles/thickness=1, bipoles/length=1cm}
        \ctikzset { label/align = straight }
        
        \draw[rounded corners=.5cm] (0,2) -- (0,4.5) -- (7,4.5) -- (7,0) -- (0,0) -- (0,2);
        \node at (3.5,2.25) {$\vb{C} = \mqty( \vb{R}_0^{-1} & -\vb{R}_0^{-1}\vb{R}^T \\ -\vb{R} \vb{R}_0^{-1} & \mathds{1}_{M\times M} + \vb{R}\vb{R}_0^{-1}\vb{R}^T)$};

        \draw[rounded corners=0.5cm] (9,2) -- (9,4.5) -- (11,4.5) -- (11,0) -- (9,0) -- (9,2);
        \draw (9,3) -- (9.6,3) to[L,l_=$\dfrac{1}{\omega_{R_1}^2}$, mirror, label distance=0.25cm] (9.6,4) -- (9,4);
        \draw (9,0.5) -- (9.6,0.5) to[L,l_=$\dfrac{1}{\omega_{R_M}^2}$, mirror, label distance=0.25cm] (9.6,1.5) -- (9,1.5);

        \draw (0,0.5) to[short, -o] (-1,0.5);
        \draw (0,1.5) to[short, -o] (-1,1.5);

        \draw (0,3) to[short, -o] (-1,3);
        \draw (0,4) to[short, -o] (-1,4);

        \draw (7,0.5) to[short, -o] (8,0.5) -- (9,0.5);
        \draw (7,1.5) to[short, -o] (8,1.5) -- (9,1.5);

        \draw (7,3) to[short, -o] (8,3) -- (9,3);
        \draw (7,4) to[short, -o] (8,4) -- (9,4);

        \node at (-0.5,2.35) {$\vdots$};
        \node at (8,2.35) {$\vdots$};
        \node at (10,2.35) {$\vdots$};

        \draw[decoration={brace}, decorate] (-1.25,0.25) -- (-1.25,4.25);
        \node[rotate=90] at (-1.75, 2.25) {$N$ ports};

    \end{circuitikz}
    \caption{Cascade synthesis for the impedance function (\ref{eq:impedance}). }
    \label{fig:cascade_impedance}
\end{figure}
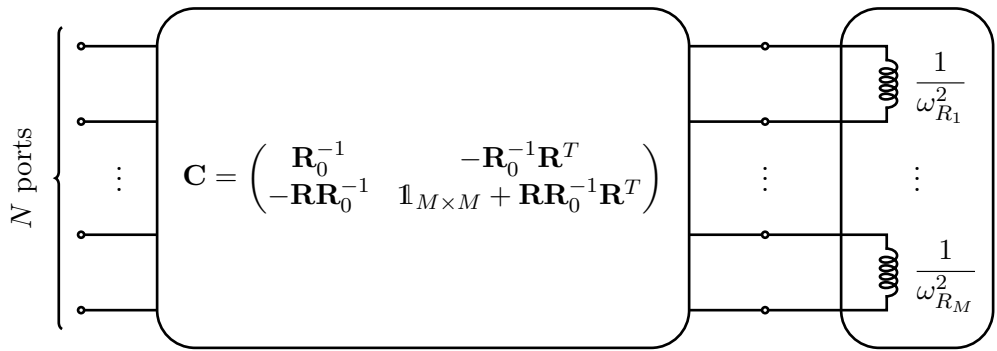

One interesting aspect of the Hamiltonian (\ref{eq:impedance_hamiltonian}) derived in the previous section is that it defines an alternative synthesis of the impedance function (\ref{eq:impedance}). Before diving into the details of the synthesis itself, we first need to discuss and define the Maxwell and mutual capacitance matrices. Given a network of $N$ nodes that each have a capacitance to ground $C_{g,i}$ and node coupling capacitances $C_{ij}$, the Maxwell capacitance matrix $\vb{C}$ gives the relationship between the voltages and charges at the nodes \cite{pos_cap, ruehli_cap}
\begin{equation}
    \vb{Q} = \vb{C} \vb{V}
\end{equation}
The matrix elements of the Maxwell capacitance matrix are defined in the following way:
\begin{equation}
    (\vb{C})_{ij} = \begin{cases}
    C_{g,i} + \sum_{k \neq i} C_{ik} & i = j\\
    -C_{ij} & i \neq j
    \end{cases}
\end{equation}
We can also define the mutual capacitance matrix $\vb{C}_{mut}$ with matrix elements
\begin{equation}
    (\vb{C}_{mut})_{ij} = \begin{cases}
        C_{g,i} & i=j\\
        C_{ij} & i \neq j
    \end{cases}
\end{equation}
In electromagnetic or circuit modeling software, this mutual capacitance matrix is sometimes called the SPICE capacitance matrix. It simply gives the mutual capacitance between the nodes of the network. It is easy to convert between the two types of matrices and the conversion is also reciprocal:
\begin{equation}
    (\vb{C}_{mut})_{ij} = \begin{cases}
        \sum_{k=0}^N(\vb{C})_{ik} & i=j \\
        -(\vb{C})_{ij} & i \neq j
    \end{cases}
\end{equation}
Now that the Maxwell and mutual capacitance matrices are defined, we can discuss the alternative synthesis of the impedance function (\ref{eq:impedance}). Disregarding the potential contribution to the Hamiltonian (\ref{eq:impedance_hamiltonian}) from the elements shunting the ports, we can see that the Hamiltonian describes a cascade network as shown in Fig.\ \ref{fig:cascade_impedance}. The first network in the cascade is a purely capacitive $(N+M)$ port network that has a Maxwell capacitance matrix $\vb{C}$ defined by (\ref{eq:impedance_cap}). A synthesis of this portion of the cascade can be easily found by converting the Maxwell form capacitance matrix to its mutual form. This purely capacitive network is then shunted by inductances $L_{R_k}=1/\omega_{R_k}^2$. We will refer to this network structure as a CL cascade.

What is interesting about the cascade synthesis presented above is that it removes the multiport transformers present in the Cauer synthesis of the impedance function. However, the caveat of this transformerless synthesis is that it is possible to end up with \textit{negative} values for the capacitances in the mutual capacitance matrix and thus negative capacitors in the cascade synthesis. Including negative inductive, capacitive or resistive circuit elements in the synthesis of a network is not a new idea. For example, in the Brune method of synthesis, negative inductances or capacitances may appear. However, these elements can be absorbed into the rest of the network to form mutual inductances or ideal transformers such that the final synthesis contains only passive elements \cite{brune_synthesis}, \cite[Chapter 9.4]{guillemin_synthesis}. This requirement that the final representation of a network should contain only passive circuit elements has shaped much of the history of network synthesis and is in general the agreed upon approach \cite[Chapter 1.2]{cauer_linear_comm}. However, for our use case and synthesized cascade network, even though there may be negative capacitances present, we have seen that the Maxwell capacitance matrix of the network is positive definite, and therefore the \textit{overall} network is passive which is what matters in our case. Having negative capacitances is a consequence of the fact that the turns ratios present in the multiport transformers can take positive and negative values. We can motivate this idea further using two simple examples.

\begin{figure}[h!]
    \centering
    \begin{circuitikz}[line width=1pt]
        \ctikzset{american}
        \ctikzset{bipoles/thickness=1, bipoles/length=1cm}
        \ctikzset { label/align = straight }
    
        \draw (2,7.5) to[L, name=TJC1] (2,6.5) -- (2,2.5) to[L, name=r11] (2,1.5);
        \draw[rounded corners=.5cm] (2,1.51) -- (2,1) -- (0,1) to[short, -o] (0,4);
        \draw[rounded corners=.5cm] (2,7.4) -- (2,8) -- (0,8) to[short, -o] (0,5);
        
        \draw[decoration={brace}, decorate] (-.25,3.75) -- (-.25,5.25);
        \node[rotate=90] at (-.75, 4.5) {Port 1};
        
        \draw[thick, double] (TJC1.core east) -- (TJC1.core west);
        \draw[thick, double] (r11.core east) -- (r11.core west);

        \node[anchor=west] at (1.97,7.65) {1};
        \node[anchor=west] at (1.97,2.65) {$1$};
    

        \draw (6,5.5) to[L, name=TJCN] (6,4.5) -- (6,2.5) to[L, name=r1n] (6,1.5);
        \draw[rounded corners=.5cm] (6,5.4) -- (6,8) -- (4,8) to[short, -o] (4,5);
        \draw[rounded corners=.5cm] (6,1.51) -- (6,1) -- (4,1) to[short, -o] (4,4);
        
        \draw[decoration={brace}, decorate] (3.75,3.75) -- (3.75,5.25);
        \node[rotate=90] at (3.25, 4.5) {Port 2};

        \draw[thick, double] (TJCN.core east) -- (TJCN.core west);
        \draw[thick, double] (r1n.core east) -- (r1n.core west);
        \node[anchor=west] at (5.97,5.65) {1};
        \node[anchor=west] at (5.97,2.65) {$-1$};
    
        \draw (8,6.25) -- (8,6.5) to[L, name=u11] (8, 7.5);
        \draw[rounded corners=.5cm] (8, 7.4) -- (8,8) -- (9.5,8) to[C={\Large $C_1$}, name=c1] (9.5,4) -- (8,4) -- (8,6.6);
        \node[anchor=east] at (8,7.65) {\small $1$};
        \draw[thick, double] (u11.core east) -- (u11.core west);
    
        \draw (11.5,5.75) -- (11.5,5.5) to[L, name=un1, mirror] (11.5,4.5);
        \draw[rounded corners=.5cm] (11.5, 5.4) -- (11.5,8) -- (13,8) to[C={\Large $C_2$}, name=c1] (13,4) -- (11.5,4) -- (11.5,4.6);
        \node[anchor=east] at (11.5,5.65) {\small $1$};
        \draw[thick, double] (un1.core east) -- (un1.core west);

        \draw (8,1.5) to[L, name=res1] (8,2.5);
        {
        \ctikzset{cute}
        \draw[rounded corners=.5cm] (8,2.4) -- (8,3) -- (11,3) -- (12,3) to[L={\Large $\;\dfrac{1}{\omega_{1}^2}$}] (12,1) -- (8,1) -- (8,1.6);
        }
        \draw (10,3) to[C={\Large $1$}] (10,1);
    
        \draw[thick, double] (res1.core west) -- (res1.core east);
        \node[anchor=east] at (8.0125,2.65) {1};
        
    \end{circuitikz}
    \caption{Cauer synthesis for the two-port impedance function (\ref{eq:impedance_ex1}).}
    \label{fig:cauer_synthesis_ex1}
\end{figure}
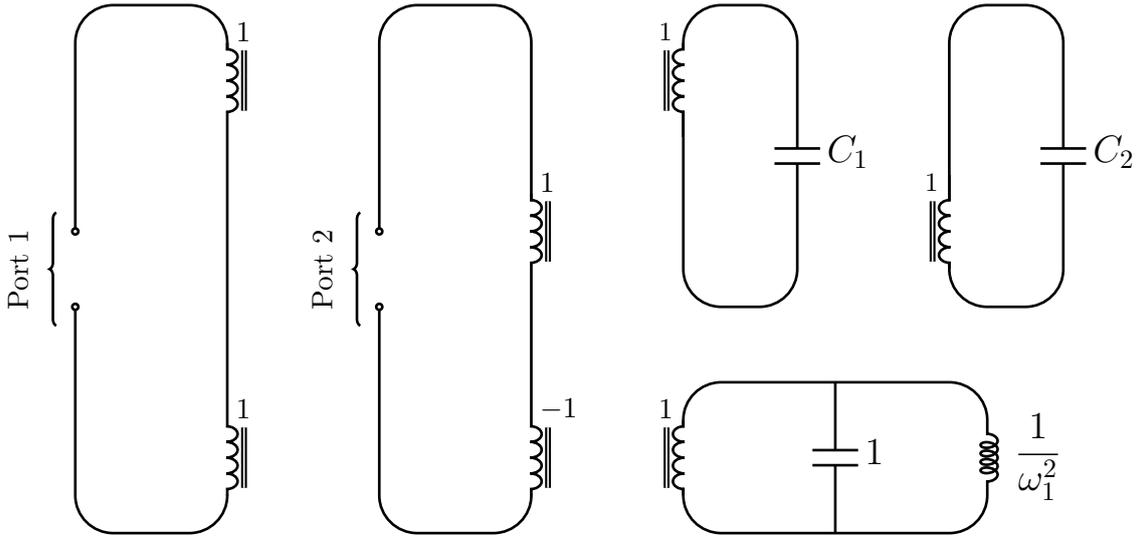

First, we consider the following two-port impedance function:
\begin{equation}\label{eq:impedance_ex1}
    \vb{Z}(s) = \frac{1}{s}\mqty(C_1 & 0 \\ 0 & C_2)^{-1} + \frac{s \vb{r}_1^T \vb{r}_1 }{s^2 + \omega_1^2} \quad \text{where}\quad \vb{r}_1 = \mqty(1 & -1), \quad \text{and}\quad C_1,C_2 > 0
\end{equation}
Note the restriction on $C_1$ and $C_2$ is in line with our earlier assumption that the DC residue of the impedance function must be positive definite. We can first look at the Cauer representation of this impedance function. It can easily be found and is shown in Fig.\ \ref{fig:cauer_synthesis_ex1} since we can read off the turns ratio matrices $\vb{U} = \mathds{1}_{2\times 2}$ and $\vb{R} = \mqty(1 & -1)$.

\begin{figure}[h!]
    \centering
    \begin{circuitikz}[line width=1pt]
        \ctikzset{bipoles/thickness=1, bipoles/length=1cm}
        \ctikzset { label/align = straight }
        
        \draw (1,2) to[short,o-] (1,2) to[C=$C_1$] (4,2) -- (5,2) to[C=$-C_2$] (8,2) to[short, -o] (9,2);
        \draw (4,0) to[C=$1+2C_2$] (4,2);
        \draw (5,0) to[L, l_=$\dfrac{1}{\omega_{1}^2}$, mirror] (5,2);
        \draw (8,0) to[C=$2C_2$] (8,2);
        
        \draw (1,0) to[short, o-o] (9,0);

    \end{circuitikz}
    \caption{Cascade synthesis for the two-port impedance function (\ref{eq:impedance_ex1}).}
    \label{fig:cascade_impedance_ex1}
\end{figure}
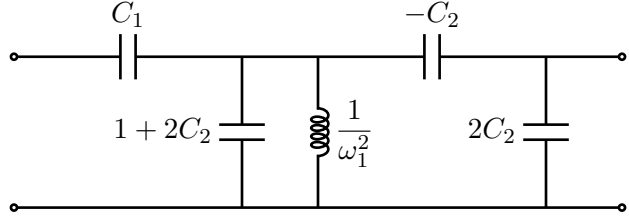

Note that in the Cauer synthesis there are no negative circuit elements, but there is a negative turns ratio on one of the ideal transformers. Next we look at the alternative cascade synthesis. Using (\ref{eq:impedance_cap}), we can write the Maxwell capacitance matrix of (\ref{eq:impedance_ex1}) and also convert it to the mutual form:
\begin{equation}
    \vb{C} = \mqty( C_1 & 0 & -C_1 \\ 0 & C_2 & C_2 \\ -C_1 & C_2 & 1 + C_1 + C_2 ) \quad\longrightarrow\quad \vb{C}_{mut} = \mqty( 0 & 0 & C_1 \\ 0 & 2C_2 & -C_2 \\ C_1 & -C_2 & 1 + 2C_2 )
\end{equation}
Immediately we see that there is a negative coupling capacitance in the cascade synthesis between port 2 and the single resonant mode. Using the mutual capacitance matrix, the circuit diagram of the cascade is presented in Fig.\ \ref{fig:cascade_impedance_ex1}.

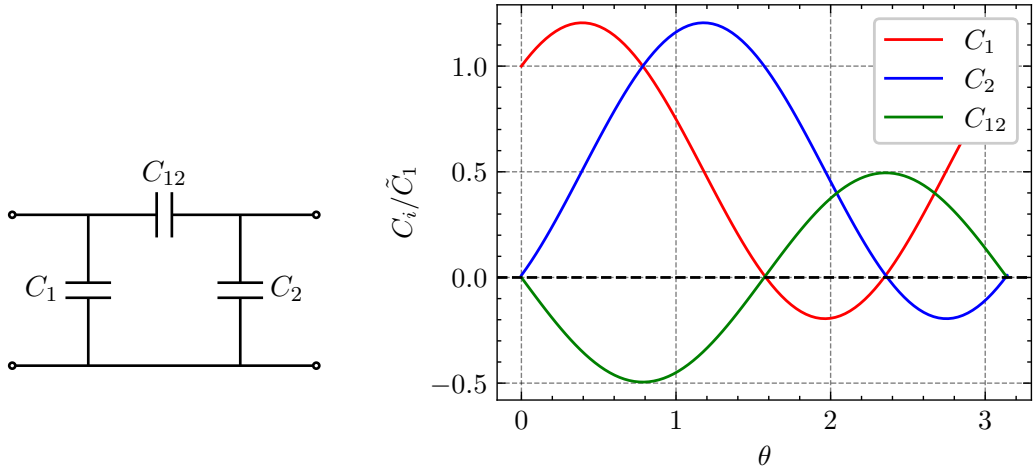
\begin{figure}[h!]
    \centering
    \begin{subfigure}{0.35\textwidth}
        \centering
        \begin{circuitikz}[line width=1pt]
            \ctikzset{bipoles/thickness=1, bipoles/length=1cm}
            \ctikzset { label/align = straight }
            
            \draw[color=white] (0,0) -- (0,-1.5);

            \draw (0,2) to[short, o-] (1,2) to[C=$C_{12}$] (3,2) to[short, -o] (4,2);
            \draw (0,0) to[short, o-o] (4,0);
            \draw (1,0) to[C, l=$C_1$] (1,2);
            \draw (3,0) to[C, l_=$C_2$] (3,2);
    
        \end{circuitikz}
    \end{subfigure}%
    \begin{subfigure}{0.55\textwidth}
        \includegraphics[width=\textwidth]{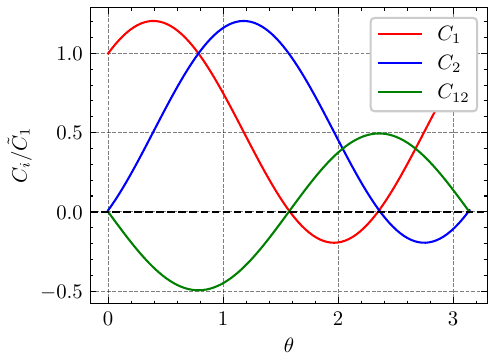}
    \end{subfigure}
    \caption{Left: Cascade synthesis of the impedance function (\ref{eq:impedance_ex2}). Right: Values of the capacitive elements in the synthesis on the left for $\tilde{C}_1/\tilde{C}_2 = 0.01$ and arbitrary $\theta$.}
    \label{fig:cascade_impedance_ex2}
\end{figure}

\newpage
Now for our second example, we show that negative capacitances can also exist if there is only a DC pole present in the impedance function. Consider the following two-port impedance function:
\begin{equation}\label{eq:impedance_ex2}
    \vb{Z}(s) = \frac{1}{s}\vb{U}\vb{C}_0^{-1}\vb{U}^T \quad\text{where}\quad \vb{C}_0 = \mqty(\tilde{C}_1 & 0 \\ 0 & \tilde{C}_2), \quad\vb{U} = \mqty(\cos\theta & -\sin\theta \\ \sin\theta & \cos\theta), \quad\text{and}\quad \tilde{C}_1,\tilde{C}_2 > 0
\end{equation}
In this case, the Maxwell form capacitance matrix is just given by $\vb{C} = \vb{U}\vb{C}_0\vb{U}^T$ and by construction, it is always positive definite for any values of $\theta$. The resulting cascade synthesis is not much of a ``cascade'' since there are no inductive shunts, so it will just be a $\pi$-network of capacitors as shown in Fig.\ \ref{fig:cascade_impedance_ex2}. Varying the ratio $\tilde{C}_1/\tilde{C}_2$ and parameter $\theta$ will change which of the capacitors in the $\pi$-network are negative. In Fig.\ \ref{fig:cascade_impedance_ex2} an example is shown where $\theta$ is swept for a fixed $\tilde{C}_1/\tilde{C}_2$ ratio. It should be noted that for this example, only one capacitance value can be negative in the synthesis for a given value of $\theta$.

While the cascade synthesis does go against the usual convention of trying to obtain a synthesis of purely passive elements, it is extremely useful since it allows us to synthesize lumped element circuits that now consist of only capacitors and inductors. We will see later how this allows us to easily interconnect networks of rational impedance functions. It will also be a crucial part of computing decay rates of resonant modes within our circuit when considering loss through the external ports of our network.

Finally, it is important to note that for all of the impedances that we deal with of the form (\ref{eq:impedance}), we always restrict ourselves to a finite number of resonant modes. If we allow for infinite resonant modes, the capacitance  and inductance matrices in the Lagrangian or Hamiltonian representation of the impedance become infinite in size. Furthermore, these infinite capacitance matrices can result in diverging values for our capacitive elements in the cascade synthesis presented here. For a direct example of this, see Appendix \ref{appendix:cascade_ideal_TL} where we consider the ideal lossless transmission line. However, our restriction to a finite number of poles should not cause any problems as long as we always make sure the truncated impedance function is a good approximation of the true impedance function within the relevant working frequency range of our device. 

\subsection{Analysis of the CL Cascade Network}\label{section:cascade_analysis}
Above we have seen how from the rational impedance function (\ref{eq:impedance}), we can obtain a synthesis in the form of a cascade circuit as shown in Fig.\ \ref{fig:cascade_impedance}. Now we show how to find the answer to the opposite question: How can we obtain the rational impedance function for a circuit of the cascade form shown in Fig.\ \ref{fig:cascade_impedance}? We want to answer this question for any general Maxwell or mutual capacitance matrix along with general values for the inductances shunting the ports of the cascade. Note that here we assume that the Maxwell capacitance matrix we start with is positive definite. Since we know that the impedance function (\ref{eq:impedance}) has a synthesis of the cascade form presented in the last section, we expect that it should be possible to find this function for a general cascade network.

To do this, we start with the Lagrangian for a general cascade with Maxwell capacitance matrix $\vb{C}$ shunted by a set of inductors $L_{R_k}$. With the ports of the cascade network left open, we have the following:
\begin{equation}
    \mathcal{L} = \frac{1}{2}\dot{\vb{\Phi}}^T \vb{C} \dot{\vb{\Phi}} - \frac{1}{2}\vb{\Phi}^T \vb{M} \vb{\Phi}^{\phantom{^T}} \quad\text{where}\quad \vb{M} = \mqty( \vb{0}_{N \times N} & \vb{0}_{N \times M} \\ \vb{0}_{M \times N} & \vb{M}_R ),
\end{equation}
$\vb{M}_R = \diag(1/L_{R_1},\dots,1/L_{R_k})$, and $\vb{\Phi} = (\Phi_{P_1}, \dots, \Phi_{P_N}, \Phi_{R_1}, \dots, \Phi_{R_M})^T$. We will also use the vector $\vb{\Phi}_P$ and $\vb{\Phi}_R$ which contain the port fluxes $\Phi_{P_i}$ and inductor fluxes $\Phi_{R_i}$, respectively. If we look back at the Lagrangian (\ref{eq:impedance_lagrangian}) or Hamiltonian (\ref{eq:impedance_hamiltonian}) corresponding to the synthesis of impedance function (\ref{eq:impedance}), we see there that the diagonal inductance matrix $\vb{M}_R$ contains the frequencies squared of the resonant modes within our network. This is different to the diagonal matrix $\vb{M}_R$ we have here for our general network that just contains the inverse inductances. We want to find a new set of transformed flux variables that also transforms the matrix $\vb{M}_R$ such that its diagonal contains the frequencies squared of the resonant modes rather than the inverse inductances shunting the network. Finding this transformation and applying it to the inductance matrix gives us the resonant modes of the network and thus the poles of the impedance function. Once the poles are found, the same transformation can be applied to the capacitance matrix from which the residues of the impedance function can be extracted using (\ref{eq:impedance_cap}). To find this transformation, we can look at the equations of motion for the flux variables in the Lagrangian:
\begin{equation}
    \vb{C}\ddot{\vb{\Phi}} = -\vb{M} \vb{\Phi} \quad\longrightarrow\quad \ddot{\vb{\Phi}} = -\vb{C}^{-1}\mqty(\vb{0}_N \\ \vb{M}_R \vb{\Phi}_R) = -\mqty( (\vb{C}^{-1})^{\phantom{RP}}_P & (\vb{C}^{-1})^{\phantom{RP}}_{PR} \\ (\vb{C}^{-1})^{\phantom{RP}}_{RP} & (\vb{C}^{-1})^{\phantom{RP}}_R )\mqty(\vb{0}_N \\ \vb{M}_R \vb{\Phi}_R)
\end{equation}
The above gives us the equations of motion for the resonator flux variable block of the inverse capacitance matrix:
\begin{equation}
    \ddot{\vb{\Phi}}_R = -(\vb{C}^{-1})_R \vb{M}_R \vb{\Phi}_R
\end{equation}
To find the transformation described above, we will need to simultaneously diagonalize the matrix product $(\vb{C}^{-1})_R \vb{M}_R$ as the eigenvalues of this product should contain the squared resonant frequencies. Since both the matrices $(\vb{C}^{-1})_R$ and $\vb{M}_R$ are real, symmetric and positive definite, these resonant frequencies will be real and positive. For more on this process, see \cite[Chapter 7.6]{horn_johnson} or \cite[Appendix B.1]{cqed_lecture_notes}.
\newpage
For our case here, we present the whole process of constructing this transformation since the proof easily follows given all the pieces. The steps are as follows:
\begin{enumerate}
    \item Diagonalize $(\vb{C}^{-1})_R = \vb{O}^{\phantom{T}}_C \vb{D} \vb{O}_C^T$.
    \item Define $\vb{T} = \vb{O}_C \vb{D}^{1/2}$.
    \item Diagonalize $\vb{T}^T \vb{M}_R \vb{T} = \vb{O}^{\phantom{T}}_M \vb{\Omega}^2 \vb{O}_M^T$.
    \item Using the above, our needed transformation can be defined as $\vb{S} = \vb{T}\vb{O}_M$.
\end{enumerate}
From the construction above, we can see that $\vb{M}_R = (\vb{S}^{-1})^T \vb{\Omega}^2 \vb{S}^{-1}$ and $\vb{S}\vb{S}^T = (\vb{C}^{-1})_R$. The matrix product then equals $(\vb{C}^{-1})_R \vb{M}_R = \vb{S} \vb{\Omega}^2 \vb{S}^{-1}$. We see that $\vb{S}$ does indeed diagonalize the matrix product. $\vb{S}$ also transforms $\vb{M}_R$ into the diagonal matrix containing the frequencies squared of the resonant modes. With this, we can now define a set of steps to follow that allow us to find the rational impedance function for the general cascade network.
\begin{framed}
\noindent \underline{Finding the Poles and Residues of the Impedance Function for a CL Cascade}
\begin{enumerate}
    \item Find the matrix $\vb{S}$ that diagonalizes $(\vb{C}^{-1})_R \vb{M}_R = \vb{S} \vb{\Omega}^2 \vb{S}^{-1}$.
    \item Define a transformed set of flux variables using $\vb{S}$:
    \begin{equation}
        \vb{\Psi} = \bar{\vb{S}}^{-1} \vb{\Phi} \quad\text{where}\quad \bar{\vb{S}} = \mqty(\mathds{1}_{N \times N} & \vb{0}_{N \times M} \\ \vb{0}_{M \times N} & \vb{S})
    \end{equation}
    This will give the transformed Lagrangian:
    \begin{equation}
        \mathcal{L} = \frac{1}{2}\dot{\vb{\Psi}}^T \bar{\vb{S}}^T \vb{C} \bar{\vb{S}} \dot{\vb{\Psi}} - \frac{1}{2}\vb{\Psi}^T \bar{\vb{S}}^T \vb{M} \bar{\vb{S}} \vb{\Psi}^{\phantom{^T}}
    \end{equation}
    \item Extract the poles of the impedance function from the diagonal matrix 
    \begin{equation}
        \vb{\Omega} = (\vb{S}^T \vb{M}_R \vb{S})^{1/2}
    \end{equation}
    \item Extract the residues of the impedance from the transformed capacitance matrix:
    \begin{equation}
        \bar{\vb{C}} = \bar{\vb{S}}^T \vb{C} \bar{\vb{S}} = \bar{\vb{S}}^T\mqty(\vb{C}_P & \vb{C}_{PR} \\ \vb{C}_{PR}^T & \vb{C}_R )\bar{\vb{S}} = \mqty(\vb{C}_P & \vb{C}_{PR} \vb{S} \\ \vb{S}^T \vb{C}_{PR}^T & \vb{S}^T \vb{C}_R \vb{S})
    \end{equation}
    Since the transformed inductance matrix $\vb{S}^T \vb{M}_R \vb{S}$ will contain the frequencies squared on the diagonal, we can say that the transformed capacitance matrix will be of the form (\ref{eq:impedance_cap}) and we can equate the two:
    \begin{equation}
        \mqty(\vb{C}_P & \vb{C}_{PR} \vb{S} \\ \vb{S}^T \vb{C}_{PR}^T & \vb{S}^T \vb{C}_R \vb{S}) = \mqty( \vb{R}_0^{-1} & -\vb{R}_0^{-1}\vb{R}^T \\ -\vb{R} \vb{R}_0^{-1} & \mathds{1}_{M\times M} + \vb{R}\vb{R}_0^{-1}\vb{R}^T)
    \end{equation}
    Equating the two matrices tells us that the DC residue of the impedance function for this network is $\vb{R}_0 = \vb{C}_P^{-1}$. Then, using equality of the port-resonator block of the matrices above, we can compute the turns ratio matrix $\vb{R}$:
    \begin{equation}
        \vb{R} = -(\vb{C}_P^{-1} \vb{C}_{PR} \vb{S})^T = -\vb{S}^T \vb{C}_{PR}^T \vb{C}_P^{-1}
    \end{equation}
    As explained earlier, the rows $\vb{r}_k$ of this turns ratio matrix can be used to obtain the residues $\vb{R}_k = \vb{r}_k^T \vb{r}_k$ of the impedance function.
\end{enumerate}
\end{framed}
\newpage
With all of the steps above, we can construct the rational impedance function for a general CL cascade network. To show that this does in fact work, we can compare the method of finding the rational impedance function to a method for computing the S-parameter of a cascade-loaded network. Here we briefly go over this method as we will see it will also be useful later on when adding elements at the ports of our networks.

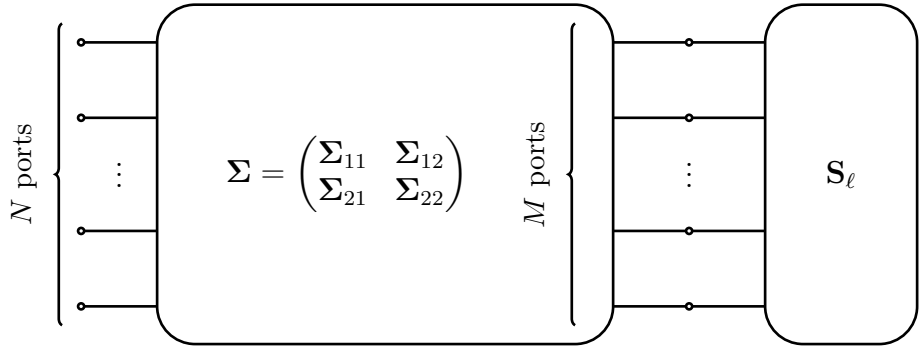
\begin{figure}[h!]
    \centering
    \begin{circuitikz}[line width=1pt]
        \ctikzset{bipoles/thickness=1, bipoles/length=1cm}
        \ctikzset { label/align = straight }
        
        \draw[rounded corners=.5cm] (0,2) -- (0,4.5) -- (6,4.5) -- (6,0) -- (0,0) -- (0,2);
        \node at (2.5,2.25) {\large $\vb{\Sigma} = \mqty(\vb{\Sigma}_{11} & \vb{\Sigma}_{12} \\ \vb{\Sigma}_{21} & \vb{\Sigma}_{22})$};

        \draw[rounded corners=0.5cm] (8,2) -- (8,4.5) -- (10,4.5) -- (10,0) -- (8,0) -- (8,2);
        \node at (9,2.25) {\large $\vb{S}_\ell$};

        \draw (0,0.5) to[short, -o] (-1,0.5);
        \draw (0,1.5) to[short, -o] (-1,1.5);

        \draw (0,3) to[short, -o] (-1,3);
        \draw (0,4) to[short, -o] (-1,4);

        \draw (6,0.5) to[short, -o] (7,0.5) -- (8,0.5);
        \draw (6,1.5) to[short, -o] (7,1.5) -- (8,1.5);

        \draw (6,3) to[short, -o] (7,3) -- (8,3);
        \draw (6,4) to[short, -o] (7,4) -- (8,4);

        \node at (-0.5,2.35) {$\vdots$};
        \node at (7,2.35) {$\vdots$};

        \draw[decoration={brace}, decorate] (-1.25,0.25) -- (-1.25,4.25);
        \node[rotate=90] at (-1.75, 2.25) {$N$ ports};

        \draw[decoration={brace}, decorate] (5.5,0.25) -- (5.5,4.25);
        \node[rotate=90] at (5, 2.25) {$M$ ports};

    \end{circuitikz}
    \caption{General cascade loaded network.}
    \label{fig:general_cascade_load}
\end{figure}

Consider the general cascade-loaded network as shown in Fig.\ \ref{fig:general_cascade_load}. The left $(N+M)$-port network in the cascade has the S-parameter $\vb{\Sigma}$ and the load has an $M$-port S-parameter $\vb{S}_\ell$. $\vb{\Sigma}$ can be broken up into blocks that correspond to external and internal ports. The S-parameter $\vb{S}$ of the loaded network can then be computed using \cite[Eq. 3.20]{newcomb}:
\begin{align}\label{eq:cascade_s}
    \vb{S} = \vb{\Sigma}_{11} + \vb{\Sigma}_{12}\vb{S}_\ell (\mathds{1}_{M \times M} - \vb{S}_{22}\vb{\Sigma}_\ell)^{-1}\vb{\Sigma}_{21}= \vb{\Sigma}_{11} + \vb{\Sigma}_{12} (\mathds{1}_{M \times M} - \vb{S}_\ell\vb{\Sigma}_{22})^{-1}\vb{S}_\ell\vb{\Sigma}_{21}
\end{align}

To use the above formula to find the S-parameter of the CL cascade, we can use the impedance functions for the purely capacitive network and the inductive shunt network. The impedance function of the $(N+M)$-port capacitive network is $\vb{Z}_C(s) = \vb{C}^{-1}/s$. For the $M$-port inductive shunt network, the impedance function is $\vb{Z}_L(s) = s\vb{L}$ where $\vb{L} = \diag(L_1,\dots,L_M)$. Converting each of these impedance functions to S-parameters allows us to use (\ref{eq:cascade_s}) to compute the S-parameter of the cascaded network discretized in frequency. The final S-parameter can be converted back to an impedance function and then compared to our expected rational impedance function.  

\begin{figure}[h!]
    \centering
    \includegraphics[width=\textwidth]{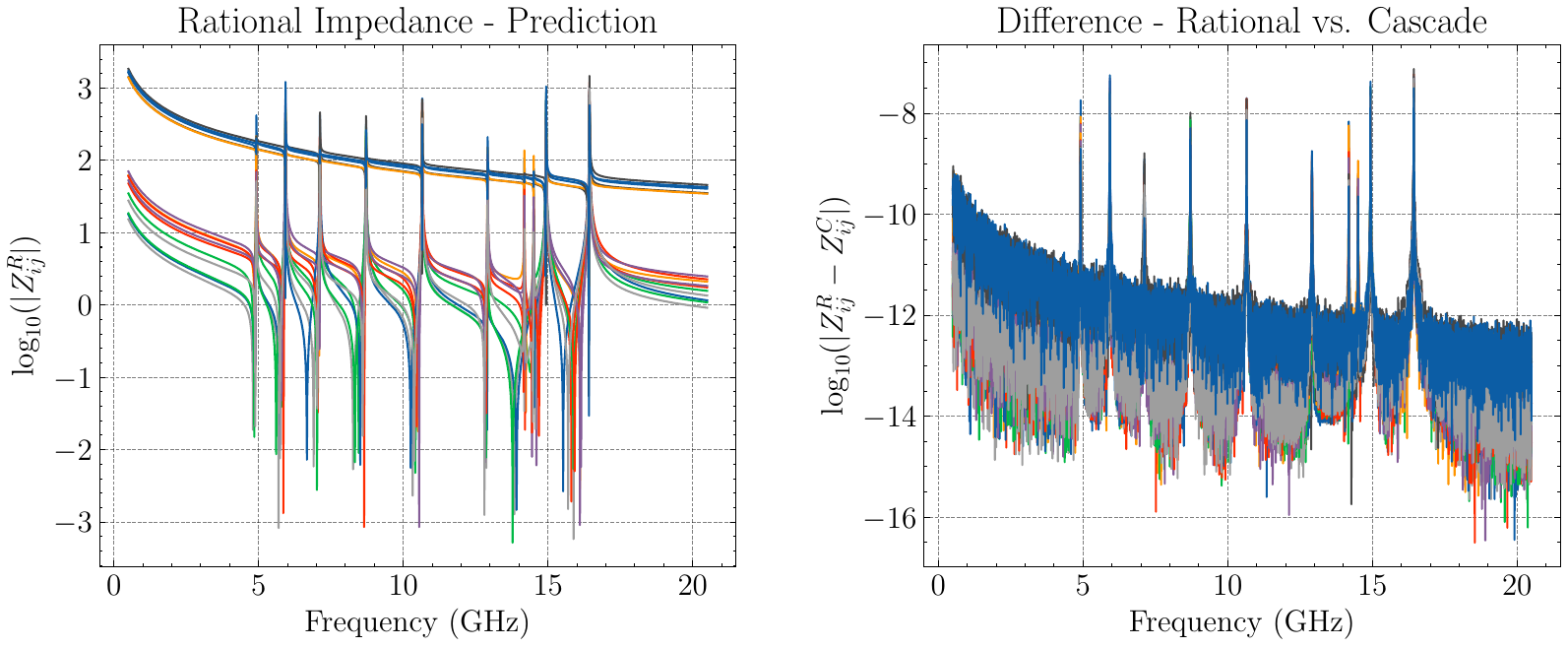}
    \caption{Left: Rational impedance function $\vb{Z}^R$ with 5 ports and 10 resonant modes. The capacitive network and inductive shunts are randomly generated. Port shunt capacitors are between 100 and 200 fF, and port coupling capacitances are between 0 and 10 fF. Shunt inductors are between 0.4 and 5 nH. Right: The difference between the predicted rational impedance function $\vb{Z}^R$ and the discretized cascade impedance $\vb{Z}^C$ computed using (\ref{eq:cascade_s}).}
    \label{fig:rand_impedance_ex}
\end{figure}

In Fig.\ \ref{fig:rand_impedance_ex}, we take a randomly generated CL cascade network and compute the poles and residues of the rational impedance function using the method discussed above. Then we compare it to the impedance function computed using (\ref{eq:cascade_s}) and the S-parameters of the capacitive and inductive networks. We see that the difference between the two is quite small in the chosen frequency range with some larger deviations around the poles. This is due to numerical error that arises in computing the pole locations and the conversions between S and Z-parameters.

If we know the capacitance matrix and shunt inductor values, then we can already write a Hamiltonian for the CL cascade network so we do not necessarily need to find the rational impedance function. However, it is useful to have a clear method for calculating the resonant frequencies of the network. Furthermore, we will see in the next section that this method can be used to recover the rational impedance function of an interconnected network of rational impedances.

%% file: mainmatter/ch2/interconnects.tex
\section{General Network Interconnection}\label{section:general_network_interconnection}
Interconnecting circuit networks will allow us to efficiently build large models of qubit networks out of smaller and simpler pieces. Many methods do exist for interconnecting networks, but for the most part, these methods are limited to specific configurations (series, parallel, cascade) for immittance or scattering parameters that are discretized in frequency \cite[Chapter 3.3]{newcomb}. There are methods that allow you to interconnect state-space representations but they are also limited to the configurations mentioned \cite[Chapter 3.7.1]{passive_macromodeling}. What we are interested in is the general interconnection of rational impedance functions that at the end gives another rational impedance function. By ``general'', we mean that we want to be able to pick and choose arbitrary and direct connections between two-terminal ports. A method for doing this has not previously been found, but here we present how this can be done for the impedance function (\ref{eq:impedance}).

\subsection{Rational Impedance Interconnection}\label{section:rational_impedance_interconnection}
To interconnect rational impedance functions, we can make use of the CL cascade synthesis presented in Section (\ref{section:cascade_synthesis}). Since we know that these impedance functions can be synthesized as a CL cascade, interconnecting the ports becomes a problem of interconnecting multiple capacitance networks.

\definecolor{connectcolor}{HTML}{2098b0}
\begin{figure}[h!]
    \centering
    \begin{circuitikz}[line width=1pt]
    \ctikzset{bipoles/thickness=1, bipoles/length=1cm}
    \ctikzset{bipoles/crossing/size=0.5}
    \ctikzset { label/align = straight }

    \draw[connectcolor] (5,2.75) -- (7,2.75);
    \draw[connectcolor] (5,1.75) -- (7,1.75);
    \draw[connectcolor] (5.5,2.75) -- (5.5,2) to[crossing, color=connectcolor, mirror] (5.5,1.5) to[short, -o, color=connectcolor] (5.5,1);
    \draw[connectcolor] (6.5,1.75) to[short, -o, color=connectcolor] (6.5,1);

    \draw[rounded corners=.5cm, connectcolor] (12.5,3) -- (13, 3) -- (13,.5) -- (12.5,.5);
    \draw[rounded corners=.5cm, connectcolor] (12.5,4) -- (13.75, 4) -- (13.75,1.5) -- (13.25,1.5);
    \draw[connectcolor] (13.25, 1.5) to[crossing, color=connectcolor, mirror] (12.75,1.5) -- (12.5,1.5);
    \draw[connectcolor] (13,1) to[short, -o, color=connectcolor] (14.5,1);
    \draw[connectcolor] (13.75, 2) to[short, -o, color=connectcolor] (14.5, 2);

    \draw[rounded corners=.5cm] (0,2) -- (0,4.5) -- (4.5,4.5) -- (4.5,0) -- (0,0) -- (0,2);
    \node at (2.25,2.25) {\LARGE $\vb{C}_1$};

    \draw (0,0.5) to[short, -o] (-0.5,0.5);
    \draw (0,1.5) to[short, -o] (-0.5,1.5);
    \draw (0,3) to[short, -o] (-0.5,3);
    \draw (0,4) to[short, -o] (-0.5,4);
    \node at (-0.25,2.35) {$\vdots$};

    \draw (4.5,2.75) to[short, -o] (5,2.75);
    \draw (4.5,1.75) to[short, -o] (5,1.75);

    \draw (0.5,4.75) -- (0.5,5) to[L] (1.5,5) -- (1.5,4.75);
    \draw (0.5,4.5) to[short,-o] (0.5,4.75);
    \draw (1.5,4.5) to[short,-o] (1.5,4.75);
    \draw (3,4.75) -- (3,5) to[L] (4,5) -- (4,4.75);
    \draw (3,4.5) to[short,-o] (3,4.75);
    \draw (4,4.5) to[short,-o] (4,4.75);
    \node at (2.25, 4.75) {$\dots$};

    \draw[rounded corners=.5cm] (7.5,2) -- (7.5,4.5) -- (12,4.5) -- (12,0) -- (7.5,0) -- (7.5,2);
    \node at (9.75,2.25) {\LARGE $\vb{C}_2$};

    \draw (12,0.5) to[short, -o] (12.5,0.5);
    \draw (12,1.5) to[short, -o] (12.5,1.5);
    \draw (12,3) to[short, -o] (12.5,3);
    \draw (12,4) to[short, -o] (12.5,4);

    \draw (7.5,2.75) to[short, -o] (7,2.75);
    \draw (7.5,1.75) to[short, -o] (7,1.75);

    \draw (8,4.75) -- (8,5) to[L] (9,5) -- (9,4.75);
    \draw (8,4.5) to[short,-o] (8,4.75);
    \draw (9,4.5) to[short,-o] (9,4.75);
    \draw (10.5,4.75) -- (10.5,5) to[L] (11.5,5) -- (11.5,4.75);
    \draw (10.5,4.5) to[short,-o] (10.5,4.75);
    \draw (11.5,4.5) to[short,-o] (11.5,4.75);
    \node at (9.75, 4.75) {$\dots$};

\end{circuitikz}
\caption{Interconnection between ports of CL cascade networks. We can have interconnection between two ports in disjoint or connected capacitance networks. Note that we can make the interconnection by defining a new port that connects two existing ports. The port can then be ``left open'' to complete the interconnection.}
\label{fig:interconnect_visual}
\end{figure}
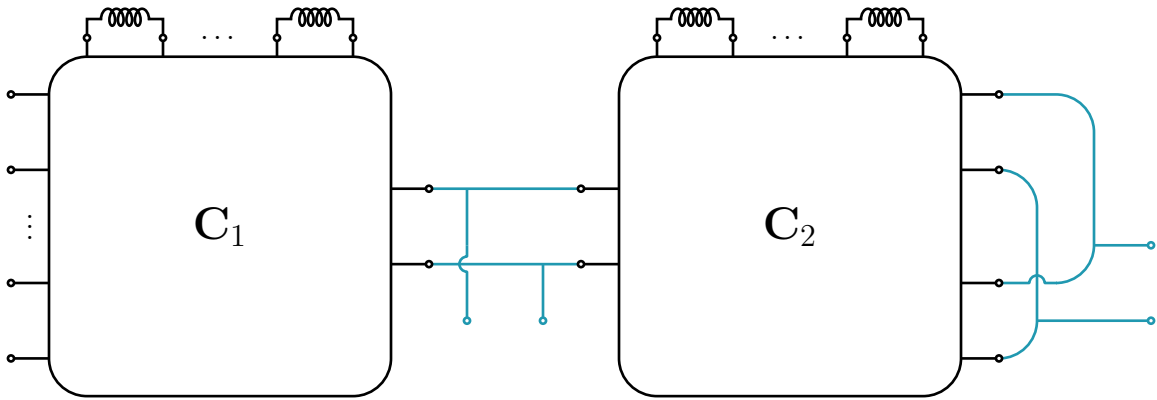

In Fig.\ \ref{fig:interconnect_visual}, we show how interconnections are defined between the two-terminal ports. Our interconnection method will work for connecting disjoint CL cascades as well as connecting 2 ports within the same cascade. The interconnection is created by forming a new port that connects two existing ports. The interconnection process is effectively completed if the new port is ``left open" which will be addressed later. Fig.\ \ref{fig:interconnect_visual} clearly shows that when performing interconnections of ports for a single or multiple CL cascade networks, the new network is also a CL cascade. This means that interconnecting CL cascades that are synthesized from impedance functions results in a CL cascade network. We can then find the rational impedance function of the interconnected CL cascade using the method of Section \ref{section:cascade_analysis}.

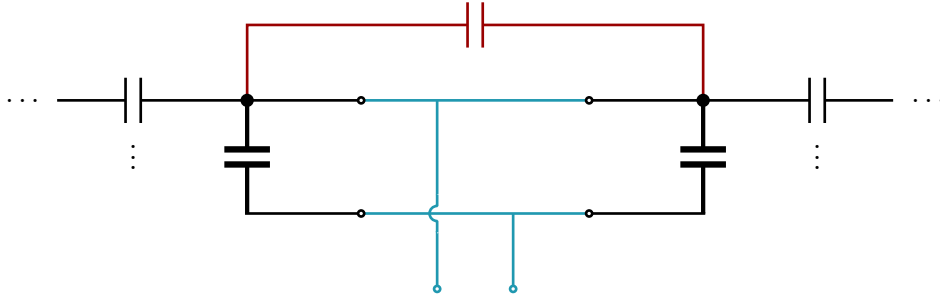
\begin{figure}[h!]
    \centering
    \begin{circuitikz}[line width=1pt]
    \ctikzset{american}
    \ctikzset{bipoles/thickness=1, bipoles/length=1cm}
    \ctikzset{bipoles/crossing/size=0.5}
    \ctikzset { label/align = straight }

    \draw[nodecolor] (-3,2) -- (-3,3) to[C, color=nodecolor] (3,3) -- (3,2);
    
    \draw[connectcolor] (-1.5, 0.5) -- (1.5,0.5);
    \draw[connectcolor] (-1.5, 2) -- (1.5, 2);
    \draw[connectcolor] (0.5, 0.5) to[short, -o, color=connectcolor] (0.5,-0.5);
    \draw[connectcolor] (-0.5, 2) -- (-0.5, 0.75) to[crossing, color=connectcolor, mirror] (-0.5,0.25) to[short, -o,color=connectcolor] (-0.5,-0.5);

    \draw (-3.028,0.5) to[short, -o] (-1.5,0.5);
    \draw (-1.5,2) to[short, o-] (-3.5,2) to[C] (-5.5,2);
    {
        \ctikzset{bipoles/thickness=1.5, bipoles/length=1cm}
        \draw[ultra thick] (-3,2) to[C] (-3,0.5);
    }

    \draw (3.028,0.5) to[short, -o] (1.5,0.5);
    \draw (1.5,2) to[short, o-] (3.5,2) to[C] (5.5,2);
    {
        \ctikzset{bipoles/thickness=1.5, bipoles/length=1cm}
        \draw[ultra thick] (3,2) to[C] (3,0.5);
    }

    \node[circle, fill=black, inner sep=0pt, minimum size=5pt] at (-3,2) {};
    \node[circle, fill=black, inner sep=0pt, minimum size=5pt] at (3,2) {};

    \node at (6,2) {\dots};
    \node at (-5.925,2) {\dots};
    \node at (-4.5,1.35) {\vdots};
    \node at (4.5,1.35) {\vdots};

\end{circuitikz}
\caption{Example interconnection of two ports for a capacitive network. Each port node can be capacitively coupled to multiple other ports (not shown). The shunt capacitance of the new port is the sum of the bold capacitances at the old ports. The red capacitance between the ports has a direct shunt across it and thus plays no role in the interconnected circuit.}
\label{fig:cap_interconnect}
\end{figure}

The central part of the interconnection procedure described above is the interconnection of the capacitance matrices. To see how this interconnection happens, we can look at the example in Fig.\ \ref{fig:cap_interconnect}. We immediately see that the shunt capacitance of the newly created port is just the sum of the capacitances shunting the old ports (shown in bold). Then, we also have that the capacitive couplings between the two newly connected ports to any other ports are added together. If there is a direct capacitive coupling between the two connected ports (as shown in red), it is then neglected since there is a short across it and which means the capacitance plays no role in the new network.

Using the rules discussed, we can go over the whole procedure. To generally interconnect $N$ rational impedance functions, we consider the cascade synthesis of each. So for each impedance function $\vb{Z}_i$ we have a corresponding Maxwell capacitance matrix $\vb{C}_i$ and a set of shunt inductors that do not affect the interconnection of the capacitance matrices. Each of these capacitance matrices has the form:
\begin{equation}
    \vb{C}_i = \mqty(\vb{C}_{i,P} & \vb{C}_{i,PR} \\ \vb{C}_{i,PR}^T & \vb{C}_{i,R} )
\end{equation}
Before interconnecting the networks, we first construct the capacitance network for the entire disconnected network:
\begin{equation}
    \vb{C} = \mqty( 
        \vb{C}_{1,P} & & \vb{0}& \vb{C}_{1,PR} & & \vb{0} \\
        & \ddots & & & \ddots & \\
        \vb{0}& & \vb{C}_{N,P} & \vb{0} & & \vb{C}_{N,PR} \\
        \vb{C}_{1,PR}^T & & \vb{0} & \vb{C}_{1,R} & & \vb{0}\\
        & \ddots & & & \ddots & \\
        \vb{0}& & \vb{C}_{N,PR}^T & \vb{0} & & \vb{C}_{N,R}
     )
\end{equation}
The upper left $P$-block corresponds to the ports of the impedance functions. This new capacitance matrix is constructed in a way such that the lower right $R$-block has the ports that are shunted by the sets of inductances for each individual CL cascade. To interconnect various ports of the impedance functions, we can perform some operations on the rows and columns that belong to the $P$-block of $\vb{C}$. After constructing $\vb{C}$, we can interconnect two ports $j$ and $k$ corresponding to rows or columns in the $P$-block with the following set of steps:
\begin{enumerate}
    \item Add row $k$ of $\vb{C}$ to row $j$. Choosing to add to row $j$ makes it so that this row will correspond to the newly formed port.
    \item Add column $k$ of $\vb{C}$ to column $j$. This step properly combines the shunt capacitances of the two ports in parallel.
    \item Delete row and column $k$ from $\vb{C}$.
\end{enumerate}
The newly formed $\vb{C}$ corresponds to the network with one port less than it started with, but now a port exists that combines the previous two ports. These row and column operations properly combine all the parallel capacitances that are present after interconnecting the two ports. After interconnecting all the port pairs needed, the method of Section (\ref{section:cascade_analysis}) can be used to obtain the rational impedance function for the new network that includes the new connection ports. Finally, you can ``leave the extra ports open'' in this final impedance function by deleting the rows and columns of the residues corresponding to these open ports. After this step, the interconnection process is done and you are left with the rational impedance function of the fully interconnected network.

\subsection{General S Parameter Interconnection}\label{section:s_interconnection}

Sometimes we may not have, or may not want, to find the rational impedance function of a network and therefore cannot use the above method for interconnection. While we will also explore how we can obtain the rational function from an impedance discretized over frequency, it is also useful to have a method for interconnecting these discretized functions. The best way to do this is to use a general interconnection algorithm for S-parameters \cite{filipsson_new_1981,subnetwork_growth}. We briefly go over this here since it will play a part in estimating the decay rates of resonant modes in our circuit.

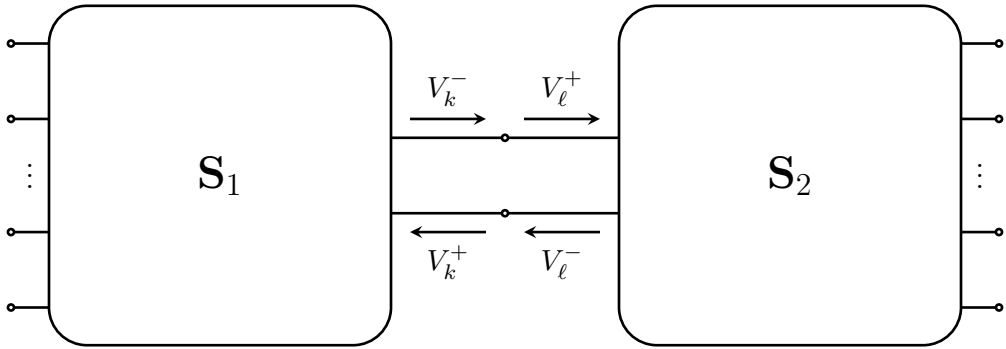
\begin{figure}[h!]
    \centering
    \begin{circuitikz}[line width=1pt]
    \ctikzset{bipoles/thickness=1, bipoles/length=1cm}
    \ctikzset{bipoles/crossing/size=0.5}
    \ctikzset { label/align = straight }

    \draw[rounded corners=.5cm] (0,2) -- (0,4.5) -- (4.5,4.5) -- (4.5,0) -- (0,0) -- (0,2);
    \node at (2.25,2.25) {\LARGE $\vb{S}_1$};

    \draw (0,0.5) to[short, -o] (-0.5,0.5);
    \draw (0,1.5) to[short, -o] (-0.5,1.5);
    \draw (0,3) to[short, -o] (-0.5,3);
    \draw (0,4) to[short, -o] (-0.5,4);
    \node at (-0.25,2.35) {$\vdots$};

    \draw (4.5,2.75) to[short, -o] (6,2.75) -- (7.5,2.75);
    \draw (4.5,1.75) to[short, -o] (6,1.75) -- (7.5,1.75);

    \draw[rounded corners=.5cm] (7.5,2) -- (7.5,4.5) -- (12,4.5) -- (12,0) -- (7.5,0) -- (7.5,2);
    \node at (9.75,2.25) {\LARGE $\vb{S}_2$};

    \draw (12,0.5) to[short, -o] (12.5,0.5);
    \draw (12,1.5) to[short, -o] (12.5,1.5);
    \draw (12,3) to[short, -o] (12.5,3);
    \draw (12,4) to[short, -o] (12.5,4);
    \node at (12.25,2.35) {$\vdots$};

    \draw [-stealth](4.75, 3) -- (5.75, 3);
    \node at (5.25,3.4) {$V_k^-$};
    \draw [-stealth](6.25, 3) -- (7.25, 3);
    \node at (6.75,3.4) {$V_\ell^+$};
    \draw [stealth-](4.75, 1.5) -- (5.75, 1.5);
    \node at (5.25,1.1) {$V_k^+$};
    \draw [stealth-](6.25, 1.5) -- (7.25, 1.5);
    \node at (6.75,1.1) {$V_\ell^-$};

\end{circuitikz}
\caption{An interconnection of two S-parameters that shows how the incident and reflected voltages at the interconnected ports are constrained. We can see that for the two ports $k$ and $\ell$, $V_k^-=V_\ell^+$ and $V_k^+=V_\ell^-$.}
\label{fig:s_interconnect_visual}
\end{figure}

Consider the two S-parameters interconnected at a single port in Fig.\ \ref{fig:s_interconnect_visual}. The S-parameter of the disjoint networks is defined by a new S-parameter such that:
\begin{equation}\label{eq:disjoint_s_params}
    \vb{V}^- = \vb{S} \vb{V}^+ = \mqty(\vb{S}_1 & \vb{0} \\ \vb{0} & \vb{S}_2) \vb{V}^+
\end{equation}
Interconnecting the networks at port $k$ of $\vb{S}_1$ and port $\ell$ of $\vb{S}_2$ places the following constraint on the incident and reflected voltages at those ports:
\begin{align*}
    V_k^- &= V_\ell^+ \\
    V_k^+ &= V_\ell^-
\end{align*}
Inserting this constraint into (\ref{eq:disjoint_s_params}) allows one to solve for the matrix elements of the interconnected S-parameter to obtain \cite{filipsson_new_1981}:
\begin{equation}
    S^{\text{int}}_{ij} = S_{ij} + \frac{S_{i\ell}S_{kj}(1-S_{\ell k}) + S_{i\ell}S_{kk}S_{\ell j} + S_{ik}S_{\ell j}(1 - S_{k \ell}) + S_{ik}S_{\ell \ell}S_{kj}}{1 - S_{k\ell} - S_{\ell k} + S_{k\ell}S_{\ell k} - S_{kk}S_{\ell\ell}}
\end{equation} 

This process will work for interconnecting two ports belonging to any general S-parameter. To connect the two ports of disconnected S-parameters, you can construct a larger S-parameter that represents the disconnected network and then apply the above formula.

%% file: mainmatter/ch2/vector_fitting.tex
\section{Vector Fitting and Lossless Enforcement}\label{section:vector_fitting}
We have covered all of the methods that we will need for synthesis and interconnection of rational impedance functions. However, our goal is to apply these methods to impedance functions obtained from electromagnetic simulations. In general, the immittance or scattering parameters obtained from electromagnetic simulation software (e.g. Ansys HFSS, COMSOL, Sonnet) will be discretized over frequency. We want to be able to obtain a rational function that approximates responses obtained from simulations.

In this section, we will go over our process for approximating the discretized impedance with a lossless impedance function. In the first step of this process, we will use the method of vector fitting \cite{gustavsen_rational_1999}. Specifically, we use the ``Matrix Fitting Toolbox" MATLAB implementation available at \cite{matrix_fitting_toolbox} that uses the methods of \cite{gustavsen_rational_1999,gustavsen_improving_2006,deschrijver_macromodeling_2008}. Unfortunately, this first step does not provide a lossless model. Here, we go over the basic vector fitting method so we can see what needs to be done to obtain a lossless model from the traditional vector fitting result.

To see how the vector fitting algorithm works, we consider a single port system with a scalar response function of the form:
\begin{equation}\label{eq:scalar_rational}
    H(s) = \sum_{n=1}^N \frac{r_n}{s-p_n} + d + se
\end{equation}
We assume that we want to fit to a set of data points $\{(s_k, \tilde{H}_k)\}$ that correspond to the above response function. The first step of the vector fitting method is to choose a set of starting poles $\bar{p}_n$ in our desired frequency range. The vector fitting method then relocates the starting poles until the algorithm converges. This iterative process uses a weighting function $\sigma(s)$:
\begin{equation}\label{eq:weight_function}
    \sigma(s) = 1 + \sum_{n=1}^N \frac{c_n}{s-\bar{p}_n}
\end{equation}
Then, for each frequency point in the set of data points, an approximation can be made using the starting poles:
\begin{equation}\label{eq:vf_approx}
    \sigma(s_k) \tilde{H}_k = \left(1 + \sum_{n=1}^N \frac{c_n}{s_k-\bar{p}_n}\right) \tilde{H}_k \approx \sum_{n=1}^N \frac{\bar{r}_n}{s_k-\bar{p}_n} + d + s_ke
\end{equation}
For more information on where this approximation comes from, see \cite{gustavsen_rational_1999}. Alternatively, we can see that the choice of the weighting function and corresponding iterative process is a Santhanan-Koerner iterative process where a set of partial fractions are used as basis functions rather than polynomials \cite[Chapter 7]{passive_macromodeling}. We can rewrite (\ref{eq:vf_approx}) in a more suggestive form:
\begin{equation}
    \left(\sum_{n=1}^N \frac{\bar{r}_n}{s_k-\bar{p}_n} + d + s_ke\right) - \tilde{H}_k\left(\sum_{n=1}^N \frac{c_n}{s_k-\bar{p}_n}\right) \approx \tilde{H}_k
\end{equation}
where we can now see that for the set of data points, the above equation defines an overdetermined set of equations $\vb{A} \vb{x} = \vb{b}$ with the rows of $\vb{A}$
\begin{equation}
    A_k = \mqty( \dfrac{1}{s_k-\bar{p}_1} & \dots & \dfrac{1}{s_k-\bar{p}_N} & 1 & s_k & -\dfrac{\tilde{H}_k}{s_k-\bar{p}_1} & \dots & -\dfrac{\tilde{H}_k}{s_k-\bar{p}_N} ),
\end{equation}
the vector of unknowns $\vb{x} = \mqty(\bar{r}_1 & \dots & \bar{r}_N & d & e & c_1 & \dots & c_N)$, and $b_k = \tilde{H}_k$. Solving this equation in the least-squares sense, we can obtain a set of residues for the current iteration. The zeros of the new weight function (\ref{eq:weight_function}) with the most recent residues $c_i$ will be used to define the next set of poles for the next iteration. The zeros of this weight function are the eigenvalues of the matrix $\vb{P} - \vb{o}\vb{c}^T$ where $\vb{P}$ is a diagonal matrix of the starting poles, $\vb{o}$ is a column vector of ones, and $\vb{c}$ is a row vector containing the residues $c_i$. Taking the zeros of $\sigma(s)$ as the new set of poles, we can iterate the above process until the found poles converge. At this point, the final set of poles can be inserted into (\ref{eq:scalar_rational}) and that equation can be solved in the least squares sense to obtain the final approximation of the residues. 

The above method describes in general terms what the vector fitting iteration consists of, but the implementation of \cite{matrix_fitting_toolbox} makes use of an improved pole relocation method \cite{gustavsen_improving_2006} as well as a faster method that doesn't compute the residues $\bar{r}_i$ that are thrown away in the pole relocation process \cite{deschrijver_macromodeling_2008}. Furthermore, the above method can then be applied to multiport response functions (treated as vector functions) with the condition that all matrix elements have the same set of poles. This is precisely what we need for our rational impedance functions.

While we can use the vector fitting methods discussed above to obtain a good rational approximation of our simulated impedance functions, the models obtained can not be immediately used to construct circuit Hamiltonians since the resulting function will not be lossless. Fitting to an $N$-port discretized impedance function, the vector fitting algorithm will provide a rational function of the form
\begin{equation}
    \vb{Z}(s) = \sum_{j=1}^{M} \left[ \frac{\vb{R}_j}{s-p_j} + \frac{\vb{R}_j^\ast}{s-p_j^\ast} \right]
\end{equation}
where $\vb{R_j} \in \mathbb{C}^{N \times N}$, $p_j \in \mathbb{C}$ and $\myRe(p_j) < 0$. Some poles from the vector fitting are real while others are complex. Since the poles and residues are allowed to be complex, a function from the vector fitting process is not necessarily passive or lossless. There are methods that can check and enforce the passivity of these rational functions \cite[Chapters 9 \& 10]{passive_macromodeling}, but we want a strictly lossless impedance function. A lossless model would have $\vb{R}_j \in \mathbb{C}^{N \times N}$ and $p_j=i\omega_j$ for $\omega_j \in \mathbb{R}$.

\begin{figure}[h!]
    \centering
    
    \resizebox{0.75\textwidth}{!}{
    \begin{tikzpicture}[rotate=90, transform shape,
        circ/.style={circle, draw, solid, fill=white, inner sep=0.75pt,
                     label=#1,
                     node contents={},
                     },
        every label/.append style = {inner sep=2pt, font=\footnotesize}
                            ]
        \draw[thick, white, fill=white!75!blue] (0,0.75) arc (90:170:0.75) -- (-2,0.130236) arc(90:270:0.130236) -- (-0.738606,-0.130236) arc(190:270:.75);

        \draw[<->]   (-2.5,0) -- (.5,0) node[above, rotate=270] {${\scriptstyle\myRe(s)}$};
        \draw[<-]   (0,-3) -- (0,-1.5) node[left, rotate=270] {$\phantom{\scriptstyle\myIm(s)}$};
        \draw[<-] (0,3) -- (0,1.5);
        \draw (0,-3) node[right, rotate=270] {${{{\scriptstyle\myIm(s)}}}$};

        \node[fill=white] at (0,1.075) {\footnotesize $\approx$};
        \node[fill=white, rotate=180] at (0,-1.075) {\footnotesize $\approx$};
        \draw (0,1.5) -- (0,1.125);
        \draw (0,0) -- (0,1.04);
        \draw (0,-1.5) -- (0,-1.125);
        \draw (0,0) -- (0,-1.04);

        \draw[line width=0.5, black!40!green, -{Latex[length=3,width=3]}] (-1.5,0) -- (-0.035,0);
        \draw[line width=0.5, black!40!green, -{Latex[length=3,width=3]}] (-.25,0.5) -- (-0.0156525,0.0313049);
        \draw[line width=0.5, black!40!green, -{Latex[length=3,width=3]}] (-.25,-0.5) -- (-0.0156525,-0.0313049);
        
        \draw[red, fill=red] (-.5,0) circle (.05);
        \draw[red, fill=red] (-1.5,0) circle (.05);
        \draw[red, fill=red] (-.25,0.5) circle (.05);
        \draw[red, fill=red] (-.25,-0.5) circle (.05);
        
        \draw[line width=0.5, black!40!green, -{Latex[length=3,width=3]}] (-.65,1.5) -- (-0.035, 1.5);
        \draw[red, fill=red] (-.65,1.5) circle (.05);
        \draw[blue, fill=blue] (0,1.5) circle (.05);
        \draw[line width=0.5, black!40!green, -{Latex[length=3,width=3]}] (-.65,-1.5) -- (-0.035, -1.5);
        \draw[red, fill=red] (-.65,-1.5) circle (.05);
        \draw[blue, fill=blue] (0,-1.5) circle (.05);
        
        \draw[line width=0.5, black!40!green, -{Latex[length=3,width=3]}] (-.35,2) -- (-0.035, 2);
        \draw[red, fill=red] (-.35,2) circle (.05);
        \draw[blue, fill=blue] (0,2) circle (.05);
        \draw[line width=0.5, black!40!green, -{Latex[length=3,width=3]}] (-.35,-2) -- (-0.035, -2);
        \draw[red, fill=red] (-.35,-2) circle (.05);
        \draw[blue, fill=blue] (0,-2) circle (.05);
        
        \draw[line width=0.5, black!40!green, -{Latex[length=3,width=3]}] (-.5,2.5) -- (-0.035, 2.5);
        \draw[red, fill=red] (-.5,2.5) circle (.05);
        \draw[blue, fill=blue] (0,2.5) circle (.05);
        \draw[line width=0.5, black!40!green, -{Latex[length=3,width=3]}] (-.5,-2.5) -- (-0.035, -2.5);
        \draw[red, fill=red] (-.5,-2.5) circle (.05);
        \draw[blue, fill=blue] (0,-2.5) circle (.05);
        
        \draw[blue, fill=blue] (0,0) circle (.05);
        
    \end{tikzpicture}}

    \caption{Pole relocation to obtain a lossless model. Poles within the shaded region get mapped to $s=0$ while the rest are shifted onto the imaginary axis.}
    \label{fig:pole_relocation}
\end{figure}
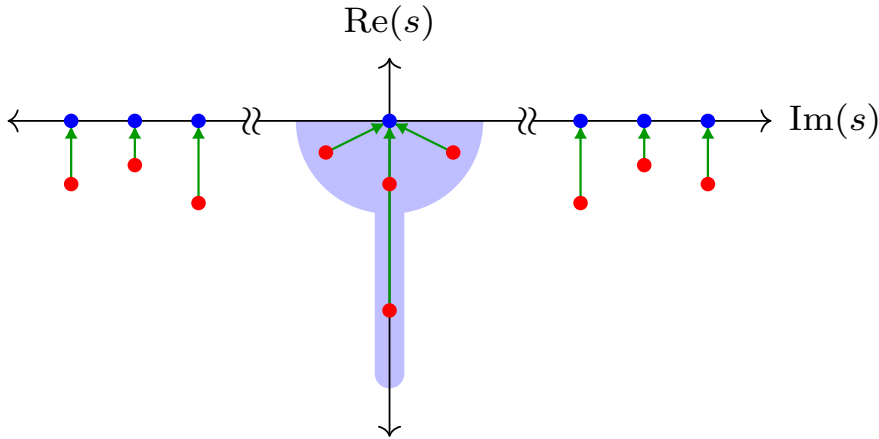

As a first step to obtaining a lossless function from the result of the vector fitting, we can extract the real symmetric parts of the residues. Then, if we shift the poles so that they all lie on the imaginary axis, the function will be purely imaginary. Since we want an impedance function of the form (\ref{eq:impedance}), we require that we end up with a DC pole at $s=0$. The poles from the fit will be complex, so we need to properly choose which poles we want to correspond to the DC pole. A range can be defined as shown in Fig.\ \ref{fig:pole_relocation} so that real poles are mapped to $s=0$, and any complex poles close to $s=0$ are also included. The rest of the poles are mapped to the imaginary axis such that they correspond to resonant modes. Then we can define a new DC residue that is the sum of the residues for poles mapped to $s=0$. The DC residue should be positive definite, and the range for shifting poles to $s=0$ should be chosen such that this is the case. After doing this, we are left with a lossless function that comes from the vector fitting process. All residues at this point are also constrained to be symmetric so that we can make sure our final result is also reciprocal.

Another problem we may encounter with the model from the vector fitting is that the residues of the resonant poles may not be rank-1 which is one of our requirements. To address this, we can split up each residue into a sum of rank-1 matrices that all correspond to the same pole. Using the eigendecomposition of a symmetric residue we can rewrite the matrix as:
\begin{equation}
    \vb{R} = \vb{Q}^T \vb{\Lambda} \vb{Q} = \sum_{n=1}^N (\vb{\Lambda})_{nn} \vb{q}^T_n \vb{q}^{\phantom{T}}_n
\end{equation}
where $\vb{q}_n$ is row $n$ of $\vb{Q}$, or in other words, $\vb{q}^T_n$ is the eigenvector corresponding to $(\vb{\Lambda})_{nn}$. The representation of the residue above shows us that if we do have a residue with general rank, we can always write it as a sum of rank-1 residues. The eigenvalues of the residue can also be used to gauge the weight of each of these rank-1 components. A threshold can then be set for the eigenvalue weight that is used to remove components that are small enough. Generally, if the pole is clearly visible in the frequency range of the discretized impedance function, the residue of the corresponding pole present in the vector fitting will be approximately rank-1. If the pole is outside of this frequency range, it is more likely that the vector fitting will return general rank residues.

After the above process, we will have transformed the rational impedance from the vector fitting into a lossless function. This comes at the cost of the new lossless function being a worse approximation of the original impedance. We can then finally apply a separate curve fitting process with the lossless part of the vector fit result as the intial guess. In this curve fitting process, the parameters of our function are the poles and residues of the impedance. For the DC residue parameters, we can use the upper triangular part of the matrix, and after each iteration of the fitting, we can check if the matrix is still positive definite. If the right poles from the vector fit result are chosen and shifted to $s=0$, this DC residue should remain positive definite at the end of this secondary fitting process. For each of the resonant residues, we only need to fit a row vector of turns ratios since we have made sure that these residues are rank-1. The resonant residues need to be positive semidefinite, which means that the turns ratios are allowed to be any real number. You can see this explicitly if you look at an arbitrary turns ratio row vector $\vb{r}_k$ containing any real numbers. The corresponding rank-1 residue $\vb{R}_k = \vb{r}_k^T \vb{r}_k$ will have one eigenvalue $\lambda_k$ such that
\begin{equation}
    \lambda_k = \Tr(\vb{R}_k) = \Tr(\vb{r}_k^T \vb{r}_k) = \vb{r}_k\vb{r}_k^T \geq 0
\end{equation}
In this secondary curve fitting, we also fit against the $\log$ of the magnitude of both the Z and S-parameters to make sure that the final rational function is as close as possible to the original impedance function. Using the $\log$ makes sure that the fitting process can accurately fit low magnitude features in the Z and S- parameters. Putting all of the above together, we have the following process for obtaining our lossless impedance function from simulation data:
\begin{enumerate}
    \item Use the traditional vector fitting method to obtain a rational impedance function that approximates the discretized impedance.
    \item Extract the real symmetric part of the residues for the result from the vector fitting and shift the poles onto the imaginary axis. Combine the set of residues corresponding to the poles close to $s=0$ into a DC residue and make sure that this is positive definite.
    \item Apply a secondary fitting process to the lossless model extracted from the vector fitting result to obtain a lossless impedance function that well approximates the original discretized impedance.
\end{enumerate}

%% file: mainmatter/ch3/chapter3.tex
\newcommand{\opa}{\hat{a}^{\phantom{\dagger}}}
\newcommand{\opb}{\hat{b}^{\phantom{\dagger}}}
\newcommand{\opad}{\hat{a}^{\dagger}}
\newcommand{\opbd}{\hat{b}^{\dagger}}

\chapter[Analysis of Impedance Functions of Qubit Networks]{Analysis of Impedance Functions\\ of Qubit Networks}
\label{chapter:analysis_impedance}

We will now see how a rational impedance function can be used to construct and analyze a circuit Hamiltonian of a superconducting multi-qubit circuit. Here we will primarily be focused on networks of transmon qubits \cite{transmon}. With the resulting Hamiltonian, we can estimate coupling rates between the qubits and resonators. We will also take this further and estimate effective qubit coupling rates that are mediated by resonant modes and tunable couplers, as well as shifts in the qubit and resonator frequencies. Finally, by including external ports (drive, readout, flux, etc.) in the impedance, we show how to estimate Purcell decay rates of the qubits.

While the focus here is on transmons, using the impedance to create circuit Hamiltonians can be extended to other qubit types (e.g. fluxonium \cite{fluxonium}). This is because you can obtain the multiport impedance of the linear part of the network and later on add Jospephson junctions or other lumped elements at the ports where they would be located in your circuit, just like we will do with the transmon circuits.

\input{mainmatter/ch3/transmon_network_hamiltonian.tex}

\input{mainmatter/ch3/external_ports.tex}

%% file: mainmatter/ch3/transmon_network_hamiltonian.tex
\section{Hamiltonian of a Transmon Network}\label{section:transmon_network_hamiltonian}
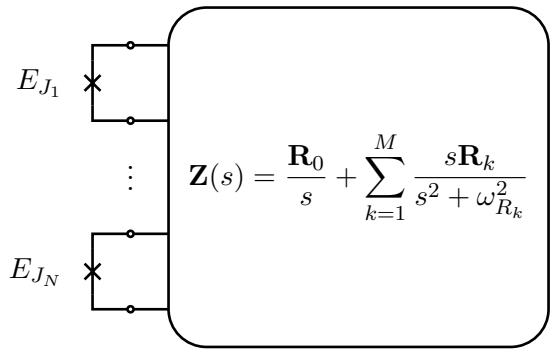
\begin{figure}[h!]
    \centering
    \begin{circuitikz}[line width=1pt]
        \ctikzset{bipoles/thickness=1, bipoles/length=1cm}
        \ctikzset { label/align = straight }
        
        \draw[rounded corners=.5cm] (0,2) -- (0,4.5) -- (5,4.5) -- (5,0) -- (0,0) -- (0,2);
        \node at (2.5,2.25) {$\vb{Z}(s) = \dfrac{\vb{R}_0}{s} + \displaystyle\sum_{k=1}^M \dfrac{s \vb{R}_k}{s^2 + \omega_{R_k}^2}$};

        \draw (0,0.5) to[short, -o] (-0.5,0.5) -- (-1, 0.5) to[barrier=$E_{J_N}$] (-1,1.5) to[short, -o] (-0.5,1.5) -- (0,1.5);

        \draw (0,3) to[short, -o] (-0.5,3) -- (-1, 3) to[barrier=$E_{J_1}$] (-1,4) to[short, -o] (-0.5,4) -- (0,4);

        \node at (-0.5,2.35) {$\vdots$};

    \end{circuitikz}
    \caption{Network of $N$ transmons represented by an arbitrary impedance function (\ref{eq:impedance}). The arbitrary impedance suggests that the transmons are coupled capacitively to each other and to resonant modes.}
    \label{fig:transmon_network}
\end{figure}

Here, we will consider an arbitrary $N$-port impedance of form (\ref{eq:impedance}) that is shunted by Josephson junctions as shown in Fig.\ \ref{fig:transmon_network}. At this point, we assume that any external ports are left open, but later on we will include them to estimate the Purcell decay rates of the qubits. We also assume that the impedance represents all of the transmon network except for the Josephson junctions, which we have added in afterwards. For example, the transmon shunt capacitances will be the shunt capacitances of the ports of the multiport impedance.

We can immediately write the circuit Hamiltoninan for Fig.\ \ref{fig:transmon_network} using the results of Section \ref{section:impedance_hamiltonian}. Now since we include the junctions, we use (\ref{eq:impedance_hamiltonian}) to find:
\begin{equation}\label{eq:transmon_hamiltonian}
    \mathcal{H} = \frac{1}{2} \vb{Q}^T \vb{C}^{-1} \vb{Q} + \frac{1}{2} \vb{\Phi}^T \vb{M} \vb{\Phi} - \sum_{i=1}^N E_{J_i} \cos(\frac{2\pi}{\Phi_0} \Phi_{J_i})
\end{equation}
Here we will represent the junction port fluxes using $\Phi_{J_i}$ instead of $\Phi_{P_i}$. Since the impedance function can also be represented as a CL cascade, we note that any analysis of Hamiltonian (\ref{eq:transmon_hamiltonian}) will also apply to transmon network built from arbitrary CL cascade. We can also promote our conjugate variables $\Phi_{i}$ and $Q_i$ to operators, such that they satisfy the commutation relation $[\hat{\Phi}_i, \hat{Q}_j] = i\hbar \delta_{ij}$. We next want to expand the Hamiltonian to obtain coupling rates between the different branches. Before doing this, we can introduce some notation that will help simplify our expressions. First, we define the effective port capacitance:
\begin{equation}\label{eq:effective_capacitiance}
    \tilde{C}_i = \frac{1}{(\vb{C}^{-1})_{ii}}
\end{equation}
With this we also define an effective charging energy $\tilde{E}_{C_i} = e^2/2\tilde{C}_i$. Generally, this charging energy is defined using just the branch shunt capacitance. However, since we have the full capacitance matrix, we can avoid this approximation. Note that from (\ref{eq:impedance_cap_inverse}) we can see that for the inductive branches, $\tilde{C}_{R_k} = 1$. We can also define an inductive energy for the inductive branches:
\begin{equation}
    E_{L_{k}} = \frac{\Phi_0^2}{4\pi^2 L_{R_k}}
\end{equation}
Next, we rescale the flux and charge variables: $\hat{\phi} = 2\pi\hat{\Phi}/\Phi_0$ and $\hat{q} = \hat{Q}/2e$. Their commutation relation is $[\hat{\phi}_i, \hat{q}_j] = i\delta_{ij}$. The Hamiltonian can now be rewritten using the above definitions:
\begin{align}
\begin{split}
    \hat{H} &= \sum_{i=1}^N \left(  4\tilde{E}_{C_i}\hat{q}_{J_i}^2 - E_{J_i}\cos(\hat{\phi}_{J_i})  + \sum_{j > i}^{j=N} 4e^2(\vb{C}^{-1})_{ij}\;  \hat{q}_{J_i}\hat{q}_{J_j} \right) \\
    &\quad + \sum_{k=1}^M \left( 4\tilde{E}_{C_i}\hat{q}_{R_k}^2 - \frac{E_{L_k}}{2}\hat{\phi}_{R_k}^2  + \sum_{\ell > k}^{\ell = M} 4e^2(\vb{C}^{-1})_{k\ell}\;  \hat{q}_{R_k}\hat{q}_{R_\ell} \right) \\ 
    &\quad + \sum_{i=1}^N \sum_{k=1}^M 4e^2(\vb{C}^{-1})_{ik}\; \hat{q}_{J_i}\hat{q}_{R_k}
\end{split}
\end{align}
Note that above and in the following, we will use the index labels $i$ and $j$ to refer to the transmon or junction branches. We will use the labels $k$ and $\ell$ to refer to the internal resonant or inductive branches. Next, we introduce creation and annihilation operators that will correspond to the transmons ($\opbd$ and $\hat{b}$) or resonator ($\opad$ and $\hat{a}$) branches. Since the operators are bosonic, they have the commutation relations $[\opb_i, \opbd_j]=\delta_{ij}$ and $[\opa_k, \opad_\ell]=\delta_{k\ell}$. We can rewrite our operators $\hat{\phi}_{J_i}$ and $\hat{q}_{J_i}$ in terms of the annihilation and creation operators $\opbd$ and $\hat{b}$:
\begin{align}
    \hat{\phi}_{J_i} &= \Bigg( \frac{2\tilde{E}_{C_i}}{E_{J_i}} \Bigg)^{\!\!1/4} (\opbd_{i} + \opb_{i}) \\
    \hat{q}_{J_i} &= \Bigg( \frac{2E_{J_i}}{\tilde{E}_{C_i}} \Bigg)^{\!\!1/4} (\opbd_{i} - \opb_{i})
\end{align}
The operators $\hat{\phi}_{R_k}$ and $\hat{q}_{R_k}$ can be rewritten in a similar way with $E_{L_k}$ instead of $E_J$. We can again rewrite the Hamiltonian in terms of these new bosonic operators:
\begin{align}
\begin{split}\label{eq:transmon_resonator_ham}
    \frac{1}{\hbar}\hat{H} &= \sum_{i=1}^N \left(  \omega_{J_i}\opbd_i\opb + \frac{\beta_{J_i}}{2}\opbd_i\opbd_i\opb_i\opb_i  + \sum_{j > i}^{j=N} g_{ij} (\opbd_i\opb_j + \opb_i\opbd_j - \opbd_i\opbd_j - \opb_i\opb_j) \right) \\
    &\quad + \sum_{k=1}^M \left( \omega_{R_k}\opad_k\opa_k + \sum_{\ell > k}^{\ell=M} g_{R_k,R_\ell} (\opad_k\opa_\ell + \opa_k\opad_\ell - \opad_k\opad_\ell - \opa_k\opa_\ell) \right) \\ 
    &\quad + \sum_{i=1}^N \sum_{k=1}^M g_{i,R_k}(\opbd_i\opa_k + \opb_i\opad_k - \opbd_i\opad_k - \opb_i\opa_k)
\end{split}
\end{align}

Above we have neglected constant terms and used the Duffing oscillator approximation to represent the transmon as a harmonic oscillator with a quartic perturbation \cite{transmon}. In the above Hamiltonian, we also have the following quantities:
\begin{alignat}{2}
    &\hbar\;\omega_{J_i} &&= \sqrt{8E_{J_i}\tilde{E}_{C_i}} + \beta_{J_i} \\
    &\hbar\;\beta_{J_i} &&= -\tilde{E}_{C_i} \\
    &\hbar\;\omega_{R_k} &&= \sqrt{8E_{L_k}\tilde{E}_{C_k}} \\
    &\hbar\; g_{ij} &&= e^2 (\vb{C}^{-1})_{ij} \left( \frac{E_{J_i} E_{J_j}}{4 \tilde{E}_{C_i} \tilde{E}_{C_j}  } \right)^{\!\!1/4} \label{eq:node_coupling}
\end{alignat}
The other coupling rates $g_{R_k,R_\ell}$ and $g_{i,R_k}$ are defined similarly but $E_J$ is replaced with $E_L$ as needed. For the impedance case, all couplings between the internal resonant modes will be zero since the resonator block in (\ref{eq:impedance_cap_inverse}) is the identity. Here, we leave the coupling term since this allows us to treat the general CL cascade case. This form of the Hamiltonian immediately gives us the coupling rates between all the qubits and resonant modes in the circuit. Thus, if we start with a CL cascade network or a rational impedance function (\ref{eq:impedance}), we can immediately compute the qubit frequencies as well as the coupling rates between the qubits and the other resonant modes present in the network.

\subsection{Effective Qubit Coupling in a Transmon Network}\label{section:effective_coupling}

In this section, we aim to find the effective coupling rates between the qubits that are mediated by the resonant modes in Hamiltonian (\ref{eq:transmon_resonator_ham}). We can even treat some of the qubits as couplers by including them in the group of resonators. To find these effective coupling rates, we want to decouple the qubits from the couplers and resonators in the Hamiltonian using a Schrieffer-Wolff transformation \cite{bravyi_SW}.

To extract the effective coupling rate, we will block diagonalize the Hamiltonian (\ref{eq:transmon_resonator_ham}) to eliminate the terms that couple the qubits and couplers. We will define an operator that will approximately block diagonalize the Hamiltonian, similar to the methods of \cite{tunable_coupler,tunable_coupler_ext}. The difference here is that we will allow for an arbitrary number of qubits and couplers. It is also possible to truncate the bosonic operators and then block diagonalize the Hamiltonian numerically \cite{bravyi_SW}. In our approach, we will instead aim to find formulas in terms of the resonant frequencies and coupling rates. This can help motivate what parameters are important to think about when designing circuits.

To approximately block diagonalize the Hamiltonian, we will make use of the following transformation (note that from this point onward we set $\hbar = 1$): 
\begin{equation}\label{eq:sw_transform}
    \hat{U} = e^{\hat{S}} = \exp\Bigg( \sum_{i=1}^N \sum_{k=1}^M \underbrace{\left[ \frac{g_{i,R_k}}{\Delta_{i,R_k}} (\opbd_i\opa_k - \opb_i\opad_k) - \frac{g_{i,R_k}}{\Sigma_{i,R_k}}(\opbd_i\opad_k - \opb_i\opa_k) \right]}_{\hat{S}_{ik}} \Bigg) 
\end{equation}
where we have defined $\Delta_{i,R_k} = \omega_{J_i} - \omega_{R_k}$ and $\Sigma_{i,R_k} = \omega_{J_i} + \omega_{R_k}$. We will see explicitly that this transformation will not exactly block diagonalize Hamiltonian (\ref{eq:transmon_resonator_ham}), but also that in the relevant parameter regimes, the resulting off-diagonal blocks can be neglected. Before applying the transformation, we first separate out the Hamiltonian into different parts:
\begin{alignat}{2}
    &\hat{H}^0 &&= \sum_{i=1}^N \omega_{J_i}\opbd_i\opb_i + \sum_{k=1}^M \omega_{R_k}\opad_k\opa_k \\
    &\hat{H}^1_Q &&= \sum_{i=1}^N \sum_{j > i}^{j=N} g_{ij} (\opbd_i\opb_j + \opb_i\opbd_j - \opbd_i\opbd_j - \opb_i\opb_j)  \\
    &\hat{H}^1_R &&= \sum_{k=1}^M \sum_{\ell > k}^{\ell = M} g_{R_k,R_\ell} (\opad_k\opa_\ell + \opa_k\opad_\ell - \opad_k\opad_\ell - \opa_k\opa_\ell) \\
    &\hat{H}^2 &&= \sum_{i=1}^N \sum_{k=1}^M g_{i,R_k}(\opbd_i\opa_k + \opb_i\opad_k - \opbd_i\opad_k - \opb_i\opa_k) \\
    &\hat{H}^{NL} &&= \sum_{i=1}^N \frac{\beta_{J_i}}{2}\opbd_i\opbd_i\opb_i\opb_i + \sum_{k=1}^M \frac{\alpha_{R_k}}{2} \opad_k\opad_k\opa_k\opa_k \label{eq:H_nonlinear}
\end{alignat}
Here we have assumed that some of the qubits in the system have been grouped with the couplers, hence there are new nonlinear terms in $\hat{H}^{NL}$ that are not explicitly present in (\ref{eq:transmon_resonator_ham}). The terms $\hat{H}^1_Q$ and $\hat{H}^1_R$ represent the interactions within the transmon and resonator blocks, respectively. The term $\hat{H}^2$ contains the interactions between the qubits and resonators, and is the term that we would like to approximately eliminate in our block diagonalization.

In this section we will not discuss the nonlinear terms. If $\beta,\alpha \ll \Delta$, then the results of this section hold. In the next section we will take a closer look at the more general case to see what effects the nonlinear terms have on the system. To find our transformed Hamiltonian, we can use the following expansion \cite{winkler2003}:
\begin{equation}\label{eq:BCH}
    \hat{U} \hat{H} \hat{U}^\dag = e^{\hat{S}} \hat{H} e^{-\hat{S}} = \hat{H} + [\hat{S},\hat{H}] + \frac{1}{2!}[\hat{S},[\hat{S},\hat{H}]] + \cdots
\end{equation}
We aim to expand to second order in the coupling rates $g$. To see what terms will be relevant, we can start by computing the commutators of the operator $\hat{S}$ with the different parts of the Hamiltonian $\hat{H}^0$, $\hat{H}^1$, and $\hat{H}^2$. For the first commutator $[\hat{S}, \hat{H}^0]$, we show how to break this up into smaller problems that can be easily handled. We break up this commutator in the following way:
\begin{equation}\label{eq:H0_comm_prelim}
    [\hat{S}, \hat{H}^0] = \Bigg[ \sum_{i=1}^N \sum_{k=1}^M \hat{S}_{ik}\;,\;\; \sum_{j=1}^N \omega_{J_j}\opbd_j\opb_j\Bigg] + \left[ \sum_{i=1}^N \sum_{k=1}^M\hat{S}_{ik}\;,\;\;\sum_{\ell=1}^M \omega_{R_\ell}\opad_\ell\opa_\ell \right]
\end{equation}
Notice how in the first term, when $i \neq j$, the commutator is 0. Thus, by computing the commutator for the case of $i=j$, we can find the total contribution of the first term:
\begin{equation}\label{eq:comm_H0_Q}
    \left[\hat{S}_{ik},\; \omega_{J_i} \opbd_i\opb_i\right] = \omega_{J_i} g_{i,R_k} \left( -\frac{1}{\Delta_{i,R_k}}(\opbd_i\opa_k + \opb_i\opad_k) + \frac{1}{\Sigma_{i,R_k}}(\opbd_i\opad_k + \opb_i\opa_k) \right)
\end{equation}
where we have used the identities $[\hat{b}, \opbd\hat{b}] = \hat{b}$ and $[\opbd, \opbd\hat{b}] = -\opbd$. We can do something similar for the second term for the case $k=\ell$:
\begin{equation}\label{eq:comm_H0_R}
    \left[\hat{S}_{ik},\; \omega_{R_k}\opad_k\opa_k \right] = \omega_{R_k} g_{i,R_k} \left( \frac{1}{\Delta_{i,R_k}} (\opbd_i\opa_k + \opb_i\opad_k) + \frac{1}{\Sigma_{i,R_k}} (\opbd_i\opad_k + \opb_i\opa_k) \right)
\end{equation}
Combining the two above terms, we find that commutator (\ref{eq:H0_comm_prelim}) is:
\begin{equation}
    [\hat{S}, \hat{H}^0] = \sum_{i=1}^N \sum_{k=1}^M g_{i,R_k}(-\opbd_i\opa_k - \opb_i\opad_k + \opbd_i\opad_k + \opb_i\opa_k) = -\hat{H}^2
\end{equation}

The transformation (\ref{eq:sw_transform}) has been chosen specifically such that the above condition is met. This is typically the desired case when attempting to block-diagonalize a Hamiltonian \cite{winkler2003, richer_masters}. Next we can begin to think about the commutator $[\hat{S}, \hat{H}^1_Q]$. Here, there will be non-zero contributions for the following case: 
\begin{equation}\label{eq:comm_H1_Q}
    [\hat{S}_{ik}, \hat{H}^1_{Q,ij}] = -g_{ij}g_{i,R_k}\left( \frac{1}{\Delta_{i,R_k}} - \frac{1}{\Sigma_{i,R_k}} \right)\left(\opbd_j\opa_k + \opb_j\opad_k - \opbd_j\opad_k - \opb_j\opa_k \right)
\end{equation}
Similar terms can be found for $[\hat{S}_{jk}, \hat{H}^1_{Q,ij}]$, where instead $i$ and $j$ are swapped on the RHS in the above expression. Not too different is the result for $[\hat{S}, \hat{H}^1_R]$. The contributing terms will be of the form:
\begin{equation}\label{eq:comm_H1_R}
    [\hat{S}_{ik}, \hat{H}^1_{R,k\ell}] = g_{R_k,R_\ell}g_{i,R_k} \left( \frac{1}{\Delta_{i,R_k}} - \frac{1}{\Sigma_{i,R_k}} \right)\left( \opbd_i \opa_\ell + \opb_i\opad_\ell - \opbd_i\opad_\ell - \opb_i\opa_\ell \right)
\end{equation}
and also the terms $[\hat{S}_{i\ell}, \hat{H}^1_{R,k\ell}]$ where the expression is the same as the above with $k$ and $\ell$ swapped on the RHS. These terms add ``corrections" to the qubit-resonator coupling terms which are the terms that we are trying to eliminate. Generally when performing Schrieffer-Wolff transformations these terms are cancelled by the inclusion of higher order additions to the operator $\hat{S}$ \cite{winkler2003,richer_masters}. Here we will only use the transformation (\ref{eq:sw_transform}), but we will see shortly that (\ref{eq:comm_H1_Q}) and (\ref{eq:comm_H1_R}) can be neglected when compared to the other terms in the expansion of the transformation.

To compute $[\hat{S}, \hat{H}^2]$, we once again break up the computation into simpler pieces and compute $[\hat{S}_{ik}, \hat{H}^2_{j\ell}]$ for different cases. For the trivial case when $i\neq j$ and $k\neq\ell$, the commutator vanishes. For the case of $i\neq j$ and $k=\ell$, we find:
\begin{equation}\label{eq:qubit_eff_coup_adj}
    [\hat{S}_{ik}, \hat{H}^2_{jk}] = g_{i,R_k}g_{j,R_k} \left( \frac{1}{\Delta_{i,R_k}} - \frac{1}{\Sigma_{i,R_k}} \right)(\opbd_i\opb_j + \opb_i\opbd_j - \opbd_i\opbd_j - \opb_i\opb_j)
\end{equation}
This term contributes to the new effective qubit-qubit coupling rates. There is a similar contributing term when $i$ and $j$ are swapped. Next, we can look at the case $i=j$ and $k\neq \ell$:
\begin{equation}\label{eq:res_eff_coup_adj}
    [\hat{S}_{ik}, \hat{H}^2_{i\ell}] = -g_{i,R_k}g_{i,R_\ell}\left(  \frac{1}{\Delta_{i,R_k}} + \frac{1}{\Sigma_{i,R_k}}  \right) (\opad_k\opa_\ell + \opa_k\opad_\ell - \opad_k\opad_\ell - \opa_k\opa_\ell)
\end{equation}
This term is similar to (\ref{eq:qubit_eff_coup_adj}) in that it contributes to effective coupling rates, but now between the resonators. Just like the qubits, the resonators have an indirect coupling between them mediated through the qubits. Finally, for the case of $i=j$ and $k=\ell$, we have (neglecting constant terms):
\begin{align}
\begin{split}\label{eq:qubit_res_shifts}
    [\hat{S}_{ik}, \hat{H}^2_{ik}] &= g^2_{i,R_k}\left( \frac{1}{\Delta_{i,R_k}} - \frac{1}{\Sigma_{i,R_k}} \right)(2\opbd_i\opb_i - \opbd_i\opbd_i - \opb_i\opb_i) \\
    & \quad - g^2_{i,R_k}\left( \frac{1}{\Delta_{i,R_k}} + \frac{1}{\Sigma_{i,R_k}} \right)(2\opad_k\opa_k - \opad_k\opad_k - \opa_k\opa_k)
\end{split}
\end{align}
This term brings shifts to the qubit and resonator frequencies due to the couplings. Now we can combine the above results to expand the transformed Hamiltonian $\hat{U} \hat{H} \hat{U}^\dag$ to second order in the couplings $g$. We assume that these couplings are small in magnitude compared to the qubit and resonator frequencies and detunings. Expanding the transformation to the second order commutator term, using $[\hat{S}, \hat{H}^0] = -\hat{H}^2$, and removing terms that are third order in $g$, we end up with:
\begin{equation}
    \hat{U} \hat{H} \hat{U}^\dag \approx \hat{H} + [\hat{S},\hat{H}] + \frac{1}{2!}[\hat{S},[\hat{S},\hat{H}]] = \hat{H}^0 + \hat{H}^1 + [\hat{S}, \hat{H}^1] + \frac{1}{2}[\hat{S}, \hat{H}^2]
\end{equation}
The resulting approximation of the transformed Hamiltonian is not purely block diagonal due to the term $[\hat{S}, \hat{H}^1]$ which is still present and contains terms that are second order in $g$. However, due to the parameter regimes for most superconducting circuit architectures, the terms from this commutator will actually contribute much less than the other terms in the expansion. Thus, the new terms that come from (\ref{eq:comm_H1_Q}) and (\ref{eq:comm_H1_R}) can be neglected. To understand why, we now explain the needed conditions for this with some physical motivation.

\tikzset{
  pics/res_node/.style args={#1,#2}{
     code={
        \draw[rounded corners=0.25cm] (-0.5,0) -- (-0.5,.5) -- (0.5,0.5) -- (0.5,-0.5) -- (-0.5,-0.5) -- (-0.5,0);
        \node (#1) at (0,0) {#2};
     }
  }
}

\begin{figure}[h!]
    \centering
    \begin{subfigure}{.5\textwidth}
        \centering
        \begin{tikzpicture}[line width=1pt]
            \draw (0,0) pic{res_node={R_k, $R_k$}};
            \draw (-2.5,0) pic{res_node={Qi, $Q_i$}};
            \draw (2.5,0) pic{res_node={Qj, $Q_j$}};
            \draw[<->] (-2,0) -- (-.5,0) node [midway, below] {$g_{i,R_k}$}; 
            \draw[<->] (2,0) -- (.5,0) node [midway, below] {$g_{j,R_k}$};
            \draw[rounded corners=0.25cm, <->] (-2.5,.5) -- (-2.5, 1) -- (2.5,1) -- (2.5,.5);
            \node at (0,1.3) {$g_{ij}$};
        \end{tikzpicture}
        \caption{$g_{ij} \ll g_{i,R_k}, g_{j,R_k}$}
        \label{fig:2Q1R_1}
    \end{subfigure}%
    \begin{subfigure}{.5\textwidth}
        \centering
        \begin{tikzpicture}[line width=1pt]
            \draw (0,0) pic{res_node={R_k, $Q_i$}};
            \draw (-2.5,0) pic{res_node={Qi, $R_k$}};
            \draw (2.5,0) pic{res_node={Qj, $R_\ell$}};
            \draw[<->] (-2,0) -- (-.5,0) node [midway, below] {$g_{i,R_k}$}; 
            \draw[<->] (2,0) -- (.5,0) node [midway, below] {$g_{i,R_\ell}$};
            \draw[rounded corners=0.25cm, <->] (-2.5,.5) -- (-2.5, 1) -- (2.5,1) -- (2.5,.5);
            \node at (0,1.3) {$g_{R_k,R_\ell}$};
        \end{tikzpicture}
        \caption{$g_{R_k,R_\ell} \ll g_{i,R_k},g_{i,R_\ell}$}
        \label{fig:1Q2R_1}
    \end{subfigure}
    \caption{}
    \label{fig:chain_layout}
\end{figure}

In Fig.\ \ref{fig:2Q1R_1}, we can see an example schematic where two qubits are coupled directly and through a resonant mode. Notably, the direct coupling between the qubits will be much smaller than the coupling between the qubits and the resonator such that $g_{ij} \ll g_{i,R_k}, g_{j,R_k}$. This type of layout is typically found in superconducting circuits with qubit chains or grids coupled by resonators \cite{solgun_sirf,rapid_multiplexed_readout} as well as systems with qubits coupled by tunable couplers \cite{tunable_coupler,high_fidelity_cz_iswap_tc,long_distance_coupler}. For these types of couplings, it is clear that the coefficients $g_{ij}g_{i,R_k}\Delta_{i,R_k}^{-1}$ and $g_{ij}g_{i,R_k}\Sigma_{i,R_k}^{-1}$ in (\ref{eq:comm_H1_Q}) generated by $[\hat{S}_{ik}, \hat{H}^1_{Q,ij}]$  will be small compared to the original term that couples the qubits and resonators, so this contribution can be neglected. If we look at the layout in Fig.\ \ref{fig:1Q2R_1}, we find the condition $g_{R_k,R_\ell} \ll g_{i,R_k},g_{i,R_\ell}$. For this layout, a similar argument can be made for why the new terms in (\ref{eq:comm_H1_R}) generated by $[\hat{S}_{ik}, \hat{H}^1_{R,k\ell}]$ can be neglected as well.

\begin{figure}[h!]
    \centering
    \begin{subfigure}{.4\textwidth}
        \centering
        \begin{tikzpicture}[line width=1pt]
            \draw (0,2.5) pic{res_node={R_k, $R_k$}};
            \draw (0,0) pic{res_node={Qi, $Q_i$}};
            \draw (2.5,0) pic{res_node={Qj, $Q_j$}};
            \draw[<->] (0.5,0) -- (2,0) node [midway, below] {$g_{ij}$};
            \draw[<->] (0,.5) -- (0,2) node [midway, left] {$g_{i,R_k}$};
            \draw[rounded corners=0.25cm, <->] (2.5,0.5) -- (2.5,1.5) -- (1.5,2.5) -- (0.5,2.5);
            \node at (2.5, 2.25) {$g_{j,R_k}$};
        \end{tikzpicture}
        \caption{$g_{j,R_k} \ll g_{i,R_k},g_{ij}$}
        \label{fig:2Q1R_2}
    \end{subfigure}%
    \begin{subfigure}{.4\textwidth}
        \centering
        \begin{tikzpicture}[line width=1pt]
            \draw (0,2.5) pic{res_node={R_k, $Q_i$}};
            \draw (0,0) pic{res_node={Qi, $R_k$}};
            \draw (2.5,0) pic{res_node={Qj, $R_\ell$}};
            \draw[<->] (0.5,0) -- (2,0) node [midway, below] {$g_{R_k,R_\ell}$};
            \draw[<->] (0,.5) -- (0,2) node [midway, left] {$g_{i,R_k}$};
            \draw[rounded corners=0.25cm, <->] (2.5,0.5) -- (2.5,1.5) -- (1.5,2.5) -- (0.5,2.5);
            \node at (2.5, 2.25) {$g_{i,R_\ell}$};
        \end{tikzpicture}
        \caption{$g_{i,R_\ell} \ll g_{i,R_k},g_{R_k,R_\ell}$}
        \label{fig:1Q2R_2}
    \end{subfigure}
    \caption{}
    \label{fig:angle_layout}
\end{figure}

While many cases are covered by Fig.\ \ref{fig:2Q1R_1}, this leaves out cases where resonant modes are not used for coupling. For example, Fig.\ \ref{fig:2Q1R_2} represents a layout where one qubit has strong direct coupling to another qubit as well as a resonator that is used for readout. You would have this layout in a capacitively coupled transmon chain where each has its own readout resonator \cite{Barends2016}. It is also clear that the coefficients in (\ref{eq:comm_H1_Q}) will be small under the condition $g_{j,R_k} \ll g_{i,R_k},g_{ij}$ compared to the original qubit-resonator coupling rates. If we consider the layout in Fig.\ \ref{fig:1Q2R_2}, we now have the condition $g_{i,R_\ell} \ll g_{i,R_k},g_{R_k,R_\ell}$ and again we can see that the new terms from (\ref{eq:comm_H1_R}) for this layout can be neglected. This layout will typically be seen in architectures that make use of Purcell filters for readout \cite{rapid_multiplexed_readout,karamlou2023probing}. In the rest of our analysis, we restrict ourselves to chip architectures with the above layouts and constraints on the couplings. While the restrictions are not ideal, it will not pose a problem as most architectures will not violate the above conditions. However, it is good to be wary of these restrictions since the approximations we will now make would break down if they are violated.

Given the restrictions on the circuit layouts that have been discussed, we can further reduce the expansion of the transformed Hamiltonian. We have determined that the new qubit-resonator coupling rates in (\ref{eq:comm_H1_Q}) and (\ref{eq:comm_H1_R}) that come from $[\hat{S},\hat{H}^1]$ can be neglected in the final Hamiltonian. Thus, the transformed Hamiltonian to second order in $g$ with the additional restrictions given our desired circuit layout is:
\begin{equation}
    \hat{U} \hat{H} \hat{U}^\dag \approx \hat{H}^0 + \hat{H}^1 + \frac{1}{2}[\hat{S}, \hat{H}^2]
\end{equation}
We have already found the contributions to the commutator $[\hat{S}, \hat{H}^2]$ in (\ref{eq:qubit_eff_coup_adj}), (\ref{eq:res_eff_coup_adj}) and (\ref{eq:qubit_res_shifts}). This allows us to write the new effective Hamiltonian, and with the approximations we've made, this Hamiltonian will have no coupling between the qubits and resonators. The effective Hamiltonian is:
\begin{align}
\begin{split}\label{eq:eff_ham}
    \hat{H}_{\text{eff}} &\approx \sum_{i=1}^N \left(  \tilde{\omega}_{J_i}\opbd_i\opb + \frac{\beta_{J_i}}{2}\opbd_i\opbd_i\opb_i\opb_i  + \sum_{j > i}^{j=N} \tilde{g}_{ij} (\opbd_i\opb_j + \opb_i\opbd_j - \opbd_i\opbd_j - \opb_i\opb_j) \right) \\
    &\quad + \sum_{k=1}^M \left( \tilde{\omega}_{R_k}\opad_k\opa_k + \frac{\alpha_{R_k}}{2} \opad_k\opad_k\opa_k\opa_k + \sum_{\ell > k}^{\ell=M} \tilde{g}_{R_k,R_\ell} (\opad_k\opa_\ell + \opa_k\opad_\ell - \opad_k\opad_\ell - \opa_k\opa_\ell) \right)
\end{split}
\end{align}
where now the qubit and resonator frequencies as well as the direct coupling rates are shifted. Note that at this point we haven't included any effect of the nonlinear terms and this assumption will hold for $\beta,\alpha \ll \Delta$. In the next section we will see some of the effects that these nonlinear terms have. The shifts to the qubit and resonator frequencies are:
\begin{align}
    \tilde{\omega}_{J_i} &\approx \omega_{J_i} + \sum_{k=1}^M g^2_{i,R_k}\left( \frac{1}{\Delta_{i,R_k}} - \frac{1}{\Sigma_{i,R_k}} \right) \label{eq:eff_qubit_freq} \\
    \tilde{\omega}_{R_k} &\approx \omega_{R_k} - \sum_{i=1}^N g^2_{i,R_k}\left( \frac{1}{\Delta_{i,R_k}} + \frac{1}{\Sigma_{i,R_k}} \right) \label{eq:eff_res_freq}
\end{align}
Note that in the above we have left out the high frequency rotating terms that arise in (\ref{eq:qubit_res_shifts}). The new effective direct coupling terms are defined as follows:
\begin{align}
    \tilde{g}_{ij} &\approx g_{ij} + \frac{1}{2} \sum_{k=1}^M  g_{i,R_k}g_{j,R_k} \left( \frac{1}{\Delta_{i,R_k}} + \frac{1}{\Delta_{j,R_k}}  - \frac{1}{\Sigma_{i,R_k}} - \frac{1}{\Sigma_{j,R_k}}\right) \label{eq:eff_qubit_coupling}\\
    \tilde{g}_{R_k,R_\ell} &\approx g_{R_k,R_\ell} - \frac{1}{2}\sum_{i=1}^N g_{i,R_k}g_{i,R_\ell} \left( \frac{1}{\Delta_{i,R_k}} + \frac{1}{\Delta_{i,R_\ell}} + \frac{1}{\Sigma_{i,R_k}} + \frac{1}{\Sigma_{i,R_\ell}} \right)
\end{align}
The results for the frequency shifts and effective coupling closely resemble the results of \cite{tunable_coupler,tunable_coupler_ext}, except now there is a sum component that adds the contributions of all the couplers or qubits that are present. The resulting expression is no real surprise, but here we have shown that a multi-qubit multi-coupler system can be treated in the above way to estimate effective couplings between the qubits.

This final result shows us that starting from a rational impedance function (\ref{eq:impedance}) or a CL cascade network representation of our multi-qubit circuit, we can estimate the effective coupling rates between qubits. If we are able to obtain the rational impedance function from an electromagnetic model of your device, then the remaining characterization process is straightforward given the above formulas. Another useful feature of the above process is that designating some of the qubits as tunable couplers does not change the process or the resulting formulas for the estimated effective coupling.

\subsection{Effects of the Nonlinear Terms}
We now find the effects of the nonlinear terms that were neglected in the last section. We are primarily concerned with finding approximations to the shifts to the qubit and coupler anharmonicities, as well as the state dependent dispersive shifts for the qubits and couplers. To do this, we find how the operators $\opb_i$ and $\opa_k$ are transformed by the operator $\hat{U}$ in (\ref{eq:sw_transform}). To do this, we expand the transformations of these operators perturbatively. Then we can obtain the relevant terms that were previously neglected. We can again use the \ref{eq:BCH} and expand to the second order commutator term which gives:
\begin{equation}
    \hat{U}\opb_i \hat{U}^\dag \approx \opb_i + [\hat{S},\opb_i] + \frac{1}{2}[\hat{S}, [\hat{S},\opb_i]]
\end{equation}
A similar expansion is used for $\opa_k$. The operator $\opb_i$ transformed with this expansion is:
\begin{equation}
    \hat{U}\opb_i \hat{U}^\dag \approx \opb_i - \sum_{k=1}^M g_{i,R_k} \left( \frac{1}{\Delta_{i,R_k}} \opa_k - \frac{1}{\Sigma_{i,R_k}} \opad_k \right) - \frac{1}{2}\sum_{j=1}^N \sum_{k=1}^M \frac{g_{i,R_k}g_{j,R_k}}{\Delta_{i,R_k}\Delta_{j,R_k}} \opb_j
\end{equation}
where we have also neglected terms of order $g^2/(\Delta\Sigma)$ and $g^2/\Sigma^2$ or higher. A similar expansion can be found for the transformed $\opa_k$:
\begin{equation}
    \hat{U}\opa_k \hat{U}^\dag \approx \opa_k + \sum_{i=1}^N g_{i,R_k} \left( \frac{1}{\Delta_{i,R_k}}\opb_i + \frac{1}{\Sigma_{i,R_k}}\opbd_i \right) - \frac{1}{2}\sum_{i=1}^N\sum_{\ell=1}^M \frac{g_{i,R_k}g_{i,R_\ell}}{\Delta_{i,R_k}\Delta_{i,R_\ell}} \opa_{\ell}
\end{equation}
We can now take the above expressions and substitute them into the original nonlinear terms in $\hat{H}^{NL}$ (\ref{eq:H_nonlinear}). To start, we show an example of finding the dispersive shifts to resonator $k$ that come from qubit $i$ and the term $(\beta_i/2)\opbd_i\opbd_i\opb_i\opb_i$. We can substitute the relevant terms from the above transformed $\opb_i$ into this nonlinear term to get:
\begin{equation}
    \frac{\beta_{J_i}}{2}\opbd_i\opbd_i\opb_i\opb_i \quad\longrightarrow\quad \frac{\beta_{J_i}}{2}\left( \opbd_i - \frac{g_{i,R_k}}{\Delta_{i,R_k}}\opad_k + \frac{g_{i,R_k}}{\Sigma_{i,R_k}}\opa_k\right)^2\left(  \opb_i - \frac{g_{i,R_k}}{\Delta_{i,R_k}}\opa_k + \frac{g_{i,R_k}}{\Sigma_{i,R_k}}\opad_k \right)^2
\end{equation} 
In the following, we will not be concerned with any non-resonant terms that come out of the expansions. The terms we will be concerned with provide contributions to the dispersive shift as well as a new additional shift to the qubit frequency. Expanding the nonlinear term for the resonator (if $\alpha_{R_k} \neq 0$), will provide another contribution to both the dispersive shift and another shift to that resonators frequency. The new dispersive shift terms that will be added to the effective Hamiltonian (\ref{eq:eff_ham}) are:
\begin{equation}\label{eq:dispersive_shifts}
    \hat{H}_{\text{eff}}^{DS} = \sum_{i=1}^N \sum_{k=1}^M 2g_{i,R_k}^2 (\beta_{J_i} + \alpha_{R_k}) \left( \frac{1}{\Delta_{i,R_k}^2} + \frac{1}{\Sigma_{i,R_k}^2} \right) \opbd_i\opb_i \opad_k\opa_k
\end{equation}
The new qubit and resonator shifts (\ref{eq:eff_qubit_freq}) and (\ref{eq:eff_res_freq}) are now shifted further (although it is a small amount compared to the first shift):
\begin{align}
    \tilde{\omega}_{J_i} &\approx \omega_{J_i} + \sum_{k=1}^M g^2_{i,R_k}\left( \frac{1}{\Delta_{i,R_k}} - \frac{1}{\Sigma_{i,R_k}} \right) + 2\beta_{J_i} \frac{g_{i,R_k}^2}{\Sigma_{i,R_k}^2} \\
    \tilde{\omega}_{R_k} &\approx \omega_{R_k} - \sum_{i=1}^N g^2_{i,R_k}\left( \frac{1}{\Delta_{i,R_k}} + \frac{1}{\Sigma_{i,R_k}} \right) + 2\alpha_{R_k} \frac{g_{i,R_k}^2}{\Sigma_{i,R_k}^2}
\end{align}
We can also now estimate the shifts to the nonlinear terms that are originally present in the Hamiltonian. To do this, consider again substituting the relevant terms from the transformed $\opb_i$ into the original nonlinear term
\begin{equation}
    \frac{\beta_{J_i}}{2}\opbd_i\opbd_i\opb_i\opb_i \quad\longrightarrow\quad \frac{\beta_{J_i}}{2}\left( \opbd_i - \frac{1}{2}\sum_{j=1}^N \sum_{k=1}^M \frac{g_{i,R_k}g_{j,R_k}}{\Delta_{i,R_k}\Delta_{j,R_k}} \opbd_j\right)^2\left(  \opb_i - \frac{1}{2}\sum_{j=1}^N \sum_{k=1}^M \frac{g_{i,R_k}g_{j,R_k}}{\Delta_{i,R_k}\Delta_{j,R_k}} \opb_j \right)^2
\end{equation} 
Doing the same for any nonlinear resonators, we get the following effective anharmonicities for the qubits and nonlinear resonators that can be substituted into the effective Hamiltonian (\ref{eq:eff_ham}):
\begin{align}
    \tilde{\beta}_{J_i} &\approx \beta_{J_i}\left( 1 - 2\sum_{k=1}^{M} \frac{g_{i,R_k}^2}{\Delta_{i,R_k}^2} \right) \\
    \tilde{\alpha}_{R_k} &\approx \alpha_{R_k}\left( 1 - 2\sum_{i=1}^N \frac{g_{i,R_k}^2}{\Delta_{i,R_k}^2} \right)
\end{align}
Using both of the above approximations for the transformed nonlinear terms, we also find another contribution to the effective Hamiltonian in the form of cross-Kerr terms for the qubits:
\begin{equation}
    \hat{H}_{\text{eff}}^{CK} = \frac{1}{2}\sum_{i=1}^{N}\sum_{j > i}^{j=N}\sum_{k=1}^M \left( \frac{g_{i,R_k}g_{j,R_k}}{\Delta_{i,R_k}\Delta_{j,R_k}} \right)^2 (\beta_{J_i} + \beta_{J_j} + 4\alpha_{R_k}) \opbd_i\opb_i\opbd_j\opb_j
\end{equation}
A similar term exists for the nonlinear resonators with the roles of the qubits and resonators swapped. Once again, these results and approximations do not differ much from the results of \cite{tunable_coupler_ext}, except that we have now shown that these formulas apply to systems with an arbitrary number of qubits and resonators. It is also possible to apply further transformations to eliminate any non-resonant terms that we have neglected as was done in \cite{tunable_coupler_ext}. Here, we will not go further as these transformations would eliminate the dispersive shift terms and what we have shown should be enough to estimate the relevant parameters when designing a chip.

%% file: mainmatter/ch3/external_ports.tex
\section{External Ports and Qubit Decay Rates}\label{section:ext_ports_decay}
Here, we will look at the effects of external ports in our system that are separate from the ports shunted by Josephson junctions as shown in Fig.\ \ref{fig:transmon_network_ext_ports}. Given the rational impedance function, we can view the network as a CL cascade as previously discussed in Section \ref{section:cascade_synthesis}. This means that we can already estimate the capacitive coupling between the external ports and the qubits. This method can potentially be used when considering designs with combined microwave and flux lines \cite{combined_xyz_line,karamlou2023probing,Moskalenko2022}. Having the capacitive coupling of the lines to the qubits allows us to investigate the control crosstalk between the different lines. Further details on estimating this crosstalk from the impedance function are given in \cite{solgun_sirf,sherbrooke_sirf}.

\begin{figure}[!p]
    \centering
    \begin{circuitikz}[line width=1pt]
        \ctikzset{bipoles/thickness=1, bipoles/length=1cm}
        \ctikzset { label/align = straight }
        
        \draw[rounded corners=.5cm] (0,2) -- (0,4.5) -- (5,4.5) -- (5,-4.5) -- (0,-4.5) -- (0,2);
        \node at (2.5,0) {$\vb{Z}(s) = \dfrac{\vb{R}_0}{s} + \displaystyle\sum_{k=1}^M \dfrac{s \vb{R}_k}{s^2 + \omega_{R_k}^2}$};

        \draw (0,0.5) to[short, -o] (-0.5,0.5) -- (-1, 0.5) to[barrier=$E_{J_N}$] (-1,1.5) to[short, -o] (-0.5,1.5) -- (0,1.5);

        \draw (0,3) to[short, -o] (-0.5,3) -- (-1, 3) to[barrier=$E_{J_1}$] (-1,4) to[short, -o] (-0.5,4) -- (0,4);

        \node at (-0.5,2.35) {$\vdots$};

        \draw (0,-0.5) to[short,-o] (-0.5,-0.5);
        \draw (0,-1.5) to[short,-o] (-0.5,-1.5);
        \node at (-1,-1) {$P_{E_1}$};

        \draw (0,-3) to[short,-o] (-0.5,-3);
        \draw (0,-4) to[short,-o] (-0.5,-4);
        \node at (-1,-3.5) {$P_{E_{K}}$};
        
        \node at (-0.5,-2.15) {$\vdots$};

    \end{circuitikz}
    \caption{Network of $N$ transmons represented by an arbitrary impedance function (\ref{eq:impedance}). Like before, the transmons and external ports are coupled capacitively and through resonant modes. Now included are any external ports of the system. This may include drive, flux, and readout lines. $K$ is the number of external ports.}
    \label{fig:transmon_network_ext_ports}
\end{figure}
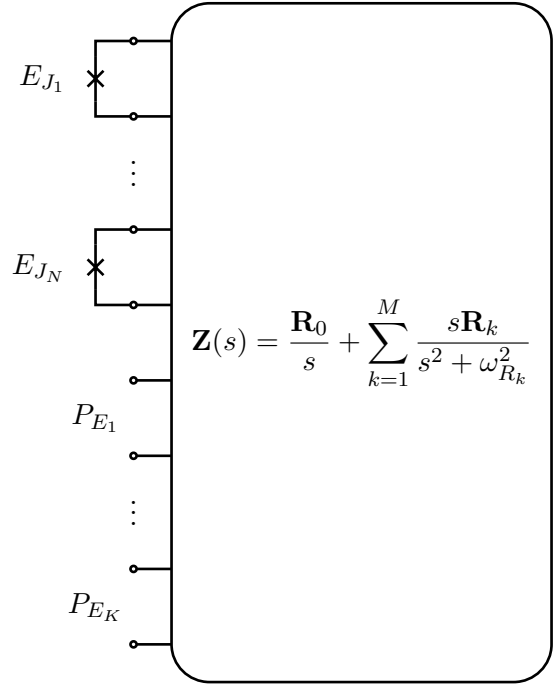

\begin{figure}[!p]
    \centering
    \begin{circuitikz}[line width=1pt]
        \ctikzset{bipoles/thickness=1, bipoles/length=1cm}
        \ctikzset { label/align = straight }
        
        \draw[rounded corners=.5cm] (0,2) -- (0,9.5) -- (3,9.5) -- (3,0) -- (0,0) -- (0,2);
        \node at (1.5,4.75) {\Large $\vb{C}$};

        \draw[rounded corners=0.5cm] (4,4) -- (4,7) -- (6,7) -- (6,2.5) -- (4,2.5) -- (4,4);
        \draw (4, 5.5) -- (4.6,5.5) to[L,l_=$L_{R_1}$, mirror, label distance=0.25cm] (4.6,6.5) -- (4,6.5);
        \draw (4,3) -- (4.6,3) to[L,l_=$L_{R_M}$, mirror, label distance=0.25cm] (4.6,4) -- (4,4);

        \draw (0,8) to[short, -o] (-0.5,8) -- (-0.5,8) -- (-1,8) to[L=$L_{J_1}$] (-1, 9) to[short, -o] (-0.5,9) -- (0,9);
        \draw (-1,8) to[short, -o] (-2.5,8);
        \draw (-1,9) to[short, -o] (-2.5,9);

        \draw (0,5.5) to[short, -o] (-0.5,5.5) -- (-1,5.5) to[L=$L_{J_N}$] (-1, 6.5) to[short, -o] (-0.5,6.5) -- (0,6.5);
        \draw (-1,5.5) to[short, -o] (-2.5,5.5);
        \draw (-1,6.5) to[short, -o] (-2.5,6.5);

        {
        \ctikzset{bipoles/thickness=1, bipoles/length=.75cm}
        \draw (0,1.5) to[short, -o] (-0.5,1.5) -- (-1,1.5) to[R, l_=$Z_{E_K}$] (-1, .5) to[short, -o] (-0.5, .5) -- (0,0.5);
        \draw (-1,1.5) to[short, -o] (-2.5,1.5);
        \draw (-1,0.5) to[short, -o] (-2.5,0.5);
        }

        {
        \ctikzset{bipoles/thickness=1, bipoles/length=.75cm}
        \draw (0,4) to[short, -o] (-0.5,4) -- (-1,4) to[R, l_=$Z_{E_1}$] (-1, 3) to[short, -o] (-0.5, 3) -- (0,3);
        \draw (-1,4) to[short, -o] (-2.5,4);
        \draw (-1,3) to[short, -o] (-2.5,3);
        }

        \draw (3,3) to[short, -o] (3.5,3) -- (4,3);
        \draw (3,4) to[short, -o] (3.5,4) -- (4,4);

        \draw (3,5.5) to[short, -o] (3.5,5.5) -- (4,5.5);
        \draw (3,6.5) to[short, -o] (3.5,6.5) -- (4,6.5);

        \node at (-0.5,2.35) {$\vdots$};
        \node at (-0.5,4.85) {$\vdots$};
        \node at (-0.5,7.35) {$\vdots$};
        \node at (3.5,4.85) {$\vdots$};
        \node at (5,4.85) {$\vdots$};        
    \end{circuitikz}
    \caption{Classical approximation to the circuit in Fig.\ \ref{fig:transmon_network_ext_ports} but now with the lossy environments represented by resistors shunting the external ports. The resistors should match the characteristic impedances of the external ports. The impedance function (\ref{eq:impedance}) has also been replaced with its equivalent CL cascade synthesis that is described in Section \ref{section:cascade_synthesis}. This circuit will have a lossy multiport immittance parameter that can be used to estimate qubit decay rates.}
    \label{fig:lossy_transmon_network}
\end{figure}
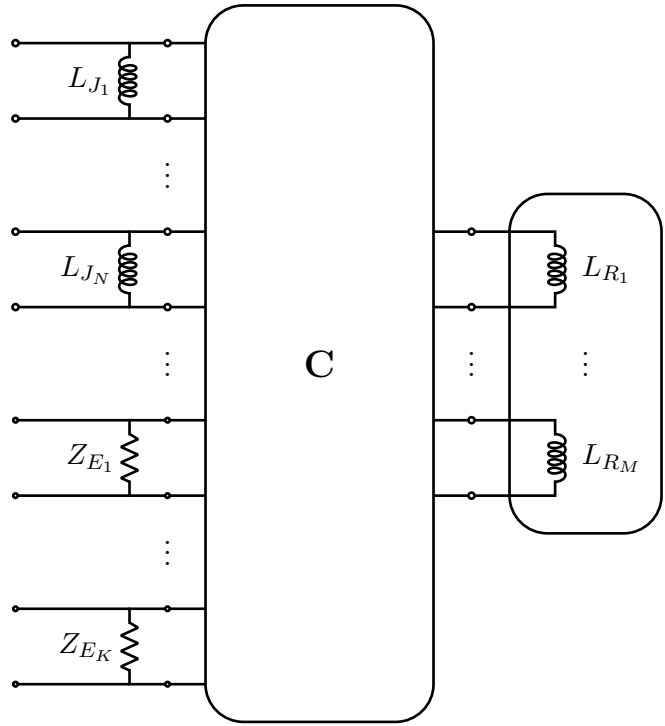

Our main concern in this section is to estimate the relaxation times of the qubits. In our analysis of the relaxation time, we will only be concerned with spontaneous emission through the circuit around the qubit to the external environment, which is also referred to as Purcell decay \cite{purcell}. In other words, the external ports shown in Fig.\ \ref{fig:transmon_network_ext_ports} will be the loss channels. In this analysis we will be neglecting other loss through channels such as two-level systems in dielectrics \cite{dielectric_loss_1,dielectric_loss_2}, non-equilibrium quasiparticles \cite{quasiparticles_transmons}, as well as many more \cite{disentangling_losses}. The goal here is to be able to estimate losses that come from the design of the circuit itself.

\newpage
In circuits with single qubits, there is a commonly used method where a single-port admittance is defined that is external to the qubit. Then, a classical approximation of the qubit lifetime is given by the $RC$ time:
\begin{equation}\label{eq:single_qubit_decay}
    T_1 = \frac{C_q}{\myRe[Y(\omega_q)]}
\end{equation}
where $C_q$ is the total qubit shunt capacitance and $\omega_q$ is the qubit frequency \cite{transformed_dissipation,controlling_spontaneous_emission,suppressing_spontaneous_emission,reducing_spontaneous_emission}. It is generally implied that you can do something similar for multi-qubit systems \cite{controlling_spontaneous_emission}. Later on, we explain how a similar formula to (\ref{eq:single_qubit_decay}) can potentially be used to approximate the qubit relaxation times in multi-qubit systems.

If we consider that we have a network of qubits as shown in Fig.\ \ref{fig:transmon_network_ext_ports}, we know that the rational impedance function can also be synthesized as a CL cascade circuit as described in Section \ref{section:cascade_synthesis}. We can treat the circuit past the ports as a purely resistive environment by shunting each external port with a resistor. This resistor should have an impedance that matches the characteristic impedance of the corresponding port. To find the relaxation times of the qubit modes, we make use of a classical approximation where the Josephson junctions are replaced by inductors. The full circuit with the impedance function replaced with the equivalent CL cascade is shown in Fig.\ \ref{fig:lossy_transmon_network}. 

Before attempting to compute the decay rates of the qubit modes, we first try to compute the resonant frequencies of the lossless circuit that excludes the resistors in Fig.\ \ref{fig:lossy_transmon_network}. To do this, we define the following vector of fluxes and the inverse inductance matrix:
\begin{align}
    \vb{\Phi} &= (\Phi_{J_1},\dots,\Phi_{J_N},\Phi_{E_1},\dots,\Phi_{E_K},\Phi_{R_1},\dots,\Phi_{R_M})^T \\[.1cm]
    \vb{M} &= \diag(L_{J_1}^{-1},\dots,L_{J_N}^{-1},\underbrace{0,\dots,0,}_{K} L_{R_1}^{-1},\dots,L_{R_M}^{-1})
\end{align}
The equations of motion for the branch fluxes of the lossless circuit including the qubit inductances are given by:
\begin{equation}\label{eq:lossless_eom}
    \vb{C}\ddot{\vb{\Phi}} = -\vb{M}\vb{\Phi}
\end{equation}
Diagonalizing the matrix $\vb{C}^{-1}\vb{M}$ will give the resonant frequencies squared of the circuit without the resistors. These equations of motion are equivalent to Kirchhoff's current law for the full circuit. Now we aim to include the resistors at the external port positions. For this, we define the matrix 
\begin{equation}
    \vb{Z} = \diag(\underbrace{0,\dots,0,}_{N}Z_{E_1},\dots,Z_{E_K},\underbrace{0,\dots,0}_{M})
\end{equation}
To include the resistors at each external port branch, we add an additional constraint to (\ref{eq:lossless_eom}) that yields the equations of motions for the lossy circuit in Fig.\ \ref{fig:lossy_transmon_network}:
\begin{equation}\label{eq:lossy_eom}
    \vb{C}\ddot{\vb{\Phi}} = -\vb{M}\vb{\Phi} - \vb{Z}^{-1}\dot{\vb{\Phi}} \quad\longrightarrow\quad \ddot{\vb{\Phi}} + \vb{C}^{-1}\vb{Z}^{-1}\dot{\vb{\Phi}} + \vb{C}^{-1}\vb{N}\vb{\Phi} = 0
\end{equation}
From this system of second order differential equations, we want to extract the complex frequencies that contain the resonant frequencies in our circuit with their corresponding decay rates. To do this, we can define the branch voltage vector $\vb{V} = \dot{\vb{\Phi}}$. Using this definition and writing it in matrix form with the equations of motion (\ref{eq:lossy_eom}), we obtain:
\begin{equation}\label{eq:matrix_eoms}
    \mqty( \dot{\vb{\Phi}}  \\ \dot{\vb{V}} ) = \mqty( \vb{0} & \mathds{1} \\ -\vb{C}^{-1}\vb{M} & -\vb{C}^{-1}\vb{Z}^{-1} ) \mqty( \vb{\Phi}  \\ \vb{V} )
\end{equation} 
Diagonalizing the matrix on the RHS of (\ref{eq:matrix_eoms}) will yield a set of complex conjugate eigenvalue pairs that correspond to the resonant modes in the circuit. These eigenvalues will also be the poles of the multiport impedance function shown in Fig.\ \ref{fig:lossy_transmon_network} which we will later verify numerically in Section \ref{section:decay_rate_example}. Adding the resistors to the network shifts the poles of the impedance function away from the imaginary axis, and they will be of the form:
\begin{equation}\label{eq:lossy_pole}
    s_i = -\frac{\kappa_i}{2} \pm i\omega_i
\end{equation}
where $\kappa_i$ is the decay rate of the resonant mode with frequency $\omega_i$. Thus, given a qubit network represented by a rational impedance function or a CL cascade, we can construct the matrix in (\ref{eq:matrix_eoms}), diagonalize it, and extract the decay rates of the qubits. Later on, we will see that if one computes the admittance of the lossy network in Fig.\ \ref{fig:lossy_transmon_network}, the qubit decay rates can be approximated by 
\begin{equation}\label{eq:qubit_decay_admittance}
    \Gamma_{J_i} = \tilde{C}_{J_i}^{-1} \myRe(\vb{Y}_{ii}(\omega_{J_i}))
\end{equation}
where the effective qubit capacitance $\tilde{C}_{J_i}$ is defined as in (\ref{eq:effective_capacitiance}). Computing this admittance is straightforward if we cascade the shunt elements with the original impedance and make use of (\ref{eq:cascade_s}) with the proper conversions between S, Y, and Z. If the full capacitance matrix is not available, we can also use the normal qubit shunt capacitance $C_{J_i}$ to approximate the decay rates. This can be useful if we only have an immittance function discretized in frequency.

%% file: mainmatter/ch4/chapter4.tex
\chapter{Examples and Applications}\label{chapter:examples}

We will now look at a number of circuits and electromagnetic models to which we apply the methods of the previous two chapters. Before presenting some examples, it is important to make note of what benefits and drawbacks the methods presented in this thesis have in comparison to the immittance formula methods of \cite{solgun_sirf,sherbrooke_sirf}.

In the immittance formula methods, an immittance function is computed at various frequency points -- typically at the desired qubit frequencies in the circuit. Using the computed immittance function at various frequencies, you can estimate the effective qubit couplings, drive crosstalk and amplitudes, and qubit decay rates. However, some issues arise with the estimates in \cite{solgun_sirf} when chips contain strong direct capacitive coupling between the qubits. This is addressed in \cite{sherbrooke_sirf} at the cost of needing to obtain the DC residue of the immittance function to compute the listed quantities. Furthermore, every time you would like to compute a parameter for a new qubit frequency, you are required to also compute the immittance function at that frequency. If you are trying to obtain some initial estimates of these parameters, this may be useful. Still, as frequency tunable qubits and couplers become more common, this may consume a lot of time if working with electromagnetic models.

When dealing with systems containing tunable qubits and couplers, it may be worth spending more time on characterizing the distributed element or electromagnetic models of a superconducting circuit. As we have discussed, we can use vector fitting methods to obtain a rational impedance function which can be translated into a Hamiltonian description of the circuit. From there, all the couplings between the qubits and the resonant modes in the circuit are available to us. Notably, the couplings between the qubits and resonant modes will not be accessible with the immittance formula methods \cite{solgun_sirf, sherbrooke_sirf}. Having this information can help in tweaking a design and estimating parameters that are dependent on these couplings (e.g.\ dispersive shifts of the resonator frequencies (\ref{eq:dispersive_shifts})). Also, this method allows you to easily ``tune'' the qubits to desired frequencies by simply changing the Josephson junction energies at the ports of your impedance. If we obtain the rational impedance through vector fitting once, this effectively removes the need to carry out simulations at a different frequency.

Unfortunately, in order to be able to use the vector fitting methods, we need to be able to compute the impedance at enough frequency points to get a good fit. For multi-qubit circuits, simulations of these electromagnetic models can be prohibitively expensive. However, with the interconnection methods of Section \ref{section:general_network_interconnection}, it may be possible to break up large electromagnetic models into smaller and simpler pieces that can be interconnected, which would decrease the overall simulation time. If the pieces are less complex and contain less resonant modes, we need to sample less points of the impedance function for the fit to accurately represent the model. In addition, these methods can be used on circuit models containing both lumped and distributed elements (e.g.\ combining capacitors, inductors and ideal transmission lines). For these types of models, computing the impedance function is simple and fast over a broad frequency range.

\input{mainmatter/ch4/lumped_distributed.tex}
\input{mainmatter/ch4/electromagnetic.tex}

%% file: mainmatter/ch4/lumped_distributed.tex
\section{Lumped and Distributed Element Models}
\subsection{Tunable Coupler}
Here we will look at a simulation of a system containing a tunable coupler \cite{tunable_coupler} to explicitly see where the approximations of Section \ref{section:effective_coupling} start to break down. Specifically, we want to see what happens to the estimate for the effective coupling rate (\ref{eq:eff_qubit_coupling}) when the coupler approaches the qubit frequency such that $g/\Delta \centernot{\ll} 1$. Note that since this system only contains two qubits and a coupler, (\ref{eq:eff_qubit_coupling}) reduces to the result for the effective coupling rate from \cite{tunable_coupler}.

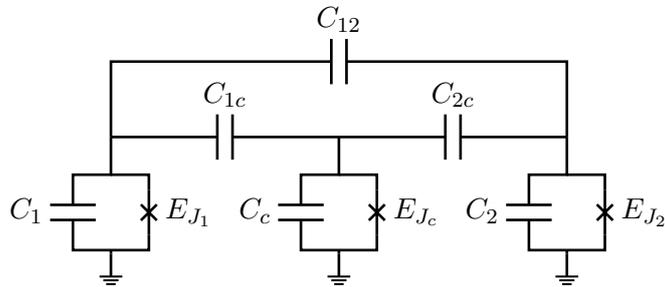
\begin{figure}[h!]
    \centering
    \begin{circuitikz}[line width=1pt]
        \ctikzset{bipoles/thickness=1, bipoles/length=1cm, monopoles/ground/thickness=0.75}
        \ctikzset { label/align = straight }

        \draw (0,0) -- (0,-0.5) -- (-0.5,-0.5) to[C,l_=$C_c$] (-0.5,-1.5) -- (0,-1.5);
        \draw (0,-0.5) -- (0.5,-0.5) to[barrier=$E_{J_c}$, label distance=-10pt] (0.5,-1.5) -- (0,-1.5);
        \draw (0,-1.5) -- (0,-1.55) node[ground] {};

        \draw (-3,0) -- (-3,-0.5) -- (-3.5,-0.5) to[C,l_=$C_1$] (-3.5,-1.5) -- (-3,-1.5);
        \draw (-3,-0.5) -- (-2.5,-0.5) to[barrier=$E_{J_1}$, label distance=-10pt] (-2.5,-1.5) -- (-3,-1.5);
        \draw (-3,-1.5) -- (-3,-1.55) node[ground] {};

        \draw (3,0) -- (3,-0.5) -- (2.5,-0.5) to[C,l_=$C_2$] (2.5,-1.5) -- (3,-1.5);
        \draw (3,-0.5) -- (3.5,-0.5) to[barrier=$E_{J_2}$, label distance=-10pt] (3.5,-1.5) -- (3,-1.5);
        \draw (3,-1.5) -- (3,-1.55) node[ground] {};

        \draw (-3,0) to[C=$C_{1c}$] (0,0) to[C=$C_{2c}$] (3,0);
        \draw (-3,0) -- (-3,1) to[C=$C_{12}$] (3,1) -- (3,0);

    \end{circuitikz}
    \caption{Model for the circuit implementing a tunable coupler (center transmon) to control the coupling between the two outer transmons. The circuit parameters we use are also used in the example given in \cite{tunable_coupler}: $C_1=$ 70 fF, $C_2=$ 72 fF, $C_c=$ 200 fF, $C_{1c}=$ 4 fF, $C_{2c}=$ 4.2 fF, $C_{12}=$ 0.1 fF. In the simulations, we treat each of the three Josephson junctions as tunable.}
    \label{fig:tc_circuit}
\end{figure}

In Fig.\ \ref{fig:tc_circuit}, we see a circuit that contains 3 transmon qubits. The two outer qubits are used as computational qubits and the center one is used as a coupler. Since the network is purely capacitive apart from the junctions, the circuit Hamiltonian can easily be found and is of the form (\ref{eq:transmon_resonator_ham}) with only transmon branches. With the circuit Hamiltonian, we can simulate the time evolution of the system. The results of the simulation can then be compared to the effective coupling rates.

\begin{figure}[!h]
    \centering
    \includegraphics[width=.75\textwidth]{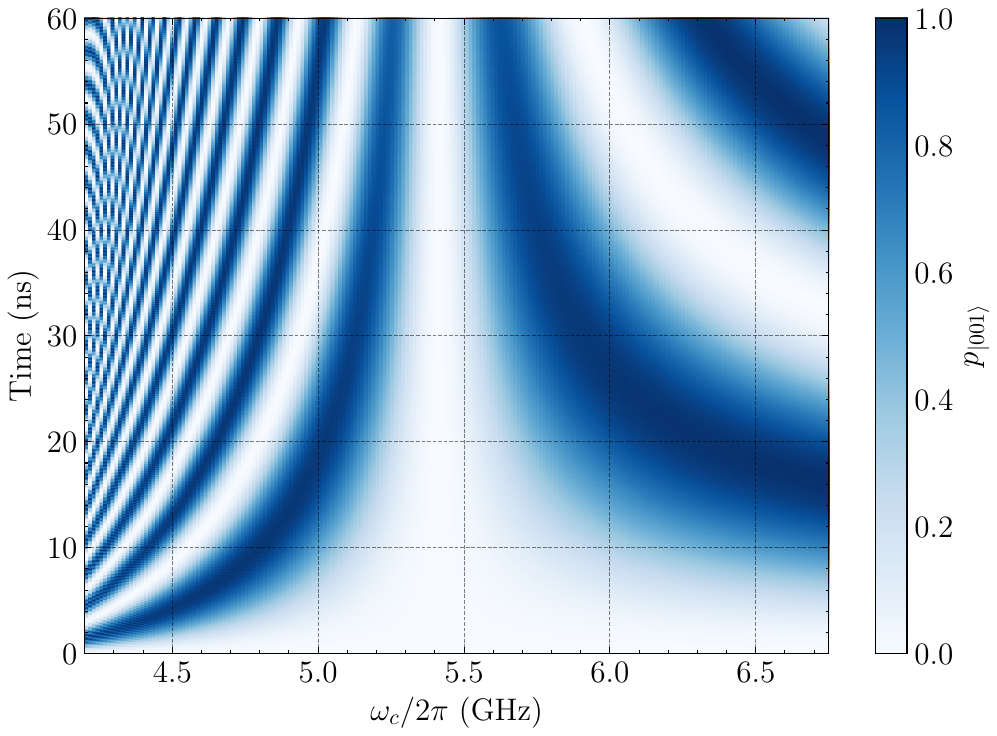}
    \caption{Time simulation results for the tunable coupler system in shown in Fig.\ \ref{fig:tc_circuit}. The system is initialized in state $\ket{100}$ and the two computational transmons are put on resonance such that $\omega_1/2\pi=\omega_2/2\pi=4$ GHz. The resulting oscillations in the population of state $\ket{001}$ are plotted above for various coupler frequencies.}
    \label{fig:tc_time_evolution}
\end{figure}

To simulate the time evolution of the tunable coupler system, we will use \texttt{sesolve} from QuTiP \cite{qutip1,qutip2} for the Hamiltonian (\ref{eq:transmon_resonator_ham}) of the circuit in Fig.\ \ref{fig:tc_circuit}. For the transmons, we use three level systems. We represent the system state using the notation $\ket{Q_1,Q_c,Q_2}$ where $Q_{1/2}$ correspond to the computational qubits and $Q_c$ to the coupler state. By initializing the system in state $\ket{100}$ and putting the states $\ket{100}$ and $\ket{001}$ on resonance, we see oscillations in the $\ket{001}$ state population from which we can extract the coupling rate. Doing this while sweeping over the coupler frequency, we obtain the results in Fig.\ \ref{fig:tc_time_evolution}.

\begin{figure}[!h]
    \centering
    \includegraphics[width=\textwidth]{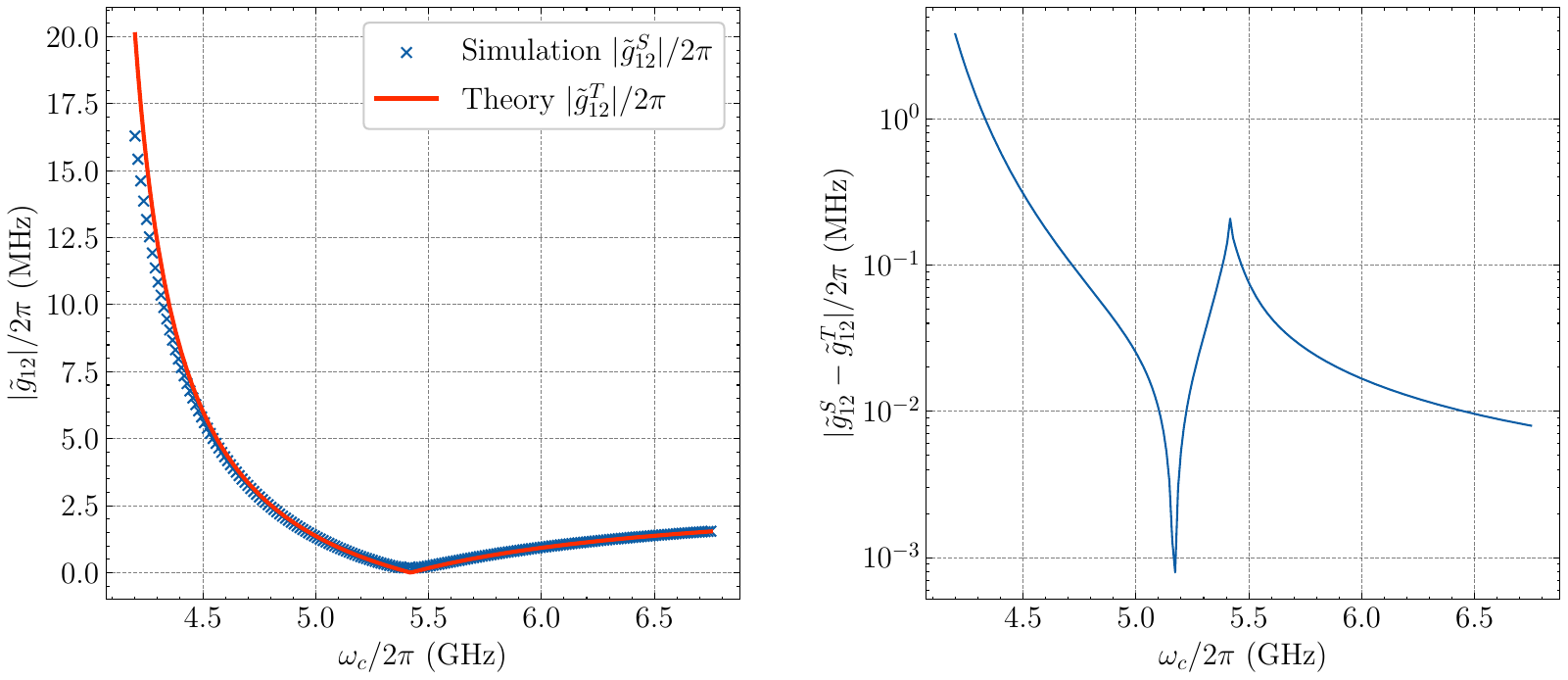}
    \caption{Left: Comparison between the theoretical and fitted effective coupling from the simulation in Fig.\ \ref{fig:tc_time_evolution}. Right: Difference between the theoretical and effective coupling extracted from the simulation.}
    \label{fig:eff_vs_sim_coupling}
\end{figure}

To extract the effective coupling rates between the two computational qubits from the simulation, we can fit the oscillations in the $\ket{001}$ population. The fitted coupling rates can then be compared to the estimate of the effective coupling rate from (\ref{eq:eff_qubit_coupling}). The results of this comparison are shown in Fig.\ \ref{fig:eff_vs_sim_coupling}. We can see that the effective coupling rates start to stray away from the theoretical estimate as the coupler gets closer to the computational transmon frequencies. In this case, the deviation is over 1 MHz once $g_{1c}/\Delta_{1c} \approx  -0.3$, and is increasing as the coupler frequency gets lower. This is to be expected, since $g_{1c}/\Delta_{1c} \centernot{\ll} 1$ and the system is approaching the nondispersive regime where the couplings can no longer be treated as perturbations. We see that when the coupler is far from the qubit frequencies, the approximation is much better. Thus, when using the formulas for the effective coupling rates, we need to take care that we are in the dispersive regime with $g \ll \Delta$. This is especially relevant in systems with tunable qubits and couplers since $g$ and $\Delta$ will depend on these frequencies. 

\begin{figure}[!h]
    \centering
    \includegraphics[width=\textwidth]{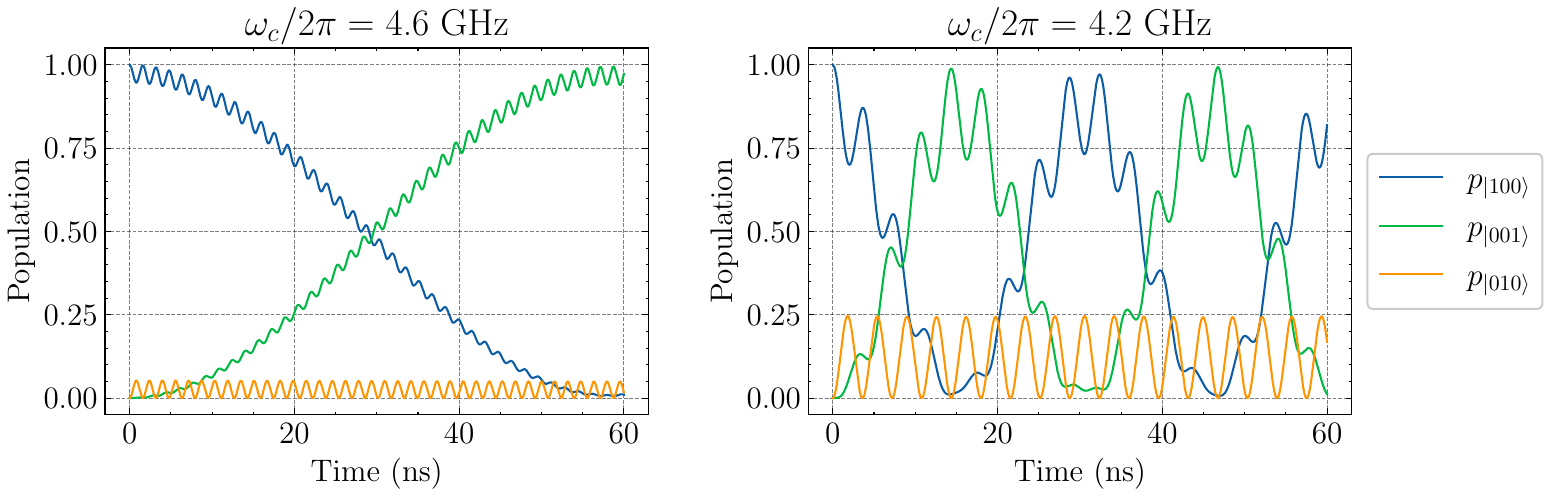}
    \caption{Time evolution of state populations for circuit \ref{fig:tc_circuit} and $\omega_1/2\pi=\omega_2/2\pi=4$ GHz and two different coupler frequencies. On the left, for the higher coupler frequency, $g_{1c}/\Delta_{1c} \approx -0.12$. On the right, we have $g_{1c}/\Delta_{1c} \approx -0.345$.}
    \label{fig:tc_time_evolve_example}
\end{figure}

When attempting to implement fast two qubit gates in the nondispersive regime, leakage into the coupler can occur due to its higher participation in the coupling compared to the dispersive regime. An example of this is shown in Fig.\ \ref{fig:tc_time_evolve_example}. Nonetheless, the potential leakage to the coupler and higher energy levels during gate operations can be suppressed by control pulse optimization, which allows for fast gates to be implemented \cite{high_fidelity_cz_iswap_tc}. However, as we've seen, the formulas for the effective coupling rates (\ref{eq:eff_qubit_coupling}) will not be accurate in this regime. While not accurate in the nondispersive regime, the formulas can still provide a good estimate for the point at which the effective coupling is ``off''.

\subsection{Qubit Decay Rates}\label{section:decay_rate_example}
We will now see how the results of Section \ref{section:ext_ports_decay} can be applied to a lumped element circuit containing two coupled qubits and four external ports. We will also use the example circuit to demonstrate that (\ref{eq:qubit_decay_admittance}) can provide a good approximation to the qubit decay rates.

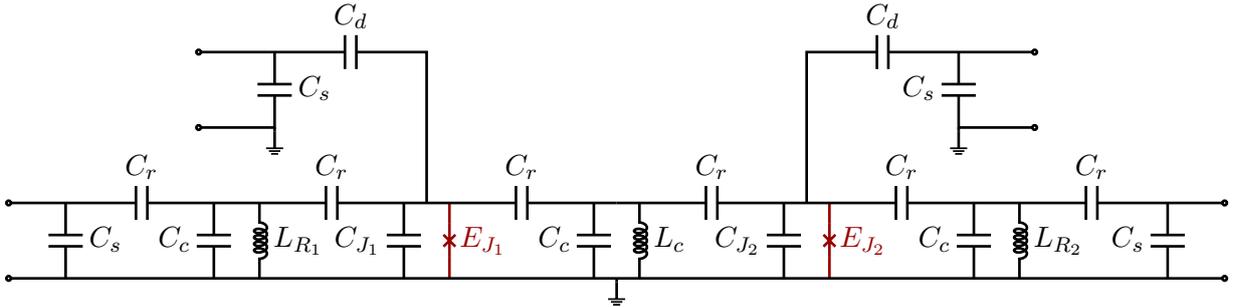
\begin{figure}[h!]
    \centering
    \begin{circuitikz}[line width=1pt]
        \normalfont
        \ctikzset{bipoles/thickness=1, bipoles/length=.75cm, monopoles/ground/thickness=0.65}
        \ctikzset { label/align = straight }

        \draw (-2.8,0) to[C,l_=$C_{J_1}$] (-2.8,-1);
        \draw[color=nodecolor] (-2.2,0) to[barrier=$E_{J_1}$, label distance=-10pt] (-2.2,-1);

        \draw (2.2,0) to[C,l_=$C_{J_2}$] (2.2,-1);
        \draw[color=nodecolor] (2.8,0) to[barrier=$E_{J_2}$, label distance=-10pt] (2.8,-1);

        \draw (0,0) -- (0.75,0) to[C=$C_r$] (1.75,0) -- (3.25,0) to[C=$C_r$] (4.25,0) -- (5.75,0) to[C=$C_{r}$] (6.75,0) to[short, -o] (8,0);
        \draw (-8,0) to[short, o-] (-6.75,0) to[C=$C_{r}$] (-5.75,0) -- (-4.25,0) to[C=$C_r$] (-3.25,0) -- (-1.75,0) to[C=$C_r$] (-0.75,0) -- (0,0);
        \draw (-8,-1) to[short, o-] (0,-1) to[short,-o] (8,-1);
        \draw (0,-1) -- (-0,-1.05) node[ground] {};

        \draw (-0.3,0) to[C,l_=$C_c$] (-0.3,-1);
        \draw (0.3,0) to[L=$L_c$] (0.3,-1);

        \draw (-5.3,0) to[C,l_=$C_c$] (-5.3,-1);
        \draw (-4.7,0) to[L=$L_{R_1}$] (-4.7,-1);

        \draw (4.7,0) to[C,l_=$C_c$] (4.7,-1);
        \draw (5.3,0) to[L=$L_{R_2}$] (5.3,-1);

        \draw (-7.25,0) to[C=$C_s$] (-7.25,-1);
        \draw (7.25,0) to[C,l_=$C_s$] (7.25,-1);

        \draw (-2.5,0) -- (-2.5,2) to[C,l_=$C_d$] (-4.5,2) to[short, -o] (-5.5,2);
        \draw (-4.5,2) to[C=$C_s$] (-4.5,1) to[short,-o] (-5.5,1);
        \draw (-4.5,1) -- (-4.5,.95) node[ground] {};

        \draw (2.5,0) -- (2.5,2) to[C,l=$C_d$] (4.5,2) to[short, -o] (5.5,2);
        \draw (4.5,2) to[C,l_=$C_s$] (4.5,1) to[short,-o] (5.5,1);
        \draw (4.5,1) -- (4.5,.95) node[ground] {};

    \end{circuitikz}
    \caption{Circuit model for two resonator-coupled transmons where each is individually coupled to an external drive port and readout resonator. Overall, there are four external ports, two for drive, and two for readout. When discussing this circuit, we sometimes refer to the two transmon or qubit ports that are located at the positions of the Josephson junctions (marked in red). Transmons 1 and 2 have the Josephson energies $E_{J_1}$ and $E_{J_2}$, respectively. The values of the circuit elements are as follows: $C_{J_1}=$ 70 fF, $C_{J_1}=$ 75 fF, $C_s=$ 100 fF, $C_c=$ 300 fF, $C_r=$ 10 fF, $C_d=$ 0.15 fF, $L_c=$ 3.25 nH, $L_{R_1}=$ 2.1 nH, and $L_{R_2}=$ 1.6 nH. The external ports are assumed to have characteristic impedances of $Z_0=$ 50 $\Omega$.}
    \label{fig:decay_lumped_circuit}
\end{figure}

\begin{figure}[!h]
    \centering
    \includegraphics[width=\textwidth]{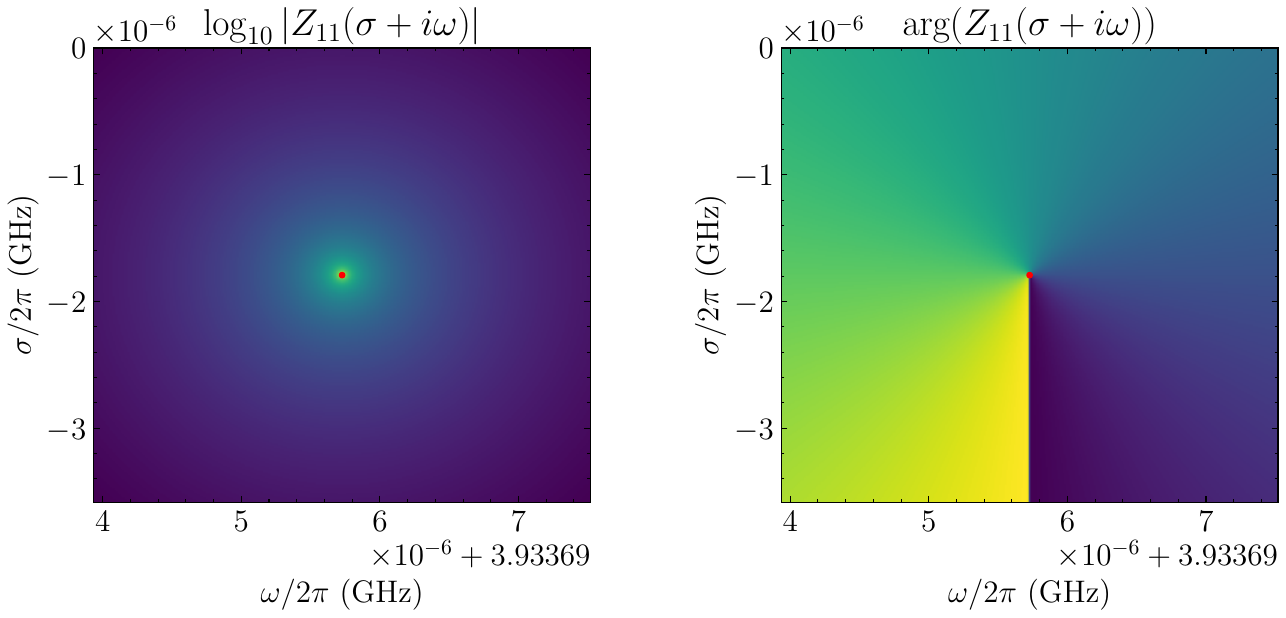}
    \caption{These plots show the magnitude and phase of the impedance function for the circuit in Fig.\ \ref{fig:decay_lumped_circuit} with inductors and resistors shunting the transmon and external ports, respectively. We have chosen $L_{J_1}=$ 18 nH and $L_{J_2}=$ 15.5 nH. In the plot, we look at a region around a complex pole that corresponds to transmon 1 in the circuit. We find that the predicted pole position computed by diagonalizing (\ref{eq:matrix_eoms}) aligns with the pole position in the numerically computed impedance.}
    \label{fig:shunted_impedance_plot}
\end{figure}

We consider the circuit in Fig. \ref{fig:decay_lumped_circuit}. If we fix the qubit inductances, we can find the eigenvalues of the matrix in (\ref{eq:matrix_eoms}) to obtain the complex poles that correspond to the resonant modes in the circuit. From these eigenvalues, we find two complex conjugate pairs that correspond to the transmon oscillator modes. Three other complex conjugate pairs are also found for the other resonant modes in the circuit. Now, we want to verify that these complex numbers are the complex poles of the impedance function for the circuit in Fig.\ \ref{fig:decay_lumped_circuit} with the external ports shunted by resistors and the transmon ports shunted by inductors (as shown in Fig.\ \ref{fig:lossy_transmon_network}). To compute the shunted impedance function at an arbitrary complex frequency, we first compute the impedance function for the circuit Fig.\ \ref{fig:decay_lumped_circuit} using the method of Section \ref{section:cascade_analysis}. Then we shunt the qubit ports with inductors and the external ports with resistors using (\ref{eq:cascade_s}). We find that the predicted pole locations from (\ref{eq:matrix_eoms}) match the positions of the poles of the lossy impedance function. This alignment is shown for one of the transmon poles in Fig.\ \ref{fig:shunted_impedance_plot}.

\begin{figure}[!h]
    \centering
    \includegraphics[width=\textwidth]{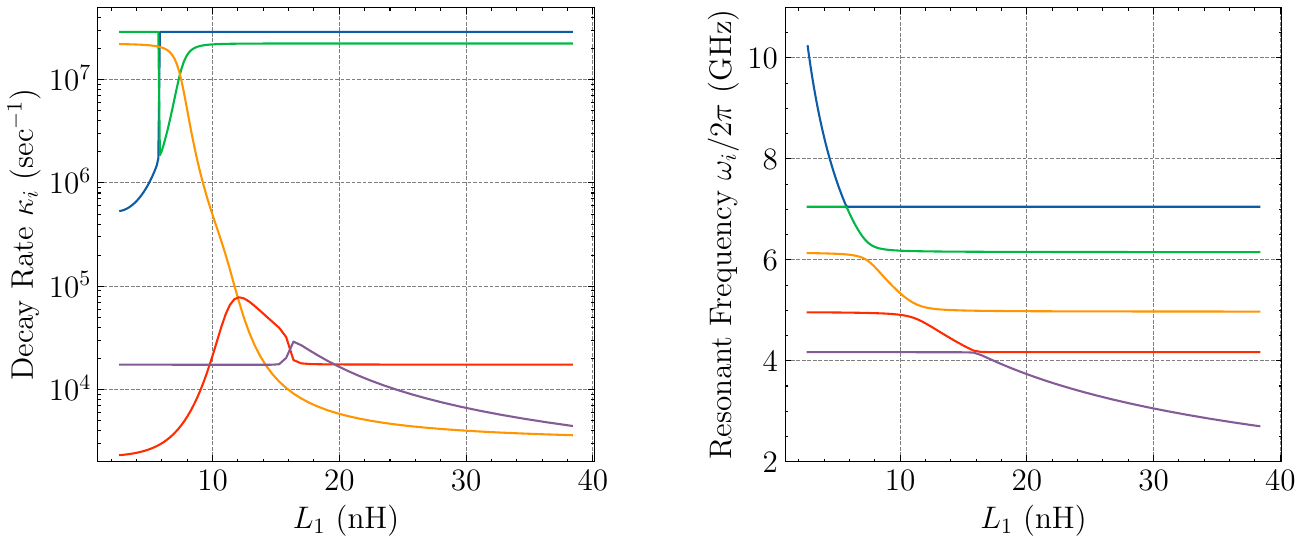}
    \caption{Decay rates and resonant frequencies of the complex poles found by diagonalizing the matrix in (\ref{eq:matrix_eoms}) while varying the inductance shunting transmon 1. The inductance of transmon 2 is fixed at $L_{J_2}=$ 13.8 nH.}
    \label{fig:poles_real_imag}
\end{figure}

We have shown that diagonalizing (\ref{eq:matrix_eoms}) does in fact provide us with the complex poles of the shunted impedance function for a network of the form shown in Fig.\ \ref{fig:lossy_transmon_network}. With the complex pole positions, we can estimate the decay rate for a given resonant frequency using (\ref{eq:lossy_pole}). We can use this to compute the decay rates of the transmons at various frequencies by varying one of the transmon inductances and computing the poles for each inductance value. As an example, we do this for transmon 1 in the circuit, and the decay rates and resonant modes for all the computed poles are plotted in Fig.\ \ref{fig:poles_real_imag}. In these plots, we can see how the resonance frequency and decay rate of the transmon oscillator changes as the shunt inductance is changed.

\begin{figure}[!h]
    \centering
    \includegraphics[width=\textwidth]{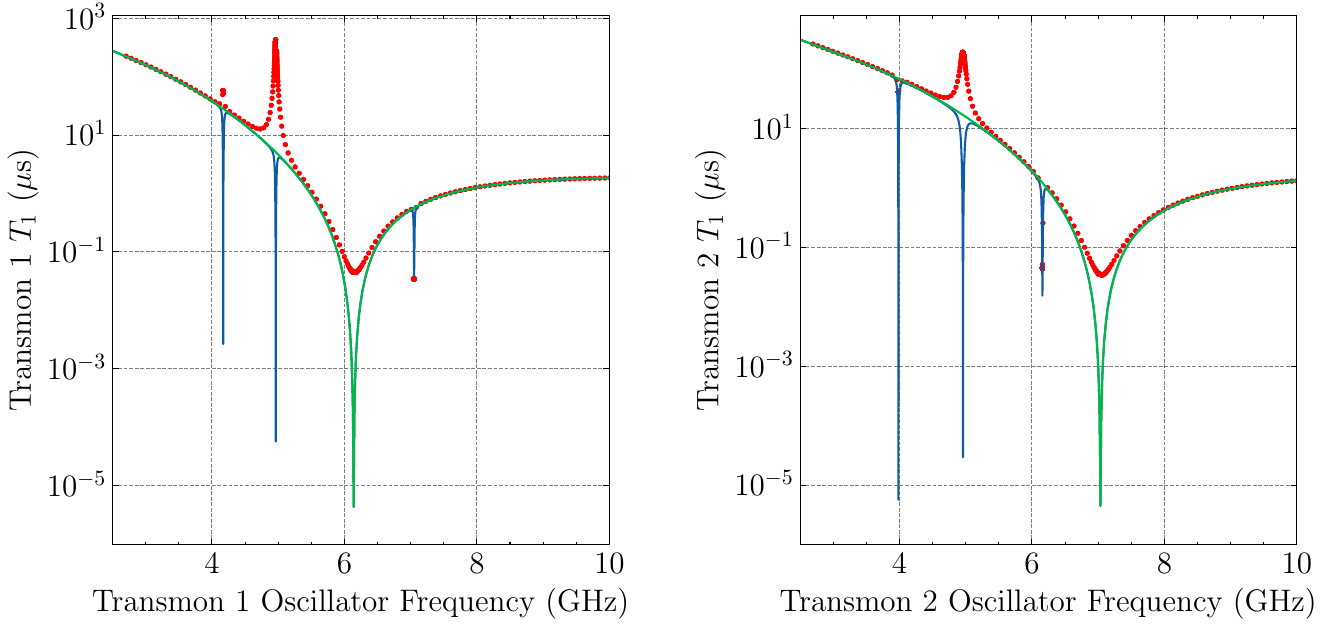}
    \caption{Above, we use three methods to plot estimates of the $T_1$ time of transmons 1 and 2 for the circuit in Fig.\ \ref{fig:decay_lumped_circuit}. \textbf{Red points:} The positions of the complex poles of the network found by diagonalizing the matrix in (\ref{eq:matrix_eoms}). As one transmon inductance is varied, the other is fixed. In the left plot, $L_{J_2}=$ 15.2 nH and on the right, $L_{J_1} =$ 17.5 nH. \textbf{Blue line:} $\Gamma_i^{-1}$ computed using the admittance in (\ref{eq:qubit_decay_admittance}) with the transmon ports shunted with inductors and the external ports shunted with resistors. \textbf{Green line:} Also using (\ref{eq:qubit_decay_admittance}), but the admittance is only shunted with resistors at the external ports.}
    \label{fig:transmons_T1}
\end{figure}

\begin{figure}[!h]
    \centering
    \includegraphics[width=\textwidth]{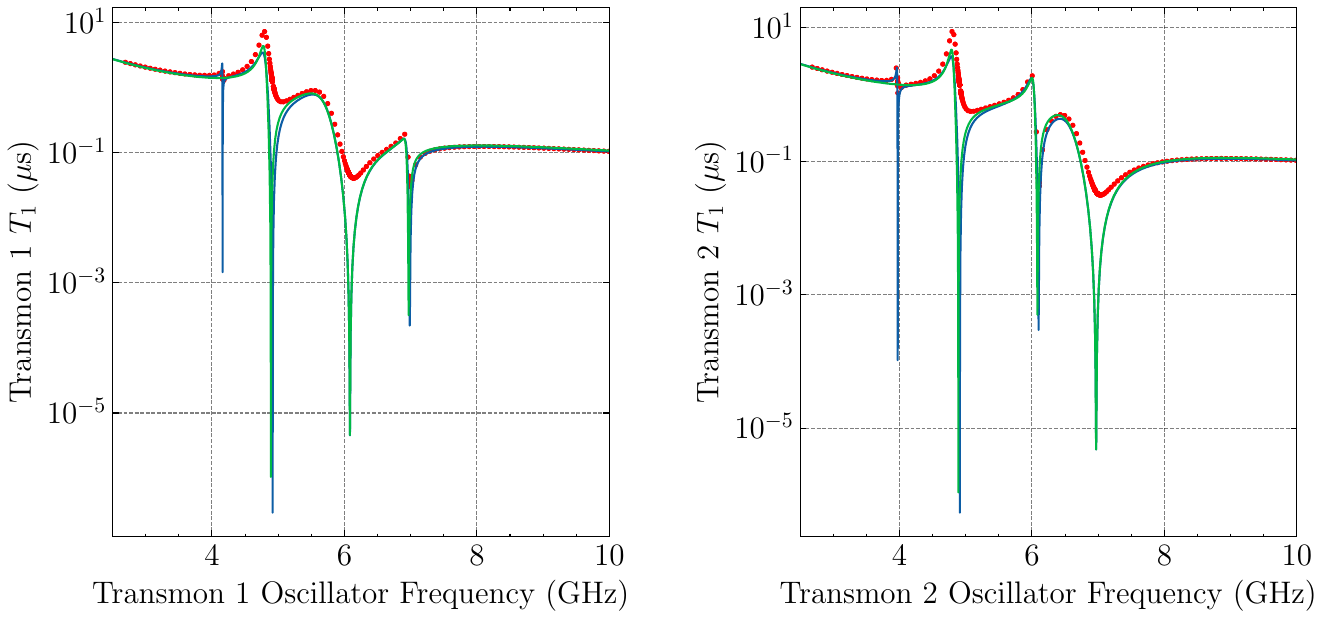}
    \caption{Same as Fig.\ \ref{fig:transmons_T1} except now there is an added 1 fF capacitance between nodes where there was no coupling present for the circuit in Fig.\ \ref{fig:decay_lumped_circuit}.}
    \label{fig:transmons_T1_ata}
\end{figure}

Taking this method further, we can then see the relationship between the transmon resonant frequency and its relaxation time $T_1 = \kappa^{-1}$. We compare this to the qubit decay rate estimate using the admittance of the lossy network (\ref{eq:qubit_decay_admittance}). For the admittance, we can also choose to include or exclude the transmon inductances. We do this for both qubits and the results of the three methods are shown in Fig.\ \ref{fig:transmons_T1}. When plotting the poles computed from (\ref{eq:matrix_eoms}), there are peaks in the transmon lifetimes that are present due to the avoided level crossings shown in Fig.\ \ref{fig:poles_real_imag}. At the resonance crossings, the transmon oscillator is not well defined on its own, and thus the peaks should not be thought of as an expected point where the transmon lifetime is increased. In the plot using the admittance with ports shunted by inductors and resistors, there are sharp dips present that are not seen when only shunting with resistors. These dips are present because of the additional resonance mode of the other transmon. Comparing the methods, we see that using the admittance estimates can provide a good lower bound on the transmon relaxation times. We can also show that if we increase the coupling complexity of the network, this is still the case. We show this by adding a 1 fF capacitance between every node where one is not already present for the circuit in Fig.\ \ref{fig:decay_lumped_circuit}. The results for the relaxation time estimates for this new network are shown in Fig.\ \ref{fig:transmons_T1_ata}.

\clearpage

\subsection{Ideal Transmission Line Coupler}
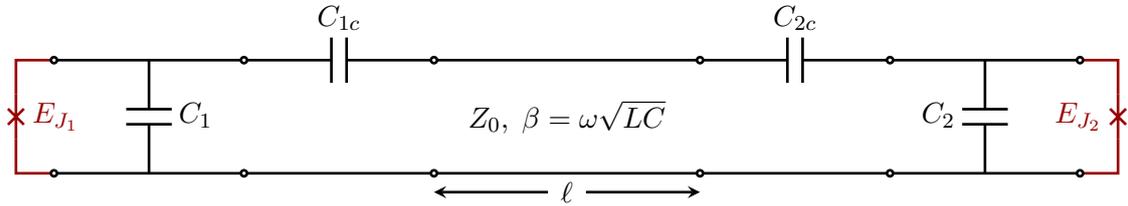
\begin{figure}[h!]
    \centering
    \begin{circuitikz}[line width=1pt]
    \ctikzset{american}
    \ctikzset{bipoles/thickness=1, bipoles/length=1cm}
    \ctikzset{bipoles/crossing/size=0.5}
    \ctikzset { label/align = straight }

    \draw[color=nodecolor] (-2.5,1) -- (-3,1) to[barrier=$E_{J_1}$, label distance=-10pt]  (-3,-0.5) -- (-2.5,-0.5);
    \draw[color=nodecolor] (11,1) -- (11.5,1) to[barrier,l_=$E_{J_2}$, label distance=-10pt]  (11.5,-0.5) -- (11,-0.5);
    
    \draw (-2.5,1) to[short,o-] (0,1) to[short, o-] (0.5,1) to[C=$C_{1c}$] (2,1) to[short, -o] (2.5,1) to[short,-o] (6,1) -- (6.5,1) to[C=$C_{2c}$] (8,1) to[short, -o] (8.5,1) to[short, -o] (11,1);
    \draw (-2.5,-0.5) to[short,o-] (0,-.5) to[short,o-] (0.5,-.5) to[short, -o] (2.5,-.5) to[short, -o] (6,-.5) to[short,-o] (8.5,-.5) to[short, -o] (11,-0.5);

    \draw (-1.25,1) to[C=$C_1$] (-1.25,-0.5);
    \draw (9.75,1) to[C,l_=$C_2$] (9.75,-0.5);

    \node at (4.25,0.25) {$Z_0,\; \beta=\omega\sqrt{LC}$};
    
    \node at (4.25,-0.75) {$\ell$};
    \draw [-stealth](4.5,-0.75) -- (6,-0.75);
    \draw [-stealth](4,-0.75) -- (2.5,-0.75);

    \end{circuitikz}
    \caption{Two qubit circuit containing two transmon qubits coupled by an ideal transmission line. The network can be viewed as a two-port system shunted with two Josephson junctions.}
    \label{fig:ideal_TL_coupler}
\end{figure}

Here, we consider a network of two qubits coupled by an ideal transmission line as shown in Fig.\ \ref{fig:ideal_TL_coupler}. Now we go through the process of obtaining the Hamiltonian from the impedance function of this two-port network.

\begin{figure}[!h]
    \centering
    \includegraphics[width=\textwidth]{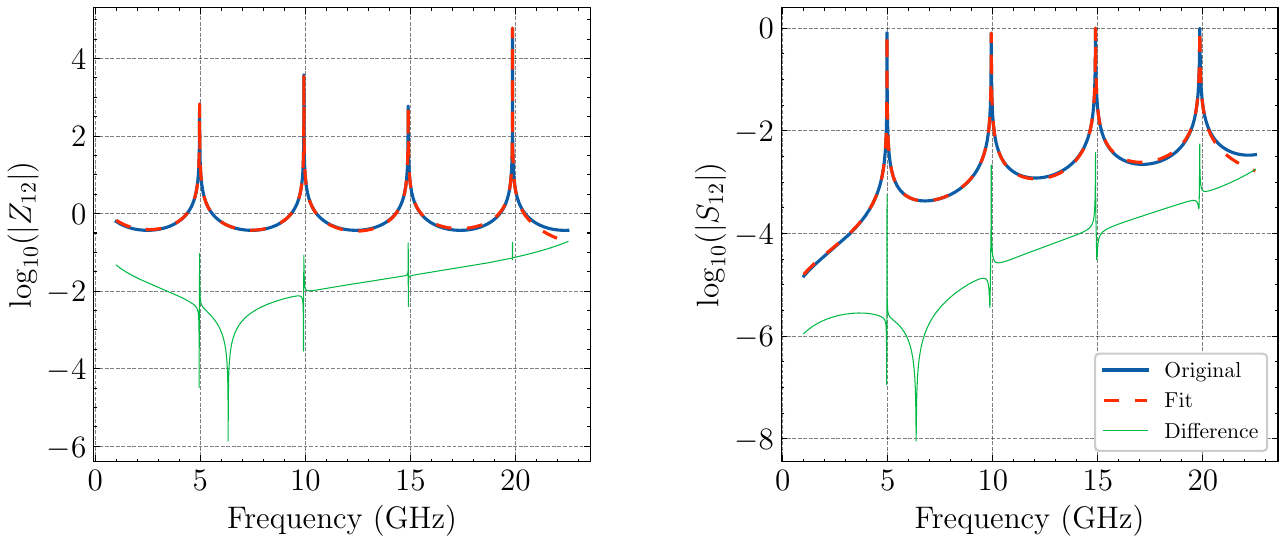}
    \caption{Fit results for the off-diagonal elements of the impedance and scattering parameters for the two-port network in Fig.\ \ref{fig:ideal_TL_coupler}. The diagonal elements of the parameters show similar or smaller difference. For the capacitors we choose $C_1 =$ 70 fF, $C_2 =$ 72 fF, $C_{1c}=C_{2c} =$ 6.5 fF. For the transmission line we choose $L = $ 0.438 $\mu$H/m, $C = $ 0.159 nF/m, $\ell =$ 12 mm.}
    \label{fig:ideal_TL_coupler_fit}
\end{figure}

We can compute the two-port impedance function over a discretized frequency range by using ABCD matrices \cite[Chapter 4.4]{Pozar_2011}. The network in Fig.\ \ref{fig:ideal_TL_coupler} is broken up into pieces containing either a capacitor or the transmission line where each has a simple ABCD matrix representation. Then, we cascade all of the ABCD matrices and convert to an two-port impedance function that is discretized in frequency. The next step is to obtain the rational impedance function that approximates the response of this network. We do this by using the vector fitting methods as described in Section \ref{section:vector_fitting}. For this circuit, taking the lossless part of the rational model obtained from the traditional vector fitting process already produces a good approximation with the results shown in Fig.\ \ref{fig:ideal_TL_coupler_fit}. For this example, we fit for a frequency range of 1 GHz to 22.5 GHz. The final fitted rational function of the form (\ref{eq:impedance}) has four resonant poles that are clearly seen in the frequency range. We could also try to fit for the next highest resonant pole that is ``invisible'' to the fitting process. Generally when doing the fitting, poles within the frequency range chosen can be restricted to be nondegenerate. Outside the frequency range, degenerate poles may appear and they may also not be where you expect (e.g. at the next highest mode of one of the resonators). For this case, we don't try to fit these poles since we want to avoid adding incorrect resonant modes to this model. 

Using the fitted rational impedance function, we can write out the Hamiltonian of the system using (\ref{eq:impedance_hamiltonian}) once we've taken into account the Josephson junctions at the ports. With both transmons at $\omega_1/2\pi=\omega_2/2\pi = $ 4 GHz, they have a direct coupling rate of $g_{12} = 0.652$ MHz. The coupling rates of the transmons to the resonant modes are shown in Table \ref{table:ideal_TL_coupling}.

\renewcommand{\arraystretch}{1.5}
\begin{table}[h!]
    \centering
    \begin{tabular}{|r|r|r|r|r|}
    \hline
    $\omega_{R_k}/2\pi$ (GHz) & $4.965470$   & $9.931947$   & $14.896434$  & $19.8619404$ \\ \hline
    $g_{1,R_k}/2\pi$ (MHz)   & $-55.113$ & $-77.924$ & $-95.422$ & $-110.154$ \\ \hline
    $g_{2,R_k}/2\pi$ (MHz)   & $54.367$  & $-76.869$ & $94.130$  & $-108.662$ \\ \hline
\end{tabular}
\caption{The resonant pole positions $\omega_{R_k}$ from the fit result in Fig.\ \ref{fig:ideal_TL_coupler_fit} and their corresponding coupling rates to the transmons.}
\label{table:ideal_TL_coupling}
\end{table}
\renewcommand{\arraystretch}{1}

\begin{figure}[!h]
    \centering
    \includegraphics[width=0.6\textwidth]{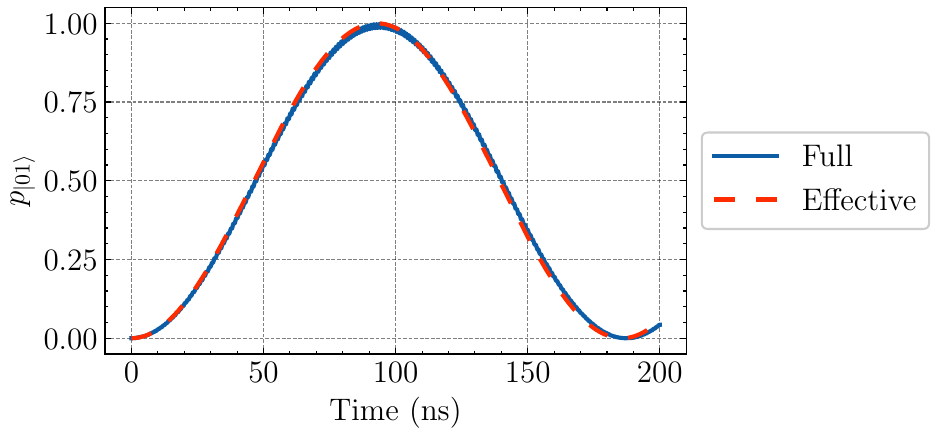}
    \caption{Time evolution of the full and effective Hamiltonians for the circuit in Fig.\ \ref{fig:ideal_TL_coupler} using the results from the fit in Fig.\ \ref{fig:ideal_TL_coupler_fit}. The transmons are at 4 GHz and the initial state is $\ket{Q_1Q_2}=\ket{10}$ with all the resonators of the full Hamiltonian in the ground state.}
    \label{fig:ideal_TL_coupler_sim}
\end{figure}

Using the resulting Hamiltonian, we can construct the effective Hamiltonian for the system that only contains the transmons. Then we can simulate the time evolution of the full Hamiltonian (\ref{eq:transmon_resonator_ham}) containing the four resonant modes and compare it to the simulation of the effective Hamiltonian (\ref{eq:eff_ham}). An example showing agreement between the two is shown in Fig.\ \ref{fig:ideal_TL_coupler_sim}.

\begin{figure}[!h]
    \centering
    \includegraphics[width=\textwidth]{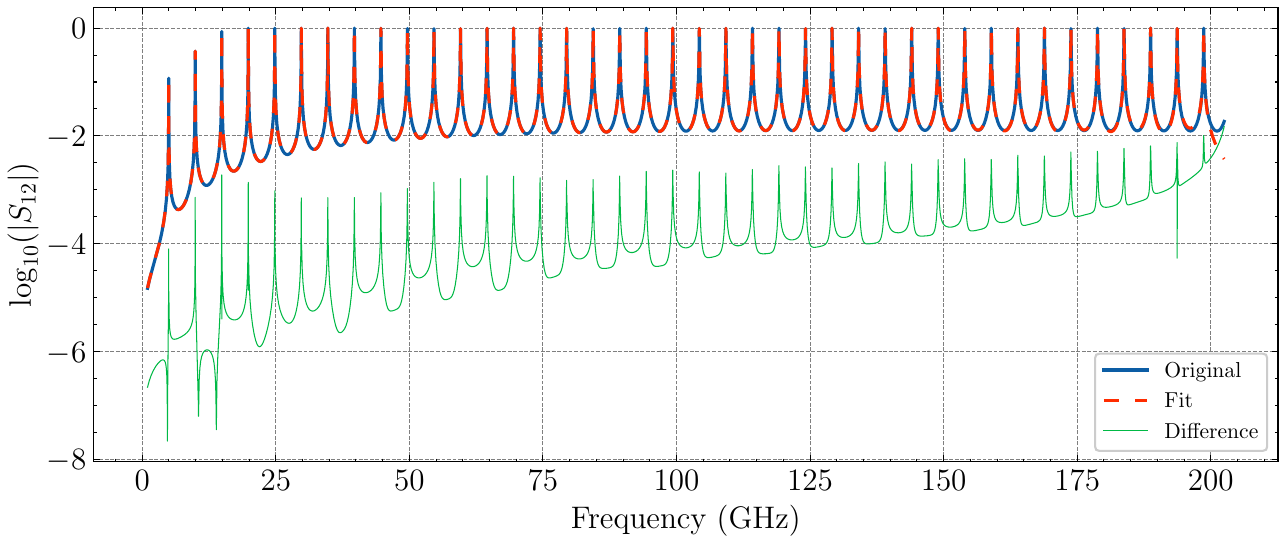}
    \caption{Fit results for the $S_{12}$ parameter for the two-port network in Fig.\ \ref{fig:ideal_TL_coupler}. Here, 40 resonant modes are included in the frequency range.}
    \label{fig:ideal_TL_coupler_fit_40_res}
\end{figure}

Using the same circuit, we can look at what happens to our estimate of the effective coupling rate between the qubits when we increase the cutoff frequency used for our fit. When increasing this range, more resonances are visible in the response and are then included as new resonance modes in the fitted rational impedance. We start by fitting a function that includes just the first resonant mode. Then for each of the next highest resonant modes, we fit new functions until we have reached a total of 40 resonant modes included. Note that this is only performed as a test of the fitting process and to see how the effective coupling formulas are affected by including these higher resonant modes. The fit for the frequency range that contains 40 resonant modes is shown in Fig.\ \ref{fig:ideal_TL_coupler_fit_40_res}. The fitting for this wide frequency range is quick (less than 30 seconds) and the results here only use the lossless part of the impedance obtained from the traditional vector fitting process.

\begin{figure}[!h]
    \centering
    \includegraphics[width=\textwidth]{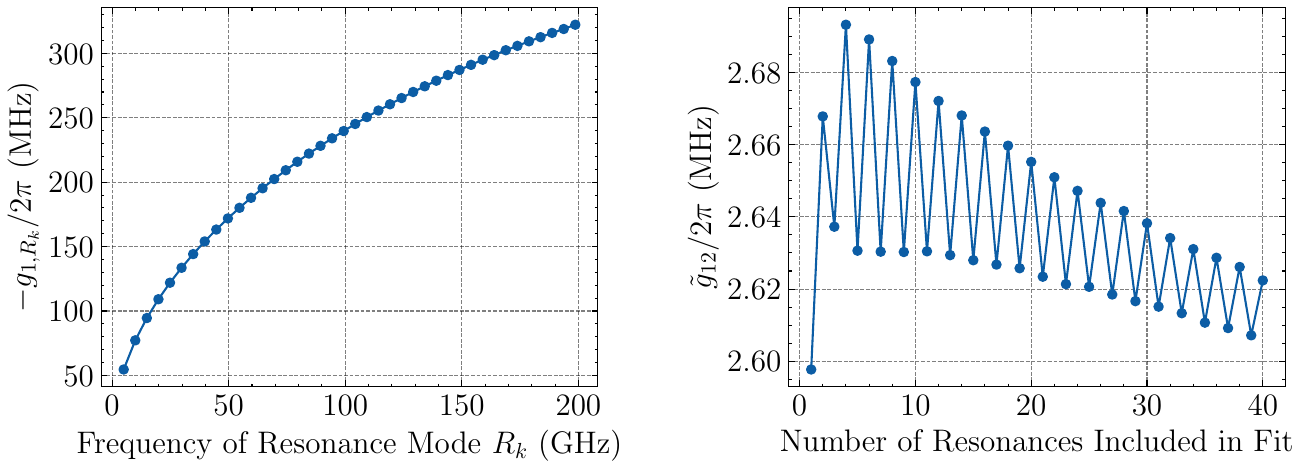}
    \caption{Left: For the fitted network in Fig.\ \ref{fig:ideal_TL_coupler_fit_40_res}, the coupling $g_{1,R_k}$ is shown for all 40 resonant modes present in the fit. Right: Effective coupling estimate (\ref{eq:eff_qubit_coupling}) as a function of the number of resonances included in the fit of the impedance function. Transmon frequencies are $\omega_1/2\pi=\omega_2/2\pi = $ 4 GHz.}
    \label{fig:ideal_TL_geff_change}
\end{figure}

First, we see that the coupling between the resonant modes and the transmons diverge as the resonant modes goes to higher frequency ($g_{i,R_k} \propto \omega_{R_k}^{1/2}$). This divergence is expected given that the Hamiltonian (\ref{eq:transmon_hamiltonian}) that we start with cannot account for the infinite number of resonant modes introduced by the ideal transmission line \cite{Parra-Rodriguez_2018}. This has the effect that the estimate of the effective coupling will also diverge if continuing to include higher resonant modes. In the effective coupling there is also an oscillation due to the alternating signs present in the coupling of one of the transmons to the resonant modes. This can be seen clearly in Table \ref{table:ideal_TL_coupling} and also in a similar example in Appendix \ref{appendix:cascade_ideal_TL}. In reality, there is a natural cutoff frequency for these resonant modes that is dependent on the superconducting gap frequency. In addition, there will be higher loss or attenuation in the material at higher frequencies that will reduce the coupling through the higher resonant modes \cite{picosecond_pulses,harmonic_superconducting}. These factors can help in choosing the cutoff frequency for our fitting.

%% file: mainmatter/ch4/electromagnetic.tex
\section{Electromagnetic Models}

In this section, we explore how the vector fitting and interconnection methods can be used for characterizing electromagnetic models of superconducting circuits. To obtain the multiport impedance parameters needed for our characterization methods, we use Ansys HFSS \cite{ansys_hfss} for the full-wave FEM electromagnetic simulations and Qiskit Metal \cite{Qiskit_Metal} for some of the device modeling. Also, when working with some of the simulation results we have used the Python package scikit-rf \cite{scikit_rf}.

\subsection{Brick Building Approach}

Simulating a full electromagnetic model of a multi-qubit superconducting circuit can be prohibitively expensive if trying to obtain the impedance parameter over a broad frequency range. To tackle this, we propose a method where circuit designs can be broken up into smaller and simpler ``bricks". Then, these pieces can be interconnected to obtain a model of a larger device. Since the number of ports and resonant modes for each of these bricks will be much smaller than for the full model, applying the fitting methods of Section \ref{section:vector_fitting} is possible. If we have the rational impedances of multiple bricks, we can then use the method of Section \ref{section:rational_impedance_interconnection} to obtain the rational impedance function of the larger interconnected model. From this rational impedance function, we can easily construct a circuit Hamiltonian.

Splitting up your circuit does not come without compromise. For example, capacitive coupling between qubits in different bricks will not be taken into account accurately. However, since we obtain the full Hamiltonian of the system, any potential cross-brick coupling that is not taken into account can be added in afterwards. For this, we would need to estimate the long range capacitive coupling of the components in the circuit, which can be done using formulas for planar electrodes \cite{planar_capacitance} or potentially with simpler capacitance only simulations.

\begin{figure}[!h]
    \centering
    \includegraphics[width=\textwidth]{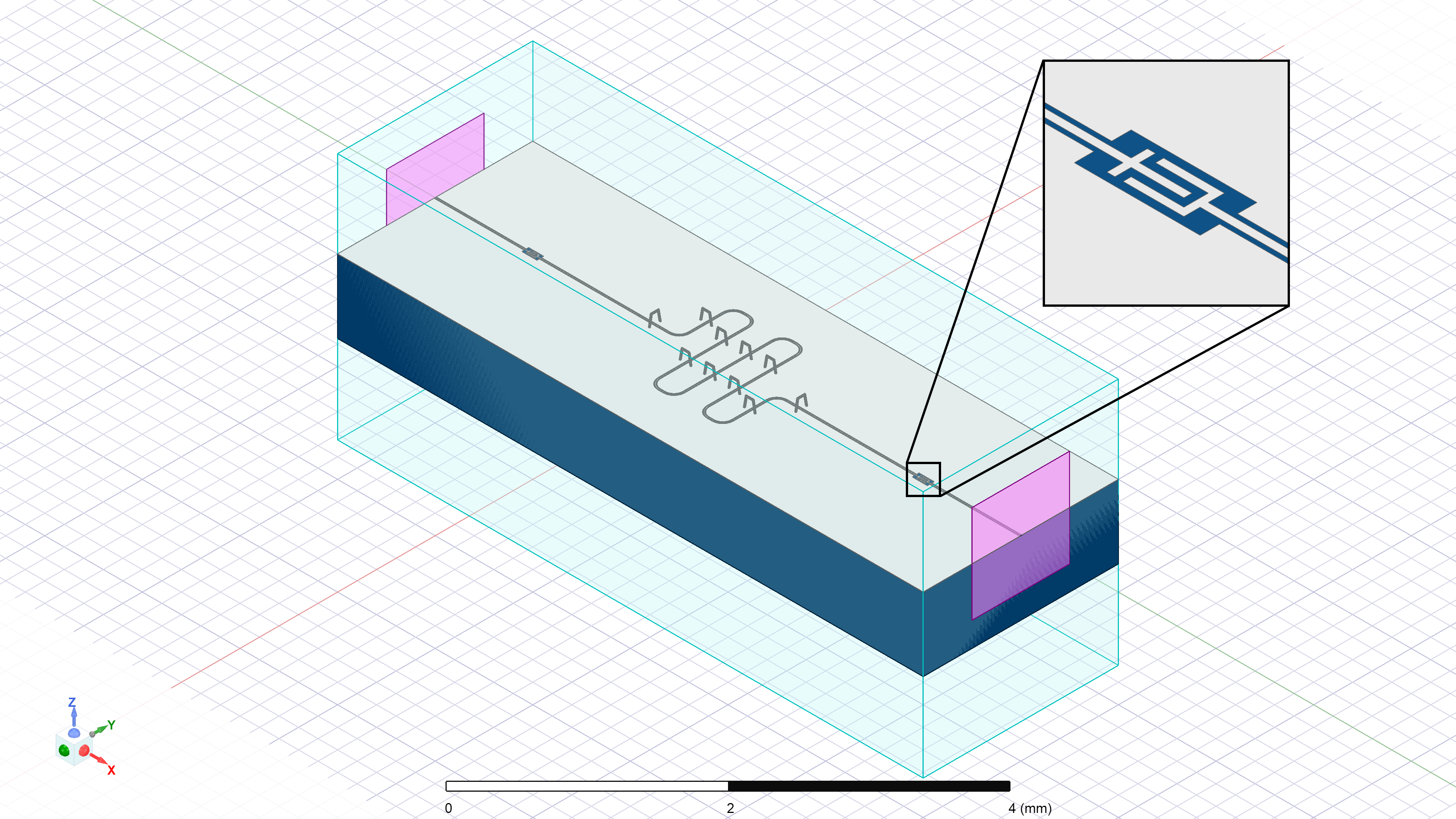}
    \caption{Two port circuit containing an 8 mm half-wave CPW resonator that is capacitively coupled to the two external ports.}
    \label{fig:cap_res_cap_full}
\end{figure}

\begin{figure}[!h]
    \centering
    \includegraphics[width=\textwidth]{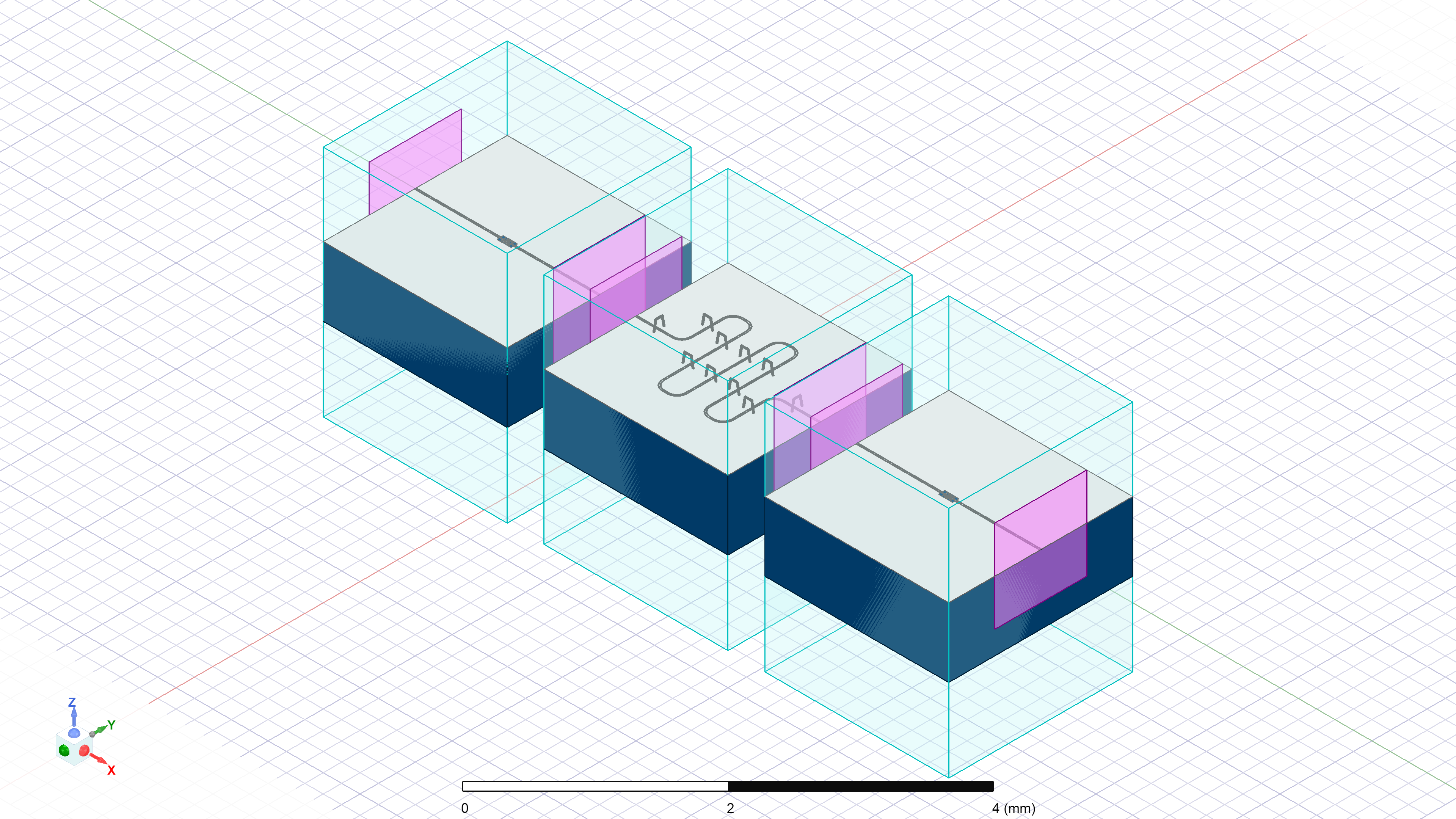}
    \caption{Split version of the model in Fig.\ \ref{fig:cap_res_cap_full}. The left and right bricks are identical and only contain the finger capacitor coupled to the two ends with CPWs. The center brick only contains a 6 mm CPW.}
    \label{fig:cap_res_cap_split}
\end{figure}

To see an example of the brick building method in action, we look at how it can be applied to the model in Fig.\ \ref{fig:cap_res_cap_full}. The circuit in Fig.\ \ref{fig:cap_res_cap_full} is a two port network which contains a half-wave coplanar waveguide (CPW) resonator that is capacitively coupled to the two ports. In this simulation and the ones that follow, the models will contain perfectly conducting sheets on top of a silicon substrate. Sometimes, wirebonds are also included over meandered CPWs. Wave ports are used for the ports located at the boundary of the model. We will compare the simulation results of this model to the split model as shown in Fig.\ \ref{fig:cap_res_cap_split}.

\begin{figure}[!t]
    \centering
    \includegraphics[width=\textwidth]{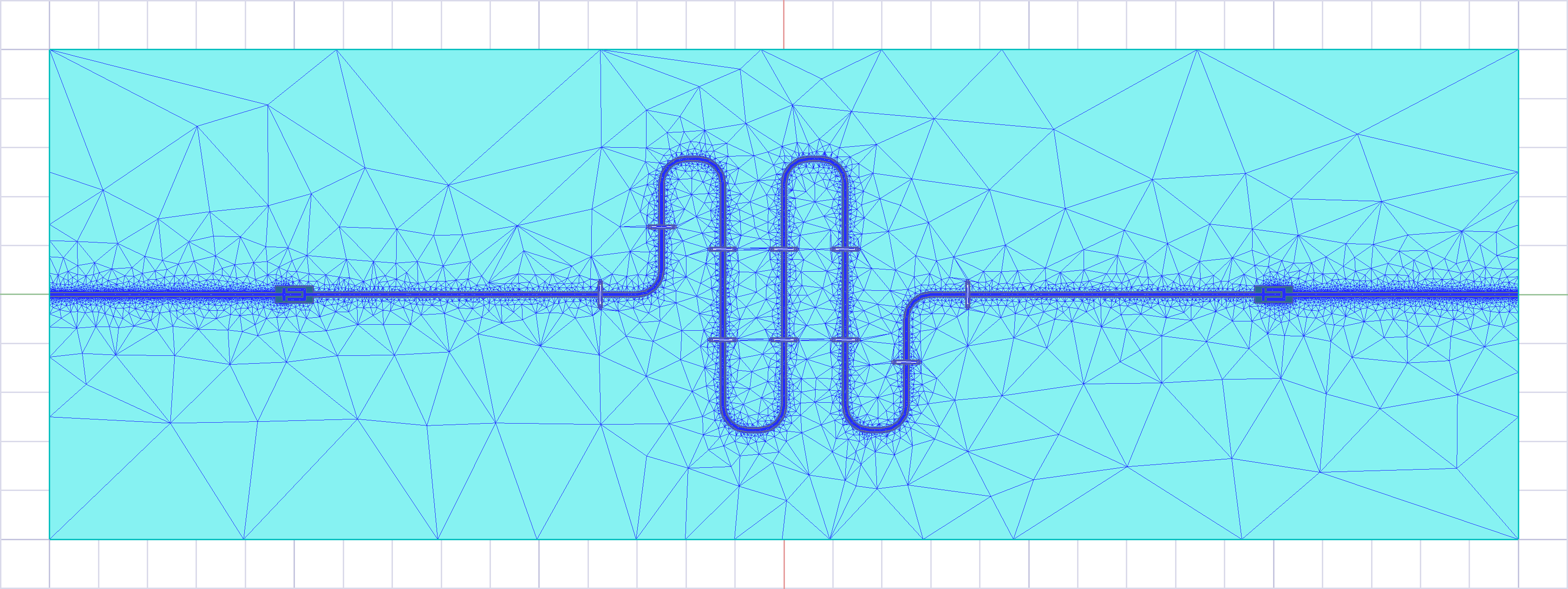}
    \caption{Mesh used for the simulation of the model in Fig.\ \ref{fig:cap_res_cap_full}.}
    \label{fig:cap_res_cap_mesh}
\end{figure}

For all of our simulations we will use the interpolating sweep options in HFSS, where the software chooses at what frequencies to solve the model, and then interpolates the solution with its own fitting methods. The model is repeatedly solved and fitted until the difference in S-parameters between runs is under a specified percentage. We have chosen an error tolerance of 0.5\% for these examples. Alternatively, we could pick these frequency points ourselves and apply the fitting methods directly. We are also using the adaptive fitting methods within HFSS to obtain a mesh that has a convergence at the high end of the our chosen frequency range (20.5 GHz). The mesh for the model in Fig.\ \ref{fig:cap_res_cap_full} is shown in Fig.\ \ref{fig:cap_res_cap_mesh}.

\begin{figure}[!t]
    \centering
    \includegraphics[width=\textwidth]{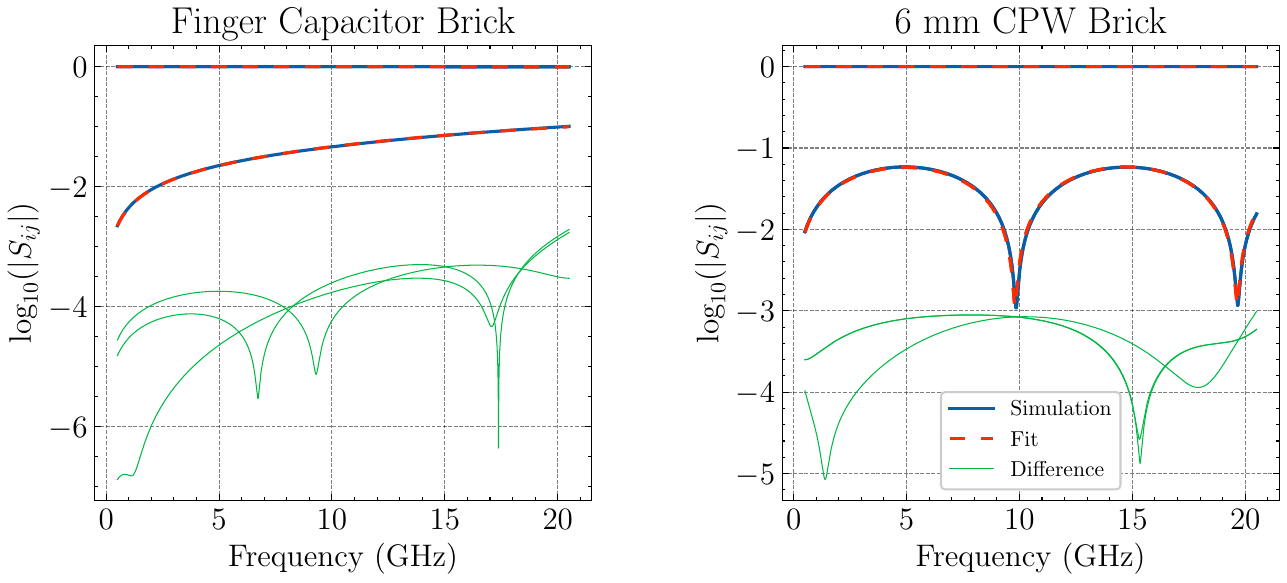}
    \caption{Fitting results for the capacitor and CPW bricks shown in Fig.\ \ref{fig:cap_res_cap_split}. After converting the fitted rational impedance function to S-parameters, it is compared to the S-parameters from the simulation. All the S-parameters and differences are plotted on the same plot.}
    \label{fig:cap_res_fit}
\end{figure}

\begin{figure}[!h]
    \centering
    \includegraphics[width=0.6\textwidth]{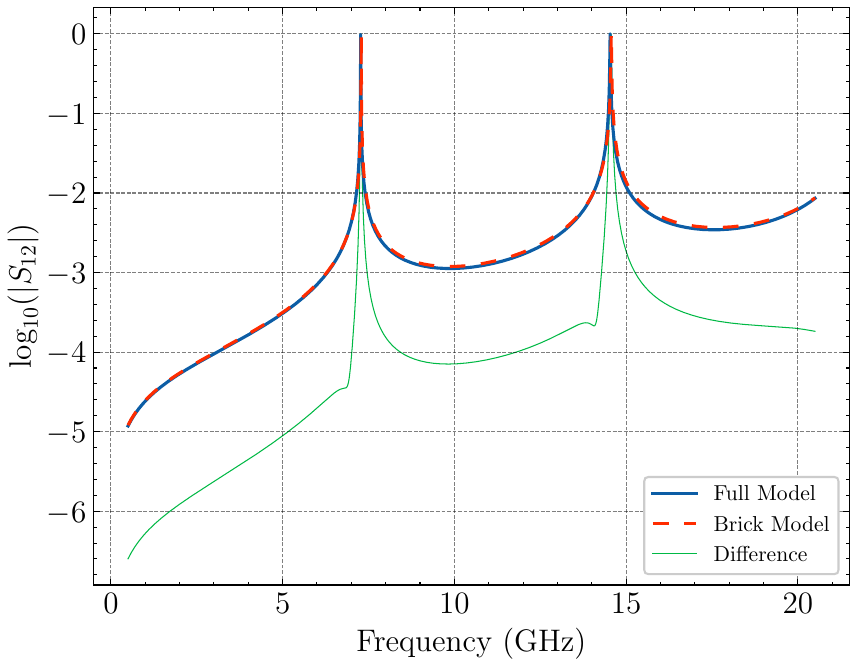}
    \caption{Difference between the $S_{12}$ parameter from the simulation of the full model in Fig.\ \ref{fig:cap_res_cap_full} and the interconnected brick model in Fig.\ \ref{fig:cap_res_cap_split}.}
    \label{fig:full_vs_brick}
\end{figure}

We then apply the fitting process from Section \ref{section:vector_fitting} to the simulations of the bricks in Fig.\ \ref{fig:cap_res_cap_split}. The results of the fitting are shown in Fig.\ \ref{fig:cap_res_fit}. Then, using the interconnection method of Section \ref{section:rational_impedance_interconnection}, we can stitch bricks together to obtain a final rational impedance function. To show the difference between the full model and the brick model, we compare the $S_{12}$ parameters in Fig.\ \ref{fig:full_vs_brick}. In this comparison, we see that primary differences are in the resonance frequencies of the full model and the brick model. We estimate that the resonance frequency of the fundamental mode in the brick model is approximately 15 MHz higher ($+0.2$\%) compared to the full model. We also confirm that this is not due to the fitting by comparing the interconnected model before and after the fit and finding a negligible difference. This suggests that the modeling of these bricks can be improved. One potential cause of the difference is the meshes for the full and split models. To improve on this, we could potentially decrease the error tolerance for the adaptive meshing process which in this case is at 1\%. We could also make the adaptive meshing process sample results from more frequency points. Another potential problem with the split model could be the dimensions of the wave ports. These points in addition to other components such as the interpolating sweeps and even comparison with other simulation software should be explored further in any implementations of this splitting method.

\begin{figure}[!h]
    \centering
    \includegraphics[width=\textwidth]{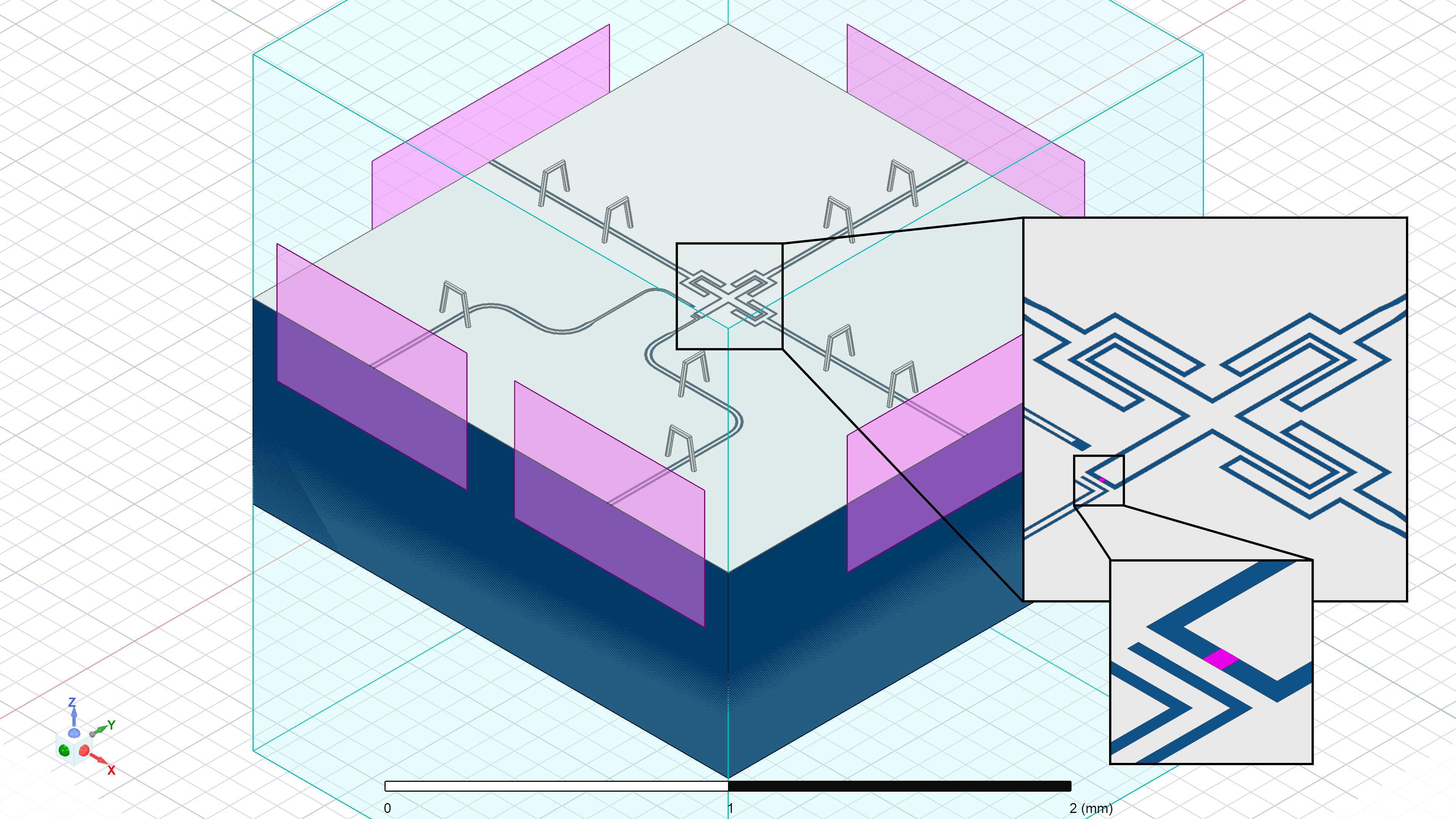}
    \caption{Brick model containing an Xmon style qubit \cite{xmon}. On the lower side of the cross, a lumped port (purple) is in the position where the SQUID would be located. Of the five wave ports on the boundary of the model, four correspond to CPW lines that are capacitively coupled to the qubit. The fifth wave port corresponds to a flux control line.}
    \label{fig:xmon_brick}
\end{figure}

\begin{figure}[!t]
    \centering
    \includegraphics[width=\textwidth]{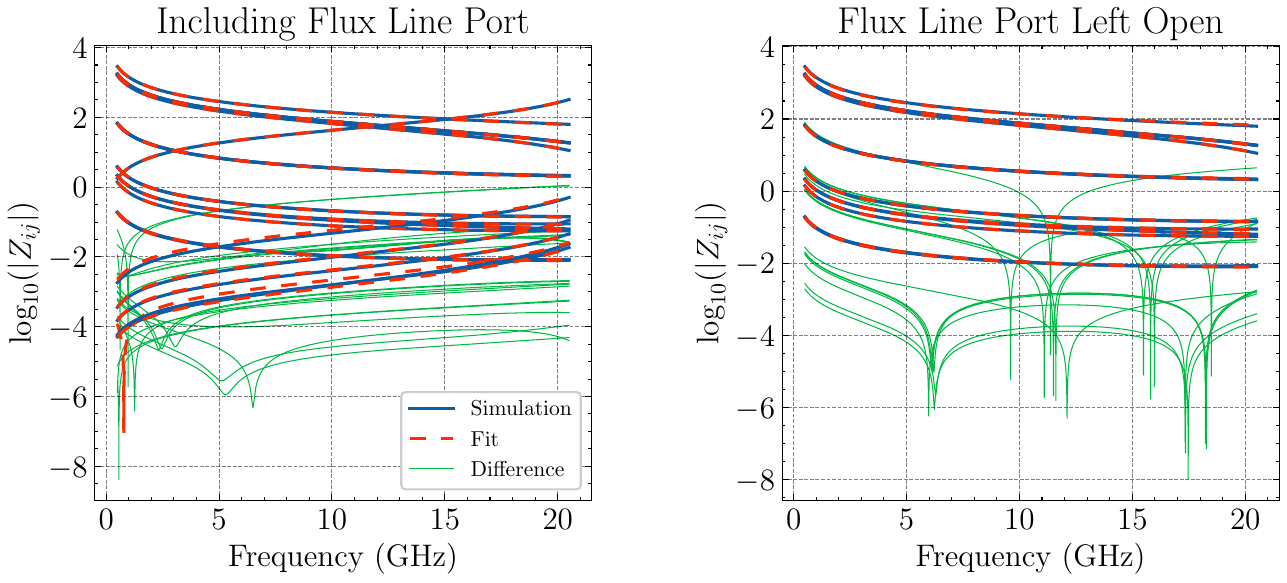}
    \caption{Impedance fit results for the simulation of the model shown in Fig.\ \ref{fig:xmon_brick}. On the left, the impedance including the flux line port is fitted. Note that the curves going to zero at low frequency on the left correspond to the parameters involving the flux line. In the fit on the right, the flux line port is left open to avoid inaccuracies brought in by trying to fit for the small residues corresponding to the flux line. All impedance parameters and differences between the fit and simulation are plotted together.}
    \label{fig:xmon_fit}
\end{figure}

Next, we look at a brick simulation that contains a transmon qubit. For qubits, we will have a lumped port that is located inside the boundaries of the simulation, unlike the wave ports previously seen. This lumped port is precisely the Josephson junction or ``qubit'' port that has been discussed in previous sections. The specific model we will discuss now is shown in Fig.\ \ref{fig:xmon_brick}. We want to use this model to discuss some more details that must be considered during the fitting process. When it comes to fitting this model, we need to be careful when dealing with the flux line port. Because the flux line is galvanically connected to the ground plane, it can be difficult to obtain a rational approximation of the form \ref{eq:impedance} with a positive definite DC residue. This is because the flux port components of the DC residue would be zero for a network like this. The fitting process can get close by including very small residue components and high frequency resonant modes, but this can cause unwanted effects in the Hamiltonian. To avoid these issues, we can fit the simulated impedance when the flux port is left open, and use this result when interconnecting with other bricks and constructing a Hamiltonian. However, we can still use the fit that includes the flux port for decay rate estimation. The results for fitting the simulated impedance for the model in Fig.\ \ref{fig:xmon_brick} are shown in Fig.\ \ref{fig:xmon_fit}. The fit including the flux line port struggles at low frequency due to the difficulty of attempting to fit the small residues and it also requires a number of poles outside of the visible frequency range. We would like to avoid including these types of poles when possible. When the flux port is left open, the fit for the remaining ports only requires one degenerate pole far out in the frequency range. We have found that allowing for a single degenerate pole when working with the electromagnetic models is sometimes needed for the fits to be accurate and this does not cause problems with the Hamiltonians. Sometimes, these degenerate poles and their residues can be thought of as an approximation to the infinite frequency pole that we have left out of (\ref{eq:impedance}).

When leaving the flux port open, we can obtain the Maxwell capacitance matrix that corresponds to the ports by inverting the DC residue. We then compare this to a Maxwell capacitance matrix obtained from a simulation in Ansys Q3D \cite{ansys_q3d}. In the Q3D simulation, the capacitance is computed between the metallic islands corresponding to the cross forming the Xmon and the CPWs leading to the boundary of the model. These matrices and differences are given in Appendix \ref{appendix:q3d_vs_hfss}. The differences are largest for the coupling capacitances that are small ($<$ 1 fF). Otherwise, estimates of the capacitance from fitting the HFSS model don't differ by more than 7\%. This difference is not caused by the fitting, and additional HFSS simulations and fits restricted to a low frequency range of 100 MHz to 1 GHz yield similar results. The differences likely come from the fact that the two simulation methods are different (HFSS is full-wave and Q3D is quasi-static) in addition to the different meshes and different port definitions in both models. There is the possibility to use Q3D within the HFSS simulation to solve the DC point, but with the combination of wave and lumped ports used in our models, this option is not available. For our examples here, we will use the models from HFSS, but it should be noted that improvements for the estimation at the DC point should be explored in the future within Ansys and potentially other simulation software.

\newpage

\begin{figure}[h!]
    \centering
    \includegraphics[width=\textwidth]{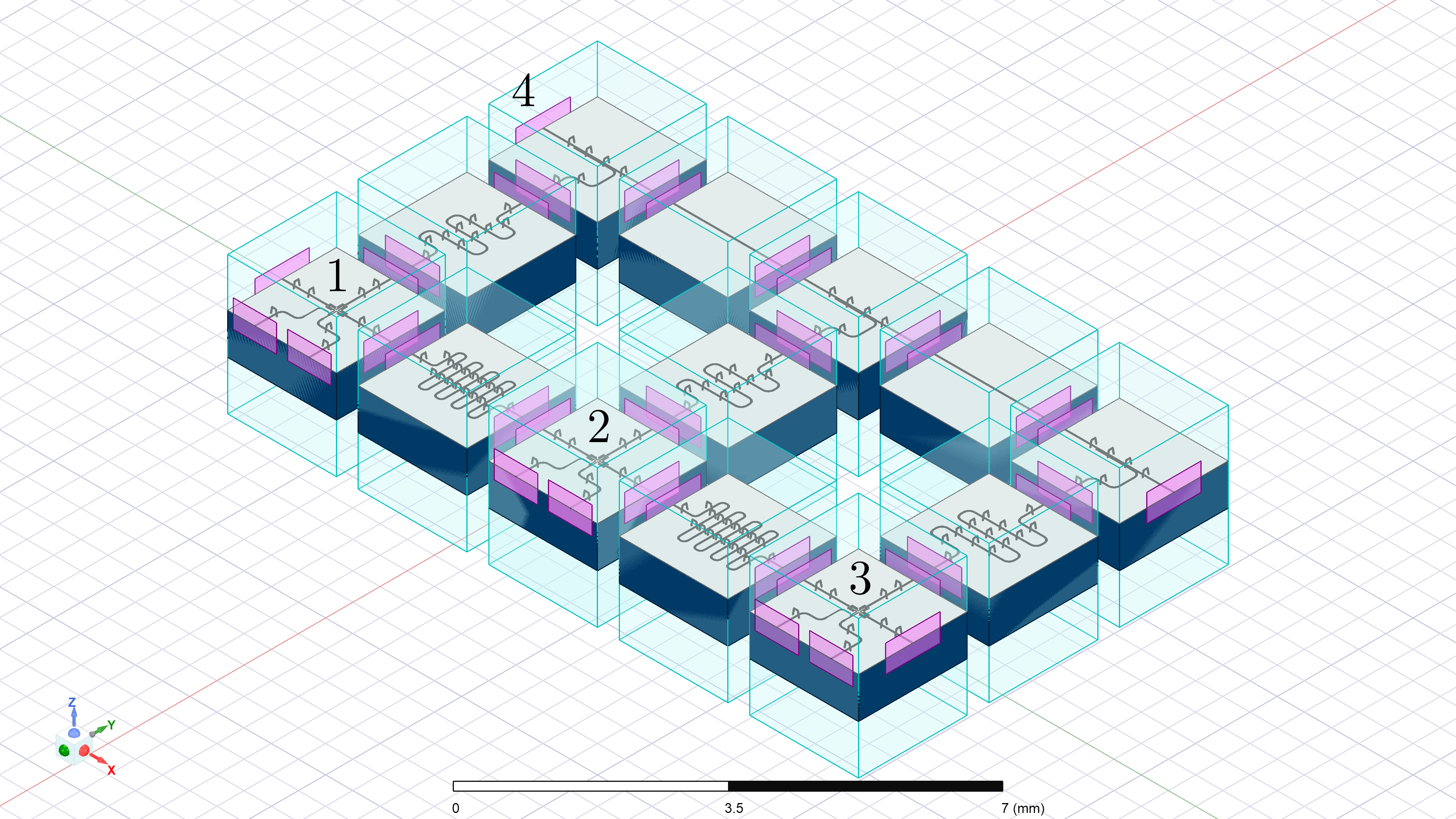}
    \caption{Brick model of a three Xmon circuit. The qubits are coupled to each other through half-wave resonators. Each qubit is also capacitively coupled to its own half-wave readout resonator that is capacitively coupled to a common readout line.}
    \label{fig:triple_xmon}
\end{figure}

\begin{figure}[h!]
    \centering
    \includegraphics[width=\textwidth]{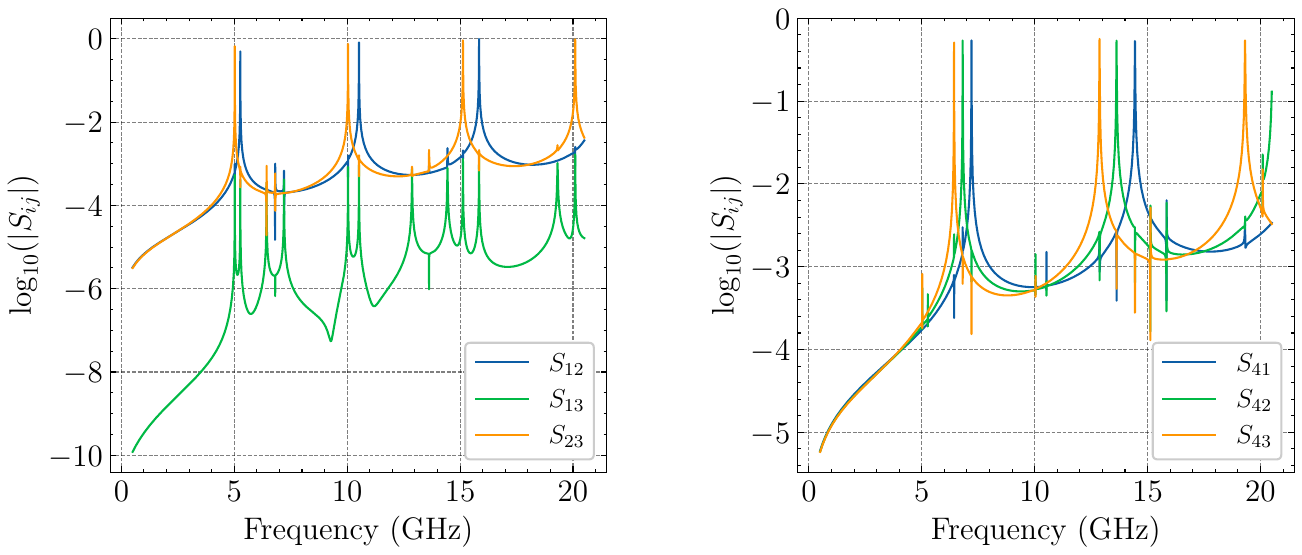}
    \caption{Some of the matrix elements of the S-parameter for the fully interconnected network shown in Fig.\ \ref{fig:triple_xmon}. Port numbers are shown in Fig.\ \ref{fig:triple_xmon}. Ports 1-3 correspond to the three qubit junction ports. Port 4 corresponds to the left port of the feedline in Fig.\ \ref{fig:triple_xmon}.}
    \label{fig:triple_xmon_S}
\end{figure}

Taking a collection of bricks and their corresponding rational impedance functions, we can interconnect them to make a larger model. As an example, we consider the model in Fig.\ \ref{fig:triple_xmon}. In this model there are three resonator-coupled Xmon qubits. Each Xmon is also coupled to its own readout resonator that is also coupled to a common readout line. By looking at the S-parameters for this network, we clearly see which resonant modes couple the qubits to each other and to the feedline. Some of the relevant S-parameters are shown in Fig.\ \ref{fig:triple_xmon_S}.

\newpage

\begin{figure}[h!]
    \centering
    \includegraphics[width=\textwidth]{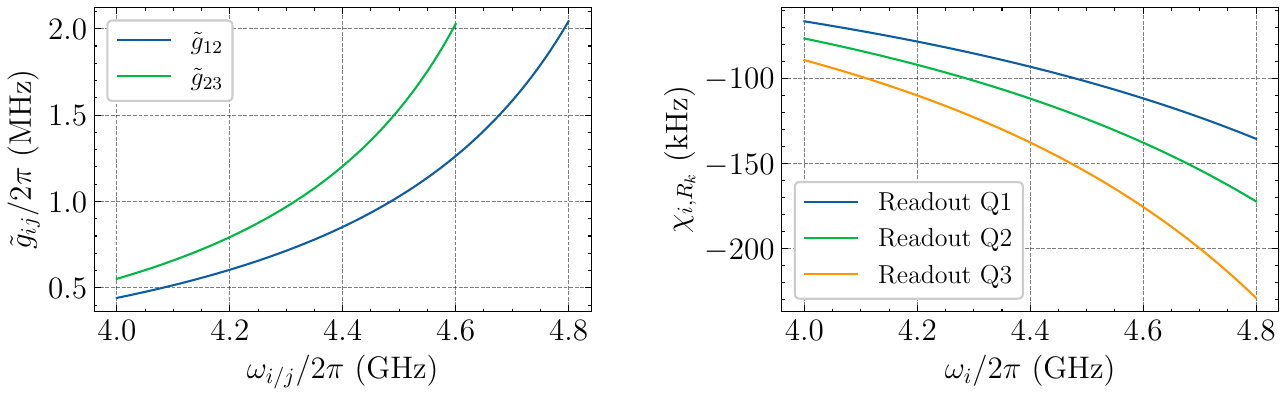}
    \caption{Effective qubit coupling rates and readout resonator dispersive shifts for the interconnected model of Fig.\ \ref{fig:triple_xmon}. Left: Effective coupling rates between the qubits (third qubit is fixed at 4 GHz while the other two qubit frequencies are varied). The fundamental frequencies for the coupling resonators are 5.268 GHz ($Q1 \leftrightarrow Q2$) and 5.023 GHz ($Q2 \leftrightarrow Q3$). A cutoff frequency of 21.5 GHz is used for this example to avoid incorrect predictions of resonant modes outside the fitting range. Right: The dispersive shift in the fundamental modes of the readout resonators. The fundamental frequencies of the readout resonators for qubits 1 to 3 are 7.205 GHz, 6.812 GHz, and 6.435 GHz, respectively.}
    \label{fig:triple_xmon_geff_chi}
\end{figure}

\begin{figure}[h!]
    \centering
    \includegraphics[width=\textwidth]{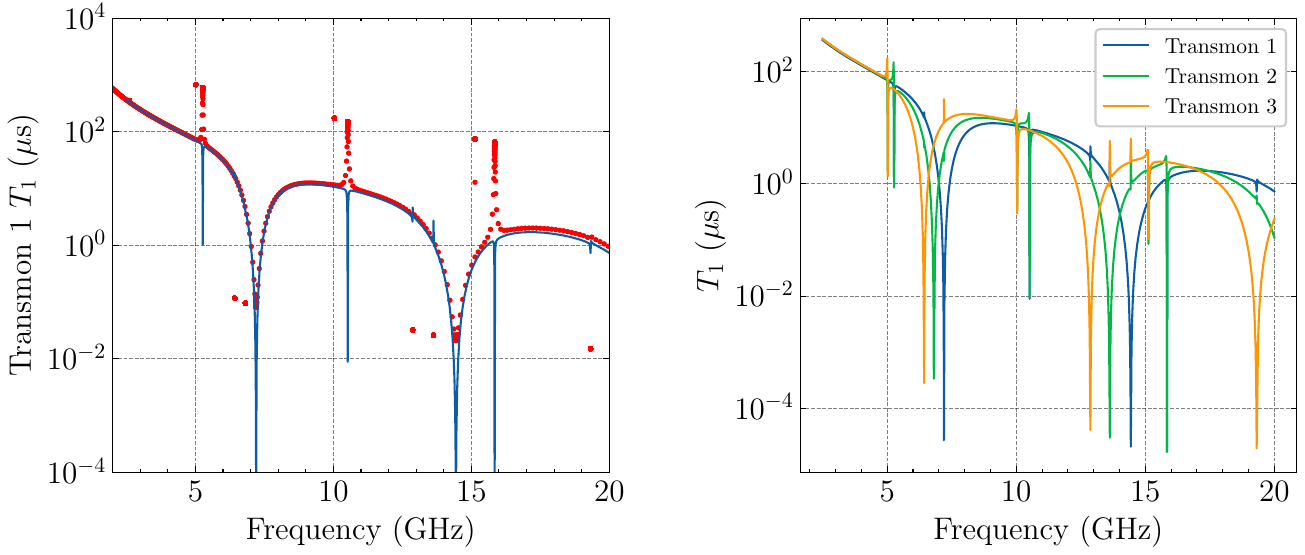}
    \caption{Relaxation times for the qubits in the circuit of Fig.\ \ref{fig:triple_xmon}. On the left plot, we focus only on qubit 1. For a sweep of the shunt inductance of qubit 1, the complex frequencies from (\ref{eq:matrix_eoms}) are plotted as red points. Isolated red points belong to the resonant modes that have a nearly fixed decay rate due to their small coupling to qubit 1. The blue line is the result of using (\ref{eq:qubit_decay_admittance}) with resistors shunting the external ports and no inductances shunting the qubit ports. On the right, we see the result from (\ref{eq:qubit_decay_admittance}) for all three qubits in the circuit.}
    \label{fig:triple_xmon_T1}
\end{figure}

With the rational impedance function corresponding to the fully interconnected model in Fig.\ \ref{fig:triple_xmon}, we can also build a circuit Hamiltonian of the form (\ref{eq:transmon_resonator_ham}). This can then be used to estimate the effective coupling rates between the qubits using (\ref{eq:eff_qubit_coupling}). We can also estimate the dispersive shift in the fundamental resonance frequency of each readout resonator using (\ref{eq:dispersive_shifts}). For the circuit in Fig.\ \ref{fig:triple_xmon}, these effective coupling rates and dispersive shifts are shown in Fig.\ \ref{fig:triple_xmon_geff_chi}. Additionally, with the fully interconnected model, we can estimate the relaxation times for each qubit by using (\ref{eq:matrix_eoms}) or (\ref{eq:qubit_decay_admittance}). The results for our example circuit in Fig.\ \ref{fig:triple_xmon} are shown in Fig.\ \ref{fig:triple_xmon_T1}.

\begin{figure}[h!]
    \centering
    \includegraphics[width=\textwidth]{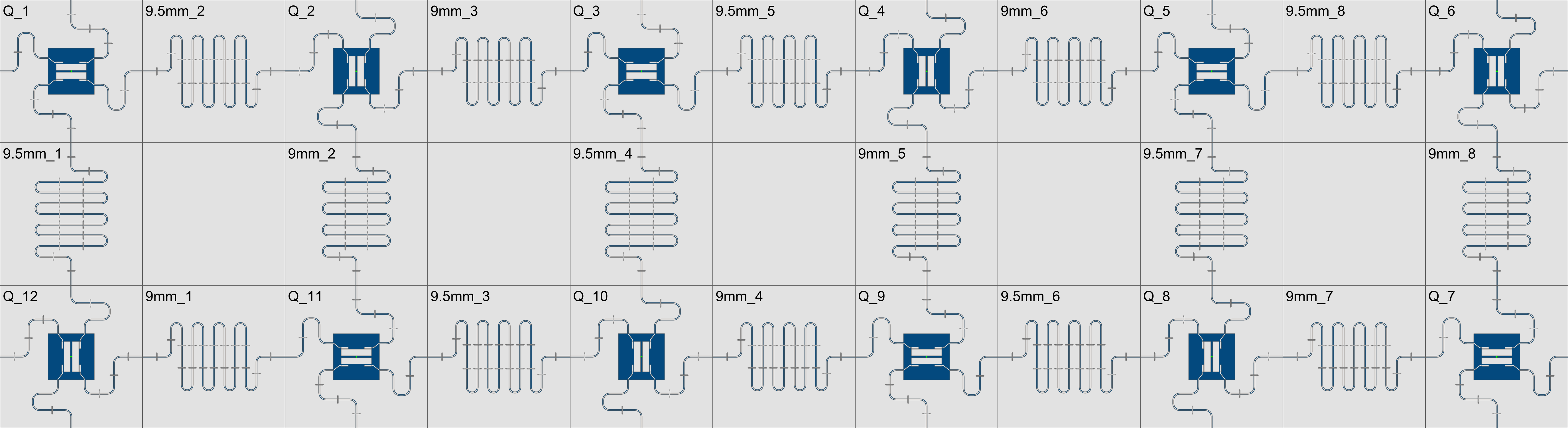}
    \caption{A grid of IBM style transmons \cite{solgun_sirf} that are coupled through bus resonators. The qubits are numbered 1 to 12 starting from the top left and moving clockwise around the lattice.}
    \label{fig:2x6}
\end{figure}

Using these brick models, we can also estimate how the effective coupling through resonant modes present in a circuit will scale for multi-qubit devices. To do this, we recreate a qubit grid circuit similar to an example from \cite{solgun_sirf} where lumped elements were used. In our case, we use various bricks that correspond to electromagnetic models to create a grid of qubits. The circuit we consider is shown in Fig.\ \ref{fig:2x6}.

\begin{figure}[h!]
    \centering
    \includegraphics[width=0.5\textwidth]{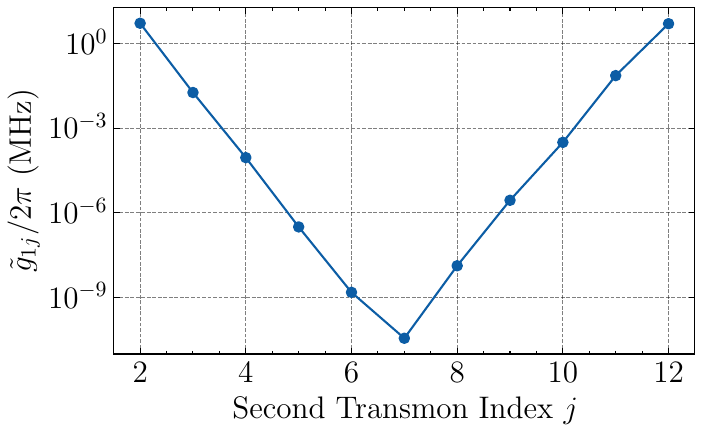}
    \caption{Effective coupling of qubit 1 of the circuit in Fig.\ \ref{fig:2x6} to all of the other qubits. All qubits are frequencies are fixed at 4.25 GHz. In practice, we would not see the influence of such small coupling rates $g/2\pi \lesssim 1$ kHz, but this can still be a useful tool for verifying that the long range coupling through resonant modes in a specific layout is negligible.}
    \label{fig:2x6_coupling}
\end{figure}

After building the model of the circuit in Fig.\ \ref{fig:2x6}, we can compute the effective coupling rates between the qubits using (\ref{eq:eff_qubit_coupling}). The effective coupling of qubit 1 to all of the other qubits is shown in Fig.\ \ref{fig:2x6_coupling}. In this figure, we can see the exponential decay of the effective coupling as we get further from qubit 1. This exponential decay was also found for networks of this type in the lumped element example of \cite{solgun_sirf}. It is also expected for networks of this form if the direct capacitive coupling is excluded (see Appendix \ref{appendix:banded_cap}).

\subsection{Parasitic Resonances}
As the number of components present within superconducting circuits increases, we may start to see parasitic resonances at frequencies closer to the operating frequency range of the qubits. These parasitic resonances at lower frequencies could have a non-negligible impact on the effective coupling between qubits. Depending on where these resonant modes arise (physically and in frequency), it is be useful to see how changes to device geometry results in changes of the couplings to these parasitic resonances.

\begin{figure}[h!]
    \centering
    \includegraphics[width=0.75\textwidth]{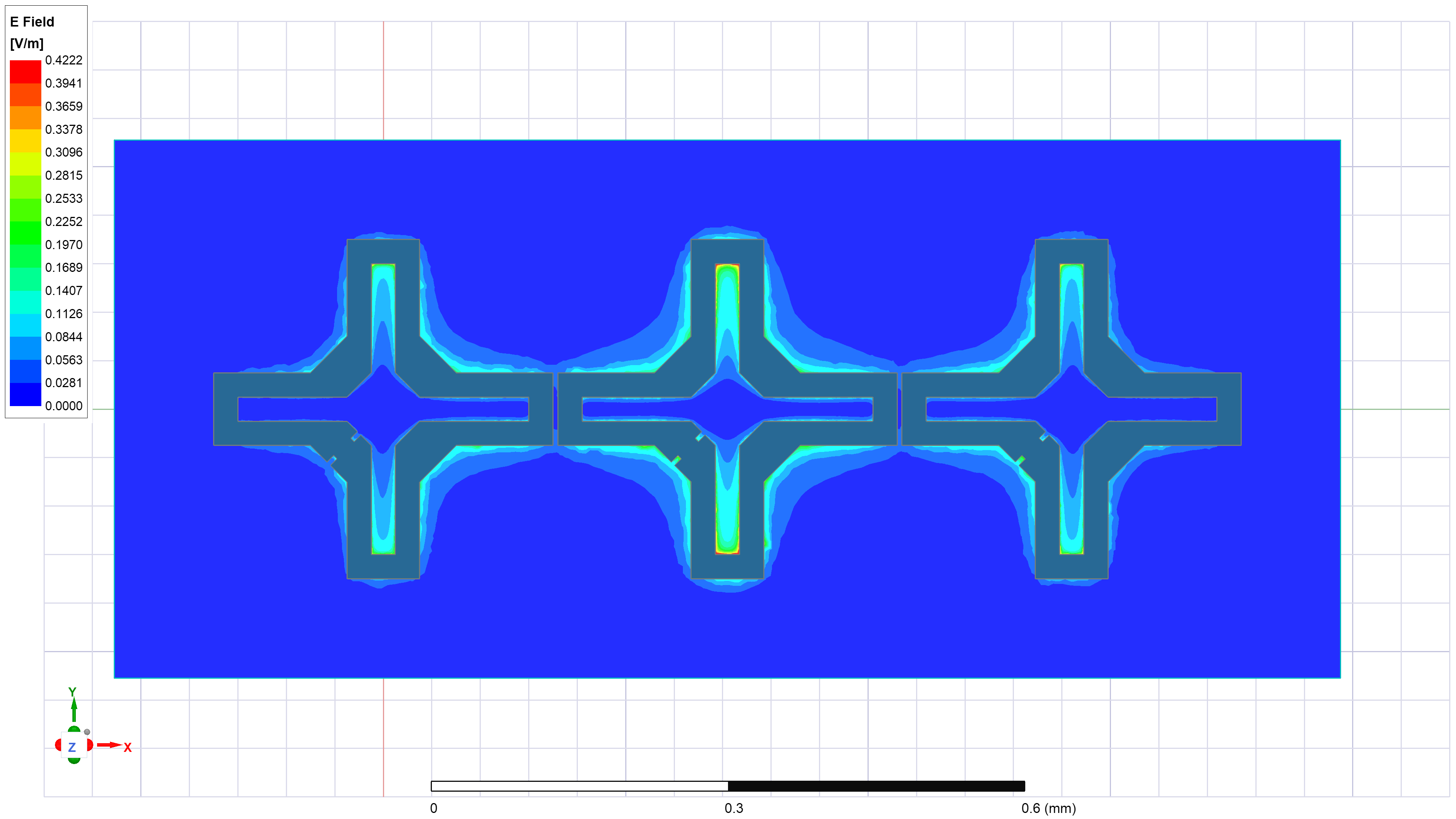}
    \caption{Magnitude of the electric field at the surface of the perfect conductor in a three Xmon chain. This is the lowest resonant mode above the operating range of the qubits. The resonant frequency is 52.5322 GHz.}
    \label{fig:xmon_1x3_eig}
\end{figure}

\begin{figure}[h!]
    \centering
    \includegraphics[width=\textwidth]{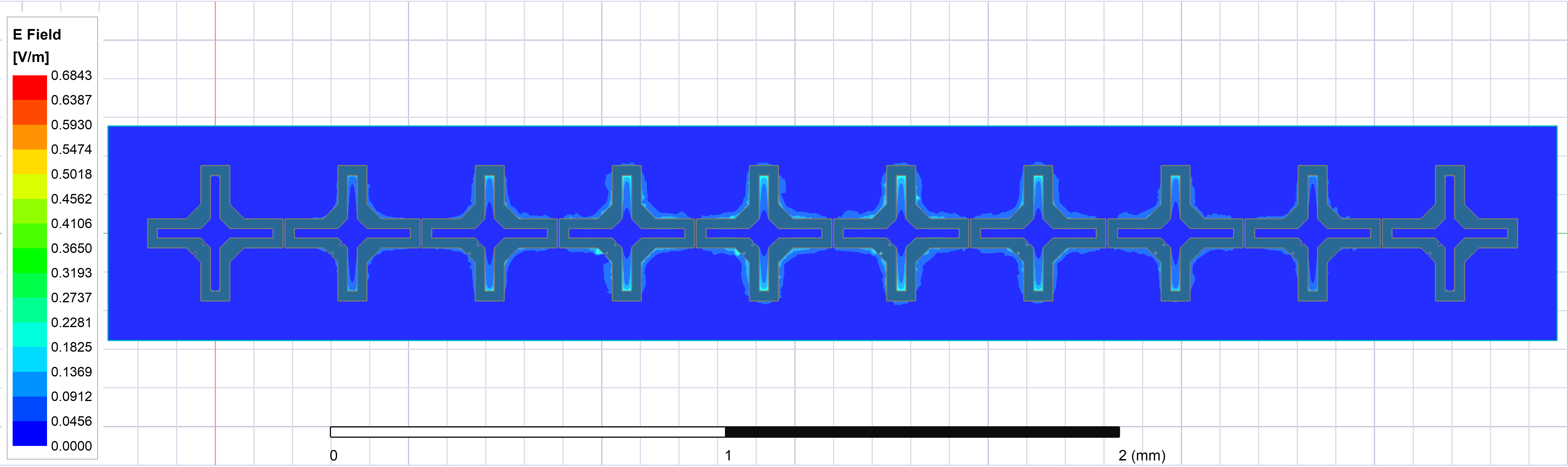}
    \caption{Magnitude of the electric field in a ten Xmon chain for the lowest resonant mode of 37.8159 GHz in the structure.}
    \label{fig:xmon_1x10_eig}
\end{figure}

\begin{figure}[h!]
    \centering
    \includegraphics[width=0.5\textwidth]{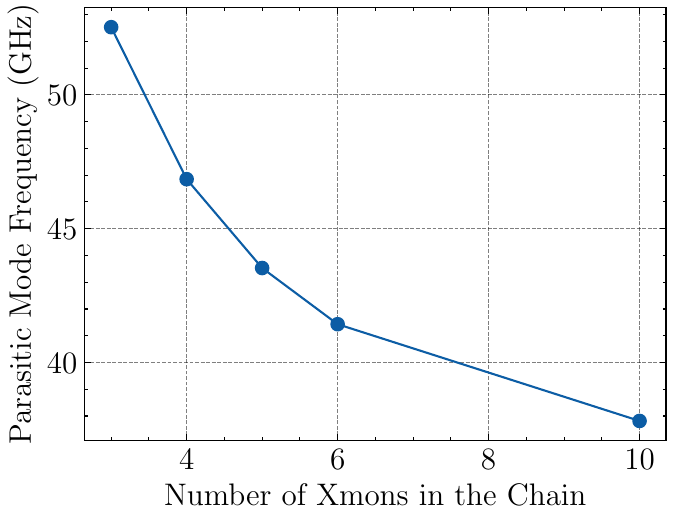}
    \caption{Lowest parasitic resonance frequencies for different numbers of Xmons in a chain. Resonance frequencies are computed using the eigenmode solver within HFSS.}
    \label{fig:xmon_chain_res}
\end{figure}

As an example, we consider a parasitic resonance that is present within a chain of Xmons. The parasitic resonance we will be concerned with is the lowest resonant mode of the chain structure that is above the operating frequencies of the qubits (around 4 to 6 GHz). Of course, there will be higher frequency parasitic resonances, but here we only consider the lowest as it should be the mode that contributes the most to any changes in effective coupling. In Fig.\ \ref{fig:xmon_1x3_eig}, we can see the electric field profile of the lowest resonant mode in a chain of three Xmons. The field profile and resonance frequency are found using the eigenmode solver within HFSS. While not super clear with just three Xmons, the outer two crosses participate less in the resonance than the center cross. This is clearer if we increase the length of the chain as shown in Fig.\ \ref{fig:xmon_1x10_eig}. We can also see that when adding more Xmons to the chain, the frequency of the parasitic resonance in the chain decreases. Using the eigenmode solver in HFSS, we obtain the results in Fig.\ \ref{fig:xmon_chain_res} that show how the frequency of the resonance decreases as we increase the number of Xmons.

\begin{figure}[h!]
    \centering
    \includegraphics[width=\textwidth]{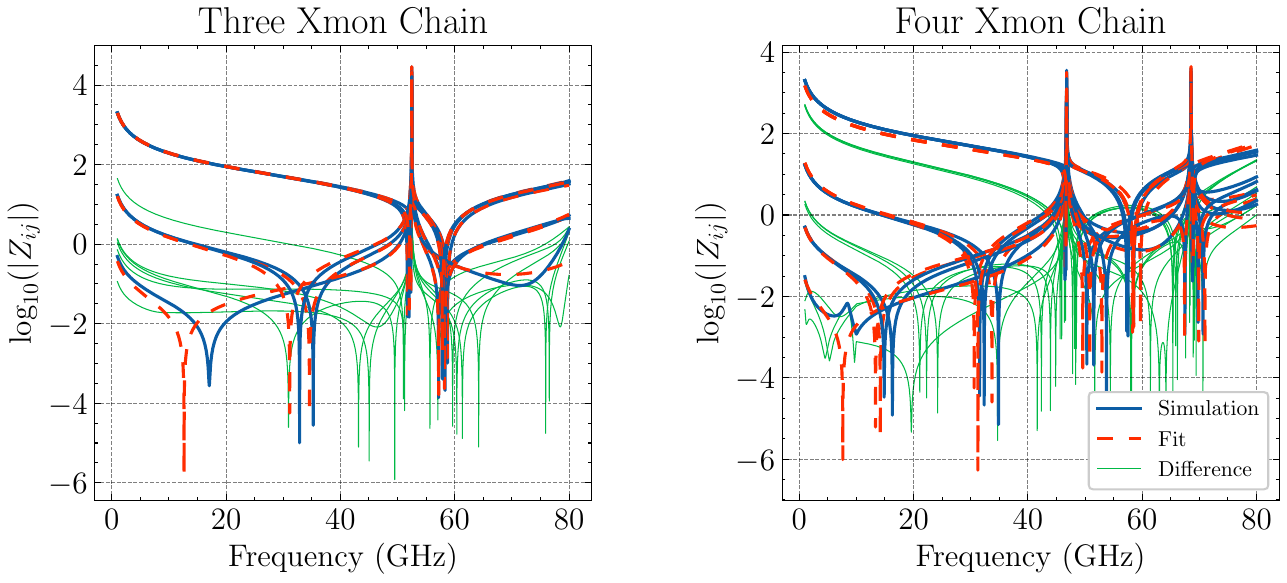}
    \caption{Attempts at fitting the impedance parameter for the three and four Xmon chains. For the three qubit chain shown in Fig.\ \ref{fig:xmon_1x3_eig}, we find that magnitudes of the coupling rates of the three qubits to the lowest parasitic resonance mode are 103, 228 and 167 MHz from left to right. For the four qubit chain, the magnitudes of the coupling rates are 129, 292, 321 and 204 MHz. This is for the qubit frequencies fixed at 4 GHz. The coupling rates are not symmetric due to the junction ports being off center on each cross. Due to the poor fitting of the DC and parasitic resonance residues, the coupling rates are likely inaccurate as we expect the coupling rates of the qubits to the resonance to be lower in the four Xmon chain (see Fig.\ \ref{fig:xmon_3_4_S}).}
    \label{fig:xmon_3_4_fit}
\end{figure}

\begin{figure}[!ht]
    \centering
    \includegraphics[width=\textwidth]{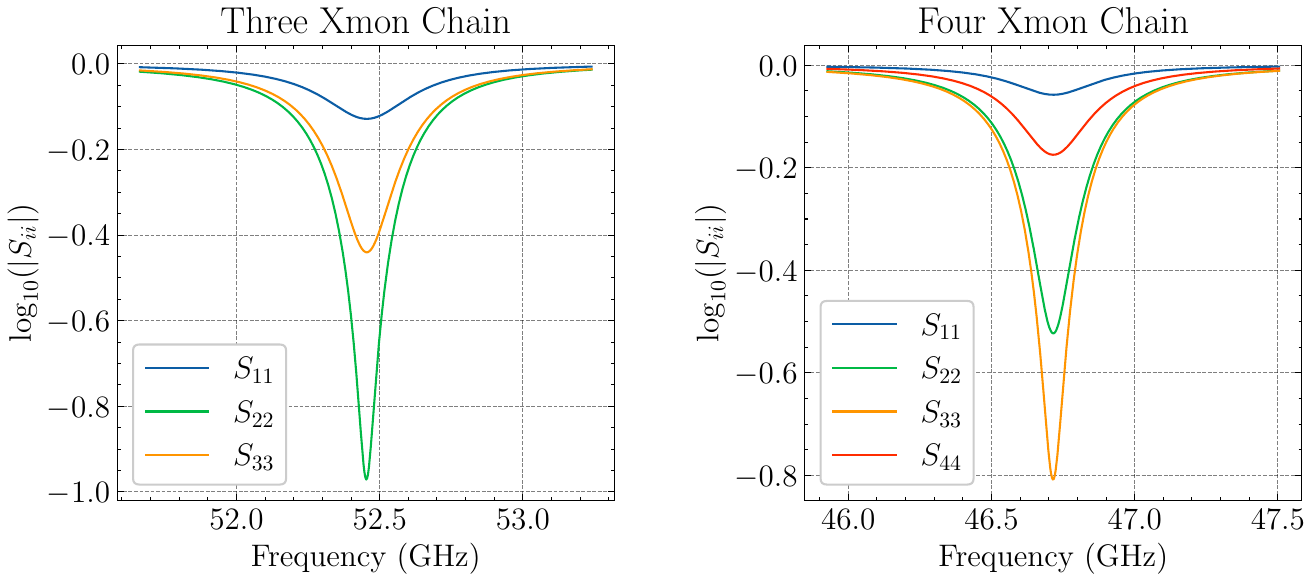}
    \caption{Diagonal matrix elements of the S-parameters for the three and four Xmon chains. The qubit ports are numbered left to right on the chain. We can see that closer to the center of the chain, the qubits are more strongly coupled to the parasitic resonance. Also, we can see that in the four Xmon chain, the qubits are coupled less to the parasitic resonance than in the three Xmon chain.}
    \label{fig:xmon_3_4_S}
\end{figure}

\begin{figure}[!ht]
    \centering
    \includegraphics[width=\textwidth]{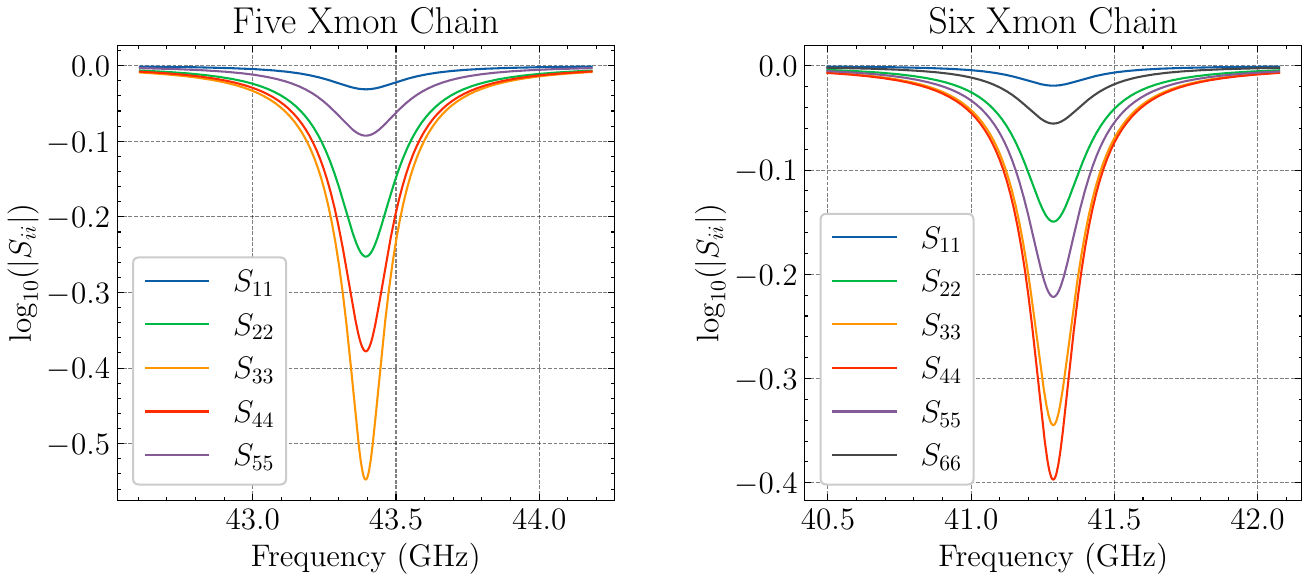}
    \caption{Similar to Fig.\ \ref{fig:xmon_3_4_S}, but for the five and six Xmon chains.}
    \label{fig:xmon_5_6_S}
\end{figure}

To estimate the coupling rates of the qubits to this parasitic resonance, we can try to use the vector fitting methods of Section \ref{section:vector_fitting}. Including lumped ports at the positions of the junctions, we obtain the impedance parameter in HFSS over a broad frequency range (1 GHz to 80 GHz) so that the lower parasitic resonance is visible. Unfortunately, the traditional vector fitting struggles to give a good fit for the whole frequency range with the results shown in Fig.\ \ref{fig:xmon_3_4_fit}. Due to the poor fit, the estimated coupling rates are likely inaccurate. We can alternatively look at the S-parameters to understand how the qubits are coupled to the parasitic resonance relative to one another. This can also be used to track how coupling to parasitic resonances changes for different device geometry. In Fig.\ \ref{fig:xmon_3_4_S} we can see the diagonal matrix elements of the S-parameters for the three and four Xmon chains. This clearly shows which qubits are most strongly coupled to the parasitic resonance. If changes are made to the device, we can potentially track how the S-parameters compare between devices to see if the coupling of the qubits to these parasitic resonances has gone down. In Fig.\ \ref{fig:xmon_5_6_S} we show similar plots for the five and six Xmon chains.

\newpage

%% file: mainmatter/ch5/chapter5.tex
\chapter{Conclusion and Outlook}

This study explored how to use multiport lossless impedance functions to construct and characterize models of multi-qubit superconducting circuits. In Chapter \ref{chapter:lossless_impedance}, we formulated the CL cascade synthesis of the lossless reciprocal impedance function. We then discussed the analysis of a CL cascade which, when paired with the synthesis, can be used to generally interconnect rational impedance functions. Then, we examined how vector fitting can be used to obtain a rational approximation of a discretized lossless impedance function. This allowed us to construct a lumped circuit model that corresponds to a distributed element model or electromagnetic simulation.

Chapter \ref{chapter:analysis_impedance} shows how to construct a Hamiltonian for a multi-qubit circuit that corresponds to a rational impedance function. We focus on transmon networks, and directly see how the coupling rates between the qubits and resonant modes are related directly to the residues of the impedance function. Then, by approximately block diagonalizing the Hamiltonian, we find an effective qubit Hamiltonian with new shifted qubit frequencies and effective coupling rates that are valid in the dispersive regime. Furthermore, we find expressions for the dispersive shifts in the resonator frequencies that are dependent on the qubit states. In this chapter, we also showed that from a classical circuit perspective, we can compute the relaxation times of all the resonant modes in the circuit due to loss through the external ports of the network.

Finally, in Chapter \ref{chapter:examples}, we present a few examples that showcase the methods discussed in Chapter \ref{chapter:lossless_impedance} and Chapter \ref{chapter:analysis_impedance}. Notably, we show how in a multi-qubit circuit, a lower bound for the qubit decay rates can be found from the diagonal elements of a lossy admittance function where the external ports are shunted by resistors. Additionally, we see how the vector fitting methods can provide us with a circuit Hamiltonian that corresponds to distributed element circuits or electromagnetic models. We also demonstrate how to use the vector fitting and the rational impedance interconnection method alongside a ``brick building” approach to construct larger models out of simpler components.

In the future, it may be worth extending the methods we have investigated to admittance parameters. This may potentially allow for a better handling of circuits that contain a positive semi-definite DC residue (e.g. port with a flux line). The technique of using the circuit Lagrangians may also demonstrate useability in other lossless circuit analysis and synthesis problems. This would allow the interconnection methods to be generalized even further. In obtaining our immittance functions from simulation results, the traditional vector fitting methods we have used do not yield passive or lossless models. For this reason, exploring new vector fitting methods specifically designed for lossless networks may be quite advantageous for the methods we've discussed.

%% file: backmatter/acknowledgements.tex
\chapter{Acknowledgements}

First and foremost, I want to thank David DiVincenzo for being an incredible mentor over the past two years of master’s studies and research. I will be forever grateful for the time you made for me and the many discussions that we have had. You allowed me great freedom, and I always enjoyed the side quests that we had along the way. Your supervision has led me to appreciate a vast range of topics and their historical significance.

I would like to express my sincere gratitude to Rami Barends for your invaluable guidance and support as my second supervisor. You have truly convinced me to pursue the dark side that is experimental physics. Your approachability and readiness to assist with any challenges I faced significantly enriched my research experience, and your willingness to let me gain hands-on experience in the lab was not only instrumental in my learning but also deeply appreciated.

I extend my heartfelt thanks to the group members of PGI-13, for their warmth and generosity. Even though I have not been with you for long, I have been met with nothing but kindness and support whenever I needed assistance in the lab. Being surrounded by such a remarkable group of talented individuals has not only been inspiring but was also a privilege. Your collective spirit and expertise have left a lasting impression on me, and I am excited to witness your future achievements and breakthroughs. 

Last, but not least, I want to thank Aubrey, my parents and my friends for their unconditional love, support, and patience. I would not have reached this point without all of you.

%% file: backmatter/appendix.tex
\chapter{Appendix}

\section[Lagrangian of the Full Lossless Reciprocal Impedance Function]{Lagrangian of the Full Lossless Reciprocal Impedance\\ Function}\label{appendix:infty_freq_residue}
Now we consider the rational impedance function (\ref{eq:impedance}) but including the infinite frequency residue:
\begin{equation}
    \vb{Z}(s) = \frac{\vb{R}_0}{s} + \sum_{k=1}^M \frac{s \vb{R}_k}{s^2 + \omega_{R_k}^2} + s\vb{R}_\infty
\end{equation}
This corresponds to the full Cauer circuit shown in Fig.\ \ref{fig:cauer_circuit}. We can construct the Lagrangian in a similar way to what was done for the circuit without the purely inductive stage. Including the purely inductive stage, the Lagrangian now reads:
\begin{equation}
    \mathcal{L} = \frac{1}{2}\dot{\vb{\Phi}}_C^T \vb{C}_0 \dot{\vb{\Phi}}_C^{\phantom{^T}} + \frac{1}{2}\dot{\vb{\Phi}}_R^T \vb{C}_R \dot{\vb{\Phi}}_R^{\phantom{^T}} - \frac{1}{2}\vb{\Phi}_R^T \vb{M}_R \vb{\Phi}_R^{\phantom{^T}} - \frac{1}{2}\vb{\Phi}_L^T \vb{M}_L \vb{\Phi}_L^{\phantom{^T}} - \sum_{i=1}^N U_i(\Phi_{P_i})
\end{equation}
where everything is defined the same as before but now we also have $\vb{M}_L=\diag(L_1^{-1},\dots,L_N^{-1})$. Now we have the following constraints for the branch and node flux vectors:
\begin{align}
    \vb{\Phi}_C &= \vb{U}^T \vb{\Phi}_{PC} \\
    \vb{\Phi}_{PC} &= \vb{\Phi}_P - \vb{R}^T \vb{\Phi}_R - \vb{T}^T \vb{\Phi}_L
\end{align}
Using the above, we can obtain a Lagrangian that is only dependent on $\vb{\Phi}_P$, $\vb{\Phi}_R$, and $\vb{\Phi}_L$. The substitution $\vb{\Phi}_C \rightarrow \vb{U}^T (\vb{\Phi}_P - \vb{R}^T\vb{\Phi}_R  - \vb{T}^T \vb{\Phi}_L)$ allows us to write the following Lagrangian:
\begin{equation}
    \mathcal{L} =  \frac{1}{2}\dot{\vb{\Phi}}^T \vb{C} \dot{\vb{\Phi}} - \frac{1}{2}\vb{\Phi}^T \vb{M} \vb{\Phi}- \sum_{i=1}^N U_i(\Phi_{P_i})
\end{equation}
where we now have
\begin{align}
    \vb{\Phi} &= (\Phi_{P_1},\dots,\Phi_{P_N},\Phi_{R_1},\dots,\Phi_{R_M},\Phi_{L_1},\dots,\Phi_{L_N})^T\\
    \vb{C} &= \mqty(
        \vb{R}_0^{-1} & -\vb{R}_0^{-1}\vb{R}^T & -\vb{R}_0^{-1}\vb{T}^T \\
        -\vb{R}\vb{R}_0^{-1} & \vb{C}_R + \vb{R}\vb{R}_0^{-1}\vb{R}^T & \vb{R}\vb{R}_0^{-1}\vb{T}^T \\
        -\vb{T}\vb{R}_0^{-1} & \vb{T}\vb{R}_0^{-1}\vb{R}^T & \vb{T}\vb{R}_0^{-1}\vb{T}^T
    ) \\
    \vb{M} &= \mqty(
        \vb{0}_{N \times N} & \vb{0}_{N \times M} & \vb{0}_{N \times N} \\
        \vb{0}_{M \times N} & \vb{M}_R & \vb{0}_{M \times N} \\
        \vb{0}_{N \times N} & \vb{0}_{N \times M} & \vb{M}_L
    )
\end{align}
Here we can see that this impedance function does have a CL representation similar to the case of the impedance without the infinite frequency pole and residue. However, we cannot construct a circuit Hamiltonian since $\vb{C}$ is singular. To show this, we can partition $\vb{C}$ into 4 blocks:
\begin{equation}
    \vb{C} = \mqty( \bar{\vb{A}} & \bar{\vb{B}} \\ \bar{\vb{C}} & \bar{\vb{D}})
\end{equation}
where
\begin{align}
    \bar{\vb{A}} &= \vb{R}_0^{-1} \\
    \bar{\vb{B}} &= \mqty( -\vb{R}_0^{-1}\vb{R}^T & -\vb{R}_0^{-1}\vb{T}^T ) \\ 
    \bar{\vb{C}} &= \mqty( -\vb{R}\vb{R}_0^{-1} \\ -\vb{T}\vb{R}_0^{-1} ) \\
    \bar{\vb{D}} &= \mqty(  \vb{C}_R + \vb{R}\vb{R}_0^{-1}\vb{R}^T & \vb{R}\vb{R}_0^{-1}\vb{T}^T \\
    \vb{T}\vb{R}_0^{-1}\vb{R}^T & \vb{T}\vb{R}_0^{-1}\vb{T}^T )
\end{align}
The matrix $\vb{C}$ is invertible if $\bar{\vb{D}} - \bar{\vb{C}}\bar{\vb{A}}^{-1}\bar{\vb{B}}$ is invertible \cite[Proposition 2.8.7]{matrix_mathematics}. Computing this matrix, we find
\begin{equation}
    \bar{\vb{D}} - \bar{\vb{C}}\bar{\vb{A}}^{-1}\bar{\vb{B}} = \mqty( \vb{C}_R & \vb{0}_{M \times N} \\  \vb{0}_{N \times M} & \vb{0}_{N \times N})
\end{equation} 
This matrix is clearly singular and thus $\vb{C}$ is also singular. 

\newpage
\section{Degenerate Resonant Modes}\label{appendix:degen_res_mode}
\begin{figure}[h!]
    \centering
    \begin{circuitikz}[line width=1pt]
        \ctikzset{bipoles/thickness=1, bipoles/length=1cm, , monopoles/ground/thickness=0.85}
        \ctikzset { label/align = straight }
        
        \draw (-3,0) -- (-2,0) to[C,l=$C_c$] (-1,0) -- (0,0) -- (1,0) to[C,l=$C_c$] (2,0) -- (3,0) to[C,l=$C_c$] (1.5,-2.25) to[C,l=$C_c$] (0,-4.5) to[C,l=$C_c$] (-1.5, -2.25) to[C,l=$C_c$] (-3,0);

        \draw (-1.5,-2.25) -- (-3,-2.25) -- (-3,-2.75) -- (-3.5,-2.75) to[C,l_=$C$] (-3.5, -3.75) -- (-2.5, -3.75) to[L,mirror,l_=$L_R$] (-2.5,-2.75) -- (-3,-2.75);
        \draw (-3,-3.75) to (-3,-4) node[ground]{};

        \draw (0,0) -- (0,-0.75) -- (-0.5,-0.75) to[C,l_=$C$] (-0.5,-1.75) -- (0.5,-1.75) to[L,mirror,l_=$L_R$] (0.5,-0.75) -- (0,-.75);
        \draw (0,-1.75) to (0,-2) node[ground]{};

        \draw (1.5,-2.25) -- (3,-2.25) -- (3,-2.75) -- (3.5,-2.75) to[C,l=$C$] (3.5, -3.75) -- (2.5, -3.75) to[L,l=$L_R$] (2.5,-2.75) -- (3,-2.75);
        \draw (3,-3.75) to (3,-4) node[ground]{};

        \draw (-3,0) -- (-5,0) -- (-5,-0.5) to[C,l=$C$] (-5,-1.5) to (-5,-1.75) node[ground]{};
        \draw (-5,-0.5) to[short, -o] (-6,-0.5);
        \draw (-5,-1.5) to[short, -o] (-6,-1.5);

        \draw (3,0) -- (5,0) -- (5,-0.5) to[C,l_=$C$] (5,-1.5) to (5,-1.75) node[ground]{};
        \draw (5,-0.5) to[short, -o] (6,-0.5);
        \draw (5,-1.5) to[short, -o] (6,-1.5);

        \draw (0,-4.5) -- (0,-5) to[C,l_=$C$] (0,-6) to (0,-6.25) node[ground]{};
        \draw (0,-5) to[short,-o] (1,-5);
        \draw (0,-6) to[short,-o] (1,-6);

        \node[circle, fill=nodecolor, inner sep=0pt,minimum size=5pt] at (0,0) {};
        \node[circle, fill=nodecolor, inner sep=0pt,minimum size=5pt] at (3,0) {};
        \node[circle, fill=nodecolor, inner sep=0pt,minimum size=5pt] at (-3,0) {};
        \node[circle, fill=nodecolor, inner sep=0pt,minimum size=5pt] at (0,-4.5) {};
        \node[circle, fill=nodecolor, inner sep=0pt,minimum size=5pt] at (-1.5,-2.25) {};
        \node[circle, fill=nodecolor, inner sep=0pt,minimum size=5pt] at (1.5,-2.25) {};

    \end{circuitikz}
    \caption{3-port circuit with 3 resonant modes (two of which are degenerate). The 6 nodes are marked at the red points.}
    \label{fig:tetrahedral_circuit}
\end{figure}
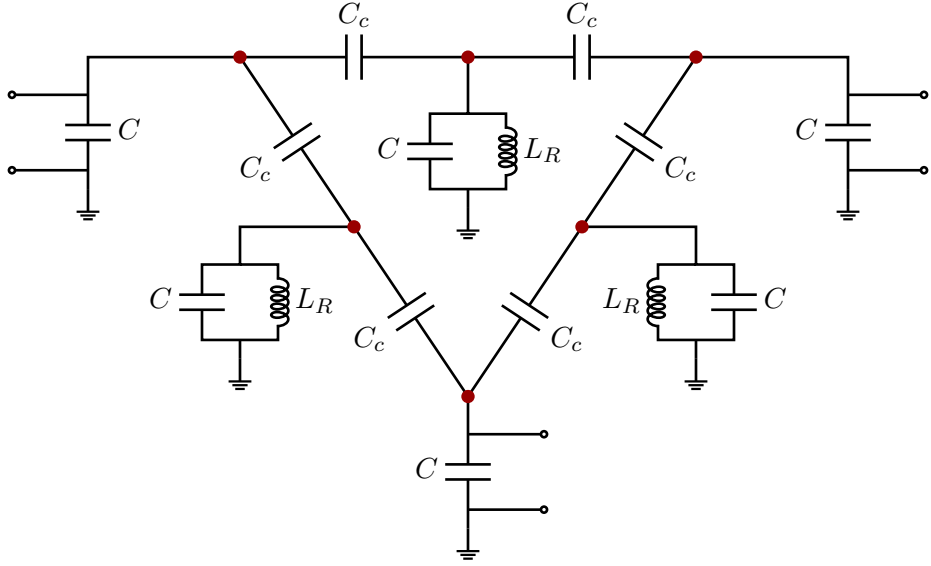

To see how degenerate resonant modes arise and how the degeneracy can be broken, we consider the 3-port network shown in Fig.\ \ref{fig:tetrahedral_circuit}. This network is based off of the tetrahedral qubit structure presented in \cite{feigelman_superconducting_2004}. 

\begin{figure}[h!]
    \centering
    \includegraphics[width=0.5\textwidth]{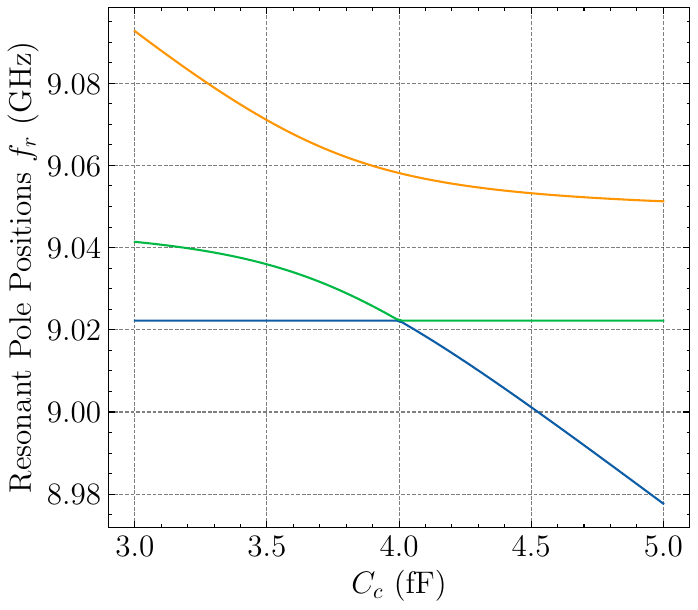}
    \caption{Resonance frequencies of the network in Fig.\ \ref{fig:tetrahedral_circuit} for the parameters $C=$ 70 fF, $C_c=$ 4 fF, and $L_R=$ 4 nH while varying a single one of the coupling capacitors $C_c$. We can see that if a single coupling capacitance is not at 4 fF, the degeneracy is broken.}
    \label{fig:tetra_poles}
\end{figure}

The network we have drawn is a CL cascade network with the capacitance matrix
\begin{equation}
    \vb{C} = \mqty(
        C & 0 & 0 & C_c & C_c & 0\\
        0 & C & 0 & 0 & C_c & C_c \\
        0 & 0 & C & C_c & 0 & C_c \\
        C_c & 0 & C_c & C & 0 & 0 \\
        C_c & C_c & 0 & 0 & C & 0 \\
        0 & C_c & C_c & 0 & 0 & C
    )
\end{equation}
and the inductance matrix $\vb{M}_R=L_R \mathds{1}_{3 \times 3}$. If we look at resonator branch block of the inverse capacitance matrix, we find:
\begin{equation}
    (\vb{C}^{-1})_R = \frac{1}{C^4 - 5C^2C_c^2 + 4C_c^4}\mqty(
        C^3 - 3CC_c^2 & CC_c^2 & CC_c^2\\
        CC_c^2 & C^3 - 3CC_c^2 & CC_c^2 \\
        CC_c^2 & CC_c^2 & C^3 - 3CC_c^2
    ) = \mqty(A & B & B \\ B & A & B \\ B & B & A)
\end{equation}
The eigenvalues of the last matrix are:
\begin{equation}
    \lambda = A - B,\; A - B,\; A + 2B
\end{equation}
and we see that two of these eigenvalues are degenerate. Recall from Section (\ref{section:cascade_analysis}) that if we diagonalize $(\vb{C}^{-1})_R\vb{M}_R$, the eigenvalues will be the squared resonant frequencies of the network. In this case, we have:
\begin{equation}
    (\vb{C}^{-1})_R\vb{M}_R = \vb{O}_C \vb{D} \vb{O}_C^T (L_R \mathds{1}_{3\times 3}) = \vb{O}_C L_R \vb{D} \vb{O}_C^T
\end{equation}
so we see that $\vb{\Omega} = (L_R \vb{D})^{1/2}$. By construction, two of the eigenvalues in $\vb{D}$ are equal, so we will have two degenerate resonant modes and a third resonant mode that is at a different frequency. If we consider the network in Fig.\ \ref{fig:tetrahedral_circuit}, we can compute the resonance frequencies for a given set of network components. We show this in Fig.\ \ref{fig:tetra_poles} while varying the value of a single coupling capacitor to show how the degeneracy is broken. In physical systems, these types of degeneracies are generally broken due to parasitic capacitances. This is why we assume in Section \ref{section:impedance_hamiltonian} that residues corresponding to resonant modes are rank-1.

\newpage
\section{Cascade Synthesis of the Ideal Transmission Line}\label{appendix:cascade_ideal_TL}
Here we consider the ideal lossless transmission line with two capacitors on either end as shown in Fig.\ \ref{fig:ideal_TL}. The result ends up being similar to the example in \cite[Appendix G]{Parra-Rodriguez_2018}, except here we have included the capacitors so that we have a positive definite DC residue.

\begin{figure}[h!]
    \centering
    \begin{circuitikz}[line width=1pt]
    \ctikzset{american}
    \ctikzset{bipoles/thickness=1, bipoles/length=1cm}
    \ctikzset{bipoles/crossing/size=0.5}
    \ctikzset { label/align = straight }

    \draw (0,1) to[short, o-] (0.5,1) to[C=$C_1$] (2,1) to[short, -o] (2.5,1) to[short,-o] (6,1) -- (6.5,1) to[C=$C_2$] (8,1) to[short, -o] (8.5,1);
    \draw (0,-.5) to[short,o-] (0.5,-.5) to[short, -o] (2.5,-.5) to[short, -o] (6,-.5) to[short,-o] (8.5,-.5);
    \node at (4.25,0.25) {$Z_0,\; \beta=\omega\sqrt{LC}$};
    
    \node at (4.25,-0.75) {$\ell$};
    \draw [-stealth](4.5,-0.75) -- (6,-0.75);
    \draw [-stealth](4,-0.75) -- (2.5,-0.75);

\end{circuitikz}
\caption{}
\label{fig:ideal_TL}
\end{figure}

The ABCD matrix of the above two-port network is \cite[Chapter 4]{Pozar_2011}:
\begin{equation}
    \mqty(A & B \\ C & D) = \mqty( 1 & 1/j\omega C_1 \\ 0 & 1) \mqty( \cos \beta\ell & jZ_0 \sin \beta\ell \\ jY_0\sin\beta\ell & \cos\beta\ell ) \mqty( 1 & 1/j\omega C_2 \\ 0 & 1)
\end{equation}
Using the ABCD parameters, we can compute the matrix elements of the two-port impedance parameter:
\begin{align}
    Z_{11} &= \frac{A}{C} = \frac{1}{j\omega C_1} - jZ_0 \cot\beta\ell \\
    Z_{12} &= Z_{21} = \frac{1}{C} = -jZ_0\csc\beta\ell \\ 
    Z_{22} &= \frac{D}{C} = \frac{1}{j\omega C_2} - jZ_0\cot\beta\ell
\end{align}
To find the cascade representation of this impedance function, we need to bring the above into partial fraction form. For this we can use the following series expansions \cite[1.421.3 and 1.422.3]{int_series_table}: 
\begin{align}
    \cot z &= \frac{1}{z} + 2z\sum_{k=1}^{\infty} \frac{1}{z^2 - \pi^2k^2} \\
    \csc z &= \frac{1}{z} + 2z\sum_{k=1}^{\infty} \frac{(-1)^{k}}{z^2 - \pi^2 k^2}
\end{align}
Putting together these expansions with the matrix elements of the impedance function for this two-port network, we find:
\begin{equation}
    \vb{Z}(s) = \frac{\vb{R}_0}{s} + \sum_{k=1}^{\infty} \frac{\vb{R}_k}{s^2 + \omega_k^2}
\end{equation}
where we have defined
\begin{align}
    \omega_k &= \frac{\pi^2 k^2}{L C \ell^2} \label{eq:ideal_TL_poles}\\
    \frac{1}{C_T} &= \frac{Z_0}{\ell\sqrt{L C}} \\
    \vb{R}_0 &= \mqty(C_1^{-1} + C_T^{-1} & C_T^{-1} \\ C_T^{-1} & C_2^{-1} + C_T^{-1}) \\
    \vb{R}_k &= \frac{2}{C_T} \mqty(1 & (-1)^k \\ (-1)^k & 1) = \frac{2}{C_T} \mqty(1 \\ (-1)^k) \Big(1 \quad (-1)^k\Big)
\end{align}

\begin{figure}[h!]
    \centering
    \includegraphics[width=\textwidth]{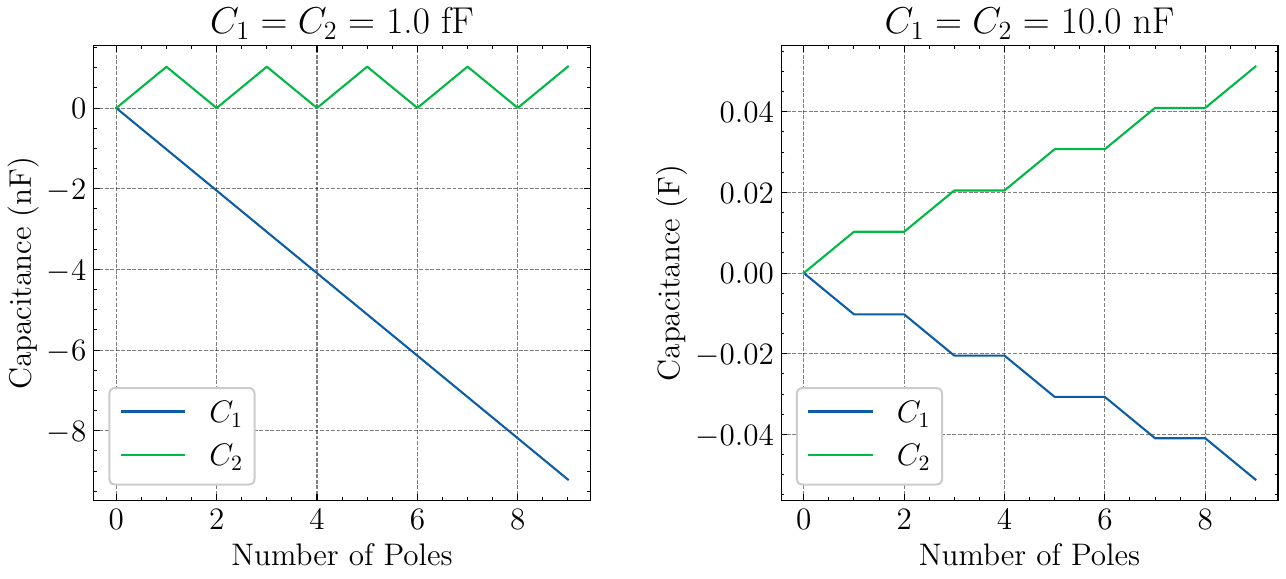}
    \caption{Cascade representation shunt capacitance divergence for a transmission line with $L = $ 0.438 $\mu$H/m, $C = $ 0.159 nF/m, and $\ell =$ 12 mm. On the left, we have the case of $C_1 \ll C_T$ and on the right, we have $C_1 \gg C_T$.}
    \label{fig:ideal_TL_shunt_cap}
\end{figure}

Now we can attempt to construct the CL cascade that corresponds to this circuit as described in Section \ref{section:cascade_synthesis}. The turns ratio matrix $\vb{R}$ for this network has the form:
\begin{equation}
    \vb{R} = \sqrt{\frac{2}{C_T}} \mqty(1 & -1 \\ 1 & 1 \\ \vdots & \vdots)
\end{equation}
where the two rows shown in the matrix are repeated infinitely. The port-resonance block of the capacitance matrix is given by (\ref{eq:impedance_cap}):
\begin{equation}
    -\vb{R}_0^{-1}\vb{R}^T = \frac{1}{C_{\Sigma}}\sqrt{\frac{2}{C_T}} \mqty( -C_1 C_T & -2C_1C_2 - C_1C_T & \dots \\ -C_2C_T & 2C_1C_2 + C_2C_T & \dots )
\end{equation}
where now the two columns are repeated infinitely. If we take this Maxwell capacitance and convert it to the mutual capacitance matrix, the capacitances in the mutual matrix correspond to the capacitive elements in the cascade representation of this network. For this case, as we take into account more poles, the capacitances shunting the ports of the cascade network will diverge. We can see this for two different parameter regimes in Fig.\ \ref{fig:ideal_TL_shunt_cap}.

In general, these divergences in the elements of the cascade representation are not encountered since we will always work with a finite number of poles. As long as we choose the poles that are within the relevant frequency range of our device, this should not cause any problems. 

\newpage
\section{Capacitance Matrix Comparison Between Q3D and HFSS}\label{appendix:q3d_vs_hfss}

Tables containing the capacitance matrix comparisons between the Q3D and HFSS simulations of the model shown in Fig.\ \ref{fig:xmon_brick}.

\renewcommand{\arraystretch}{1.5}
\begin{table}[h!]
    \centering
    \begin{tabular}{|r|r|r|r|r|r|}
    \hline
    Island  & Xmon    & Control & CPW1    & CPW2    & CPW2    \\ \hline
    Xmon    & 107.677 & -0.161  & -4.591  & -4.599  & -4.591  \\ \hline
    Control & -0.161  & 205.059 & -0.172  & -0.020  & -0.024  \\ \hline
    CPW1    & -4.591  & -0.172  & 188.247 & -0.278  & -0.078  \\ \hline
    CPW2    & -4.599  & -0.020  & -0.278  & 188.096 & -0.274  \\ \hline
    CPW3    & -4.591  & -0.024  & -0.078  & -0.274  & 188.211 \\ \hline
    \end{tabular}
    \caption{Maxwell capacitance matrix in units of fF for the Q3D simulation.}
\end{table}

\begin{table}[h!]
    \centering
    \begin{tabular}{|r|r|r|r|r|r|}
    \hline
    Port    & Xmon    & Control & CPW1    & CPW2    & CPW2    \\ \hline
    Xmon    & 115.226 & -0.150  & -4.323  & -4.353  & -4.350  \\ \hline
    Control & -0.150  & 203.097 & -0.155  & -0.015  & -0.017  \\ \hline
    CPW1    & -4.323  & -0.155  & 181.583 & -0.233  & -0.061  \\ \hline
    CPW2    & -4.353  & -0.015  & -0.233  & 183.722 & -0.234  \\ \hline
    CPW3    & -4.350  & -0.017  & -0.061  & -0.234  & 184.450 \\ \hline
    \end{tabular}
    \caption{Maxwell capacitance matrix in units of fF extracted from the HFSS simulation using the fit of the impedance shown in Fig.\ \ref{fig:xmon_fit}.}
\end{table}

\begin{table}[h!]
    \centering
    \begin{tabular}{|r|r|r|r|r|r|}
    \hline
    Island/Port & Xmon   & Control & CPW1    & CPW2    & CPW2    \\ \hline
    Xmon        & 7.011 \% & -6.933 \%&     -5.857 \%  &  -5.356 \%  &  -5.246 \%  \\ \hline
    Control     & -6.933 \% & -0.957 \% &   -9.738 \%  & -24.805 \% &  -31.562 \% \\ \hline
    CPW1        & -5.857 \% & -9.738 \% &   -3.540 \%  & -16.344 \% &  -22.103 \% \\ \hline
    CPW2        & -5.356 \% & -24.805 \% & -16.344 \% &   -2.326 \%  & -14.519 \% \\ \hline
    CPW3        & -5.246 \% & -31.562 \% & -22.103 \% &  -14.519 \% &   -1.998 \%  \\ \hline
    \end{tabular}
    \caption{Percent increase/decrease in the Maxwell capacitance elements from HFSS compared to Q3D.}
\end{table}

\newpage
\section[Decay of Coupling Rates for Networks with Banded Capacitance Matrices]{Decay of Coupling Rates for Networks with\\ Banded Capacitance Matrices}\label{appendix:banded_cap}

\begin{figure}[h!]
    \centering
    \includegraphics[width=\textwidth]{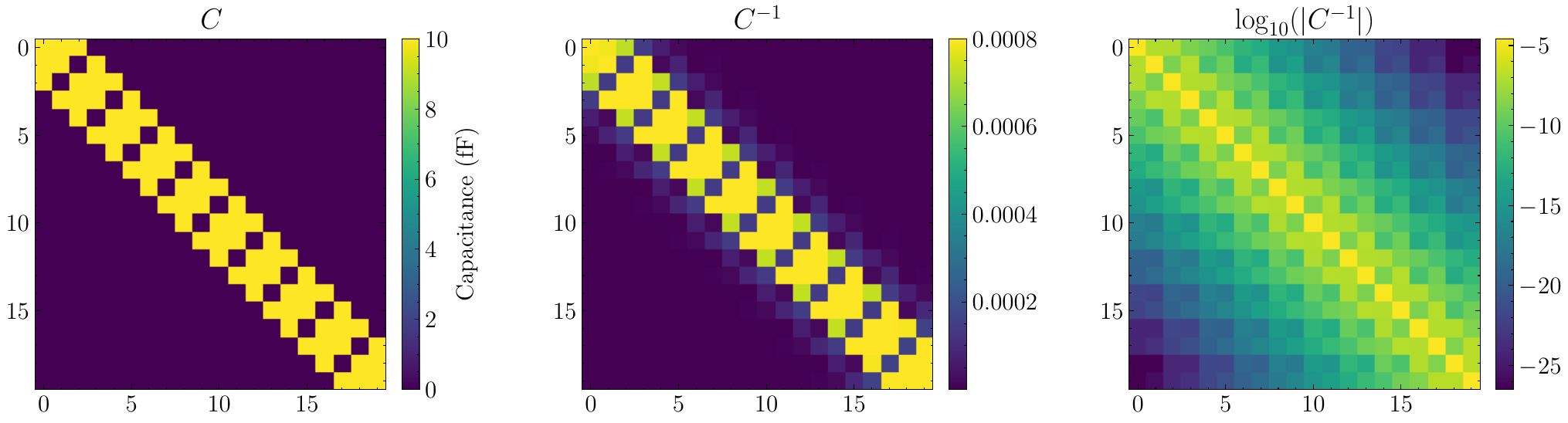}
    \caption{Capacitance matrix and its inverse of a $2\times 10$ grid of nodes. The ground shunt capacitances of the nodes are $C_s=$ 100 fF, and the coupling capacitance to nearest neighbor nodes is $C_c =$ 10 fF.}
    \label{fig:2x10_grid_cap}
\end{figure}

\begin{figure}[h!]
    \centering
    \includegraphics[width=\textwidth]{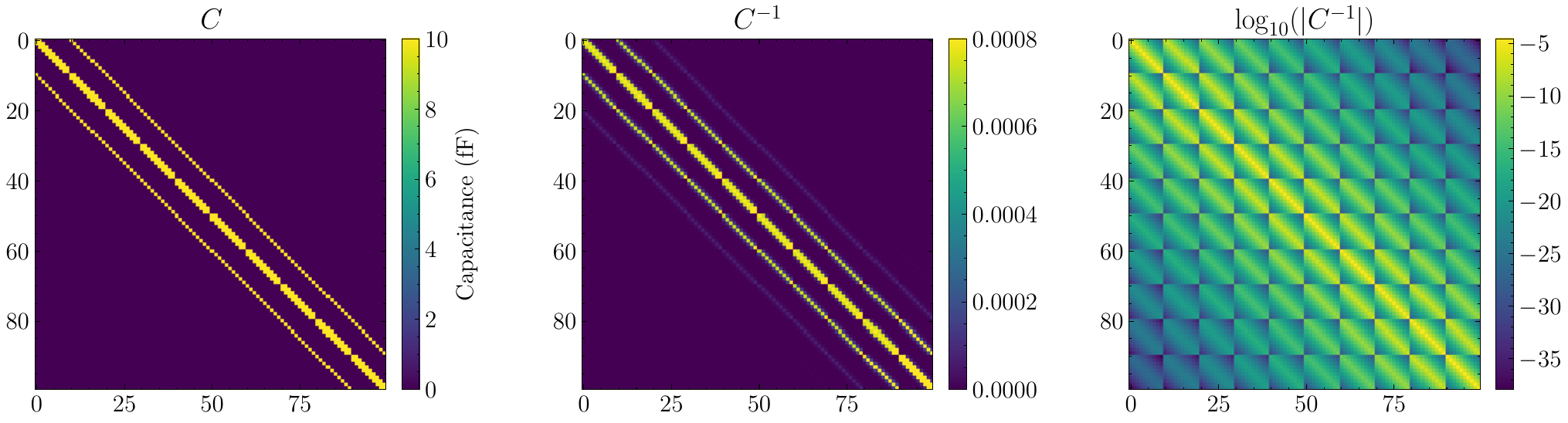}
    \caption{Same as Fig.\ \ref{fig:2x10_grid_cap} but now for a $10 \times 10$ grid of nodes.}
    \label{fig:10x10_grid_cap}
\end{figure}

Consider a rectangular $N \times M$ grid of nodes. Each node has it's own shunt capacitance to ground as well as capacitors connecting it to its nearest neighbors. Notably, the capacitance matrix for such a network is of size $NM \times NM$ and is banded with bandwidth of $k=\min\{N,M\}$. Banded matrices have the property that the matrix elements of the inverse decay exponentially to zero as you move further from the diagonal \cite{banded_1, banded_2}. Since the coupling rates between the nodes are dependent on the inverse of the capacitance matrix (\ref{eq:node_coupling}), we expect that as you move further from a node in this type of network, the coupling rates will decay exponentially. Two examples of this are shown in Fig.\ \ref{fig:2x10_grid_cap} and Fig.\ \ref{fig:10x10_grid_cap}. In reality, the presence of long range capacitive coupling will change this, and it should be taken into account when modeling grids of qubits.

